\documentclass[11pt,fleqn]{article}
\usepackage{amsmath,amssymb,amsthm,enumerate}
\usepackage[mathscr]{eucal}

\newcommand{\vspacebefore}{\raisebox{0ex}[2.5ex][0ex]{\null}}
\newcommand{\p}{\partial}
\newcommand{\const}{\mathop{\rm const}\nolimits}
\newcommand{\sign}{\mathop{\rm sign}\nolimits}
\newcommand{\rank}{\mathop{\rm rank}\nolimits}
\newcounter{tbn}
\newcommand{\cc}{\medskip\par\noindent\refstepcounter{tbn}\thetbn. }

\newtheorem{theorem}{Theorem}
\newtheorem{lemma}{Lemma}
\newtheorem{corollary}{Corollary}
\newtheorem{proposition}{Proposition}
{\theoremstyle{definition}
\newtheorem{definition}{Definition}
\newtheorem{example}{Example}

\newtheorem{note}{Note}
\newtheorem*{note*}{Note}
}

\begin{document}

\par\noindent {\LARGE\bf
Admissible Transformations and Normalized Classes\\ of Nonlinear Schr\"odinger Equations
\par}

{\vspace{5mm}\par\noindent {\it
Roman O. POPOVYCH~$^\dag$, Michael KUNZINGER~$^\ddag$ and Homayoon ESHRAGHI~$^\S$
} \par\vspace{2mm}\par}


{\vspace{2mm}\par\it\small
\noindent $^{\dag,\ddag}$Fakult\"at f\"ur Mathematik, Universit\"at Wien, Nordbergstra{\ss}e 15, A-1090 Wien, Austria
\par}

{\vspace{2mm}\par\noindent \it\small
$^\dag$~Institute of Mathematics of NAS of Ukraine, 3 Tereshchenkivska Str., Kyiv-4, Ukraine
 \par}

{\vspace{2mm}\par\noindent \it\small
$^\S$\,\,Physics Department, Iran University of Science and Technology (IUST), Narmak, Tehran, Iran,\\
$\phantom{^\S}$\,\,Institute for Studies in Theor. Physics and Mathematics (IPM), Tehran P.O.\,Box: 19395-5531, Iran
\par}

{\vspace{2mm}\par\noindent\small $\phantom{^{\dag,\S}}$\rm E-mail: \it
$^\dag$rop@imath.kiev.ua, $^\ddag$michael.kunzinger@univie.ac.at, $^\S$eshraghi@iust.ac.ir
\par}

{\vspace{6mm}\par\noindent\hspace*{10mm}\parbox{140mm}{\small
The theory of group classification of differential equations is analyzed, 
substantially extended and enhanced based
on the new notions of conditional equivalence group and normalized class of differential equations.
Effective new techniques are proposed. 
Using these, we exhaustively describe admissible point transformations
in classes of nonlinear $(1+1)$-dimensional Schr\"odinger equations, in particular,
in the class of nonlinear $(1+1)$-dimensional Schr\"odinger
equations with modular nonlinearities and potentials and some subclasses thereof.
We then carry out a complete group classification
in this class, representing it as a union of disjoint normalized subclasses and
applying a combination of algebraic and compatibility methods. Moreover, we
introduce the complete classification of $(1+2)$-dimensional cubic Schr\"odinger equations with potentials.
The proposed approach can be applied to studying symmetry properties of a wide range
of differential equations.
}\par\vspace{2mm}}

\section{Introduction}

Group classification is an efficient tool for choosing physically relevant models from parametrized classes of 
differential equations. 
In fact, invariance with respect to certain groups of transformations is a fundamental principle 
of many physical theories. 
Thus, all non-relativistic theories have to satisfy the Galilean principle of relativity, 
i.e., their underlying physical laws have to be invariant with respect to the Galilei group. 
In relativistic theories the Galilean principle is replaced by the principle of special relativity 
corresponding to invariance under the Poincar\'e group. 
Hence physical models are usually constrained by a priori requirements as to their symmetry properties. 
This naturally leads to the so-called \emph{inverse problems of group classification} 
for systems of differential equations, which are generally formulated in the following way: 
\emph{Given a group of point transformations in a space of a fixed number of dependent and independent variables, 
describe those systems of differential equations admitting the given group as their symmetry group.} 
The solution of such inverse problems typically involves methods of the theory of differential invariants. 

Differential equations modelling real world phenomena often contain numerical or functional parameters 
(so-called \emph{arbitrary elements})
which are determined experimentally and, therefore, are not fixed a priori. 
At the same time, the chosen models should be accessible to a systematic study. 
The symmetry approach offers a universal method for solving such nonlinear models. 
Moreover, a large symmetry group of a system of differential equations indicates 
the presence of further interesting properties of this system and paves the way
for applying other more specific methods adapted to the respective concrete problem. 
Hence invariance properties of a model can serve for assessing its feasibility and, moreover,  
the strategy of choosing, among all possible ones, the model displaying the maximal  
symmetry group can be used to single out the `right' 
values of the arbitrary elements in the modelling equations~\cite{Ovsiannikov1982}. 
Within the framework of group analysis of differential equations, this strategy is called the (direct) \emph{problem of group classification}.
Its general formulation is the following: 
\emph{Given a parametrized class of systems of differential equations 
in a space of a fixed number of dependent and independent variables, 
first determine the common symmetry group of all systems from the class
and then describe those systems admitting proper extensions of this
common group.} 
This description is carried out up to the equivalence relation generated by the equivalence group of 
the class. 
(A rigorous definition of the problem of group classification for a class of system of differential equations 
is given in Section~\ref{SectionOnGroupClassificationProblems}.)
Some problems of group analysis 
(e.g., the selection of equations from a parametrized class which admit realizations of a fixed Lie algebra)
display features of both direct and inverse group classification problems. 
A~classical example of solving a group classification problem is 
Lie's classification of second order ordinary differential equations~\cite{Lie1891} 
that allowed Lie to describe the equations from this class which can be integrated by quadratures. 

The group classification in a class of differential equations is a much more complicated problem 
than finding the Lie symmetry group of a single system of differential equation since 
it usually requires the integration of a complicated overdetermined system
of partial differential equations with respect to both the coefficients of the infinitesimal symmetry operators and the arbitrary elements 
(\emph{determining equations}).

Summing up, group classification of differential equations is an important tool for real world applications
and at the same time displays interesting mathematical structures in its own right.

Various methods for solving group classification problems have been developed over the past decades. 
The infinitesimal invariance criterion~\cite{Olver1986,Ovsiannikov1982} produces
overdetermined systems of determining equations for these problems. These typically
are analyzed with respect to compatibility and direct integrability
up to the equivalence relation generated by the corresponding equivalence group.
While this is the most common approach, its applicability is 
confined to classes of a relatively simple structure, e.g.\ 
those which have only few arbitrary elements of one or two arguments. 
A number of classification results obtained within the framework of this approach are 
collected in \cite{BlumanKumei1989,Ibragimov1994V12,Ovsiannikov1982}.

Another approach, more algebraic in character, focuses on the algebraic properties of 
the solution sets of the corresponding determining equations~\cite{Olver1986,Ovsiannikov1982}. 
Thus, for certain classes (now called normalized \cite{Popovych2006b,Popovych&Eshraghi2004Mogran}), 
the problem of group classification is reduced to the subgroup analysis of the 
corresponding equivalence groups.  

Before the notion of normalization was defined in a rigorous and precise form, it had been used implicitly for a long time.
The best known classical group classification problems such as Lie's classifications
of second order ordinary differential equations~\cite{Lie1891} and
of second order two-dimensional linear partial differential equations~\cite{Lie1881}
were solved essentially relying on the strong normalization of the above classes of differential equations.
Similar classification techniques implicitly based on properties of normalized classes were recently applied by a number of authors
to solving group classification problems of important classes of differential equations arising in physics and other sciences
(see e.g.~\cite{Abramenko&Lagno&Samojlenko2002,Gagnon&Winternitz1993,Lagno&Samojlenko2002,Lahno&Spichak2007,Lahno&Zhdanov2007,
Lahno&Zhdanov&Magda2006,Popovych&Ivanova&Eshraghi2004Gamma,Popovych&Ivanova&Eshraghi2004Cubic,Zhdanov&Lahno1999,Zhdanov&Roman2000}) 
although the underlying reason for the effectiveness of these techniques and the scope of their applicability long remained unclear. 
The so-called method of preliminary group classification involving subgroup analysis of equivalence groups and construction of differential invariants 
of the resulting inequivalent subgroups was independently developed~\cite{Akhatov&Gazizov&Ibragimov1989,Ibragimov&Torrisi&Valenti1991}. 
The question about when results of preliminary and complete group classifications coincide also remained open. 

In the most advanced versions, the above approaches are combined with each other. 
To extend both their range of applicability and their efficiency, different notions and techniques
(extended and generalized equivalence groups, additional equivalence transformations, conditional equivalence group, 
gauging of arbitrary elements by equivalence transformations, partition of a class to normalized subclasses etc.)
were introduced
\cite{Ibragimov1994V12,Meleshko1994,Popovych2006b,Popovych&Ivanova2003a,
Vaneeva&Johnpillai&Popovych&Sophocleous2006,Vaneeva&Popovych&Sophocleous2007}.
  
The purpose of this paper is twofold. First, we develop a systematic new 
approach to group classification problems 
which is based on the central notion of normalized classes of differential equations. 
To this end we build on the first presentation of this approach in \cite{Popovych2006b,Popovych&Eshraghi2004Mogran}.
The corresponding elements of the framework of group classification are analyzed, starting 
with the basic notion of classes of differential equations. 
A rigorous definition of sets of admissible point transformations is given. 
Special attention is paid to the recently proposed notion of conditional equivalence groups 
and its applications in different classification problems for classes of 
differential equations.
A number of relevant notions (point-transformation image of a class, maximal conditional equivalence group, 
maximal normalized subclass, etc.) are introduced here for the first time. 

Our second goal is to give a nontrivial example for the application of this new framework. 
For this purpose we choose the class~$\mathscr F$ of $(1+1)$-dimensional nonlinear Schr\"odinger 
equations of the general form 
\begin{equation}\label{vgNSchE}
i\psi_t+\psi_{xx}+F=0,
\end{equation}
where $\psi$ is a complex dependent variable of the two real independent variables~$t$ and~$x$ and 
$F=F(t,x,\psi,\psi^*,\psi_x,\psi^*_x)$ is an arbitrary smooth complex-valued function of its arguments. 
(The coefficient of~$\psi_{xx}$ is assumed to be scaled to~1.)
It is proved that this class is normalized, and a hierarchy of nested normalized subclasses of the class~$\mathscr F$ is constructed. 
Then  we describe the admissible point transformations
in the class~$\mathscr V$ of $(1+1)$-dimensional nonlinear Schr\"odinger equations with modular nonlinearities and potentials, 
i.e., the equations with $F=f(|\psi|)\psi+V\psi$,
where $f$ is an arbitrary complex-valued nonlinearity depending only on $\rho=|\psi|$, $f_\rho\not=0$, and
$V$~is an arbitrary complex-valued potential depending on $t$ and $x$.
Using the techniques proposed in this paper, we also carry out the complete group classification for the class~$\mathscr V$. 
This encompasses and extends the results of \cite{Ivanova&Popovych&Eshraghi2005,Popovych&Eshraghi2004Mogran,Popovych2006a,
Popovych&Ivanova&Eshraghi2004Cubic,Popovych&Ivanova&Eshraghi2004Gamma} on group classification in subclasses of the class~$\mathscr V$.

To demonstrate the effectiveness of these new techniques also in the multidimensional case, 
we consider the normalized class~$\mathscr C$ of $(1+2)$-dimensional cubic Schr\"odinger 
equations with potentials of the general form
\begin{equation}\label{CSchEP12D}
i\psi_t+\psi_{11}+\psi_{22}+|\psi|^2\psi+V(t,x)\psi=0.
\end{equation}
Here $\psi$ is a complex dependent variable of the real independent variables~$t$ and~$x=(x_1,x_2)$ 
and $V$~is an arbitrary complex-valued potential depending on $t$ and $x$. 
The problem of group classification is solved for the class~$\mathscr C$ by relying on
the normalization property of this class and its subclasses.

Our paper is organized as follows. 
The first substantial part of this work 
(Sections~\ref{SectionOnPointTransformationsInClassesOfDifferentialEquations} and~\ref{SectionOnNormalizedClasses})
is devoted to the theoretical foundations of symmetry analysis in classes of (systems of) differential equations in general.
First of all, in Section~\ref{SectionOnPointTransformationsInClassesOfDifferentialEquations} 
we analyze notions and objects pertinent to the framework of group classification 
(classes of differential equations and their properties, admissible transformations, 
different kinds of equivalence and symmetry groups, etc.). 
The classical formulation of group classification problems is presented in a rigorous way and  
a number of possibilities for modifications and extensions are indicated.  
New notions associated with classes of differential equations
naturally arise under these considerations, in particular as analogues of corresponding
notions for single systems of differential equations.
Thus, similarity of equations is extended to similarity and point-transformation imaging of classes of equations. 
The notion of conditional symmetry groups is a motivation for introducing the notion of conditional equivalence groups.
Normalized classes of differential equations are studied in Section~\ref{SectionOnNormalizedClasses}. 
We give rigorous definitions for (several variants of) normalized classes and provide examples for these notions. 
We also derive results which are necessary for concrete applications of normalization in group classification 
and for the determination of admissible transformations.

The class~$\mathscr F$ of $(1+1)$-dimensional nonlinear Schr\"odinger equations of the general form~\eqref{vgNSchE} is the main subject of 
the second part of the paper (Sections~\ref{SectionOnKnownResultsOnLieSymmetriesOfNSchEs}--\ref{SectionOnApplicationsOfEquivTransForNschEs}).
Known results on Lie symmetries of nonlinear Schr\"odinger equations are reviewed in Section~\ref{SectionOnKnownResultsOnLieSymmetriesOfNSchEs}.
The strong normalization of the class~$\mathscr F$ is proved in Section~\ref{SectionOnNestedNormalizedClassesOfNSchEs}. 
Ibidem we single out two important nested normalized subclasses of~$\mathscr F$. 
The smaller subclass~$\mathscr S$ still contains the class~$\mathscr V$ of 
nonlinear $(1+1)$-dimensional Schr\"odinger equations with modular nonlinearities and potentials.  
That is why normalization properties of~$\mathscr S$ are significant for the group classification in the class~$\mathscr V$, 
carried out in Section~\ref{SectionOnGroupClassificationOfNSchEsWithMNP}. 
Since the class~$\mathscr V$ is not normalized, for completing the classification we partition it into a family 
of normalized subclasses and then classify each of these subclasses separately. 
In order to demonstrate the effectiveness of the approach based on normalization properties 
in the case of more than two independent variables, 
in Section~\ref{SectionOnGroupClassificationOfCSchEP12D} we additionally carry out the group classification of the class~$\mathscr C$ 
of $(1+2)$-dimensional cubic Schr\"odinger  equations with potentials of the general form~\eqref{CSchEP12D}.
To point out possible applications, in Section~\ref{SectionOnApplicationsOfEquivTransForNschEs} 
the results of the second part of the paper are connected with previous results from the literature.

\section{Point transformations in classes of differential equations}\label{SectionOnPointTransformationsInClassesOfDifferentialEquations}

In this section we provide the necessary background for a rigorous exposition of the notion of normalized classes of differential equations 
in Section~\ref{SectionOnNormalizedClasses}.
A number of notions of group analysis are revised.
We start by analyzing the basic notion of a general class (of systems) of  differential equations.
Then the notion of point transformations in such classes is formalized and different types of equivalence groups are defined.
The problem of group classification in a class of differential equations
as well as the problem of classification of admissible transformations are rigorously stated.
This forms the basis for the development of group analysis methods based on the notion of normalized classes of differential equations 
and for the further investigation of classes of nonlinear Schr\"odinger equations.

The following conventions will be active throughout the paper: all transformations are assumed to act 
from the left-hand side. The term `point transformation group' is used as an abbreviation
for `local (pseudo)group of locally defined point transformations'. 
We do not explicitly indicate mapping domains and suppose that all 
functions are sufficiently smooth (as a rule, real analytic) on suitable open sets.

\subsection{Classes of systems of differential equations}

Let~$\mathcal{L}_\theta$ be a system
$
L(x,u_{(p)},\theta(x,u_{(p)}))=0
$ 
of $l$~differential equations 
for $m$~unknown functions $u=(u^1,\ldots,u^m)$
of $n$~independent variables $x=(x_1,\ldots,x_n).$
Here $u_{( p)}$ denotes the set of all the derivatives of~$u$ with respect to $x$
of order not greater than~$ p$, including $u$ as the derivatives of order zero.
$L=(L^1,\ldots,L^l)$ is a tuple of $l$ fixed functions depending on~$x,$ $u_{( p)}$ and~$\theta$.
$\theta$~denotes the tuple of arbitrary (parametric) functions
$\theta(x,u_{(n)})=(\theta^1(x,u_{( p)}),\ldots,\theta^k(x,u_{( p)}))$
running through the set~$\mathcal S$ of solutions of the auxiliary system
$ 
S(x,u_{( p)},\theta_{(q)}(x,u_{( p)}))=0.
$ 
This system consists of differential equations with respect to $\theta$,
where $x$ and $u_{( p)}$ play the role of independent variables
and $\theta_{(q)}$ stands for the set of all the partial derivatives of $\theta$ of order not greater than $q$
with respect to the variables $x$ and $u_{( p)}$.
Often the set $\mathcal S$ is additionally constrained by the non-vanish condition
$\Sigma(x,u_{( p)},\theta_{(q)}(x,u_{( p)}))\ne0$ with another tuple $\Sigma$ of differential functions.
(This inequality means that no components of~$\Sigma$ equal 0. For simplicity the tuple $\Sigma$ can be
replaced by a single differential function coinciding with the product of its components.)
In what follows we call the functions $\theta$ arbitrary elements.
Also, we denote \emph{the class of systems~$\mathcal{L}_\theta$ with the arbitrary elements $\theta$ running through $\mathcal S$}
as~$\mathcal L|_{\mathcal S}$.

Let $\mathcal{L}_\theta^i$ denote the set of all algebraically independent differential consequences of $\mathcal{L}_\theta$,
which have, as differential equations, orders not greater than~$i$. We identify~$\mathcal{L}_\theta^i$ with
the manifold determined by~$\mathcal{L}_\theta^i$ in the jet space~$J^{(i)}$.
In particular, $\mathcal{L}_\theta$ is identified with the manifold determined by~$\mathcal{L}_\theta^p$ in~$J^{(p)}$.
Then $\mathcal L|_{\mathcal S}$ can be interpreted as a family of manifolds in~$J^{(p)}$, parametrized with
the arbitrary elements $\theta\in\mathcal S$.

It should be noted that the above definition of a class of systems of differential equations is not complete.
The problem is that the correspondence $\theta\to\mathcal{L}_\theta$
between arbitrary elements and systems (treated not as formal algebraic expressions
but as real systems of differential equations or manifolds in~$J^{(p)}$) may not be one-to-one.
Namely, the same system may correspond to different values of arbitrary elements.
A reason for this indeterminacy is that different values $\theta$ and $\tilde\theta$ of arbitrary
elements can result
after substitution into $L$ in the same expression in $x$ and $u_{( p)}$.
Moreover, it is enough for $\mathcal{L}_\theta^p$ and $\mathcal{L}_{\tilde\theta}^p$ to coincide if
the associated systems completed with independent differential consequences differ from one another
by a nonsingular matrix being a function in the variables of~$J^{(p)}$.

The values $\theta$ and $\tilde\theta$ of arbitrary elements are called \emph{gauge-equivalent}
($\theta\mathrel{\smash{\stackrel{\mathrm{g}}{\sim}}}\tilde\theta$)
if $\mathcal{L}_\theta$ and $\mathcal{L}_{\tilde\theta}$ are the same system of differential equations.
For the correspondence $\theta\to\mathcal{L}_\theta$ to be one-to-one,
the set $\mathcal S$ of arbitrary elements should be factorized with respect to the gauge equivalence relation.
We formally consider $\mathcal{L}_\theta$ and $\mathcal{L}_{\tilde\theta}$  as different representations
of the same system from $\mathcal L|_{\mathcal S}$.
It is often possible to realize gauge informally via changing the chosen representation of the class
under consideration (changing the number $k$ of arbitrary elements and the differential functions $L$ and $S$
although this may result in more complicated calculations).

\begin{definition}\label{DefOnSimilarClasses}
The classes $\mathcal L|_{\mathcal S}$ and $\mathcal{L}'|_{\mathcal S'}$ are called \emph{similar} if
$n=n'$, $m=m'$, $p=p'$,  $k=k'$ and
there exists a point transformation $\Psi\colon(x,u_{(p)},\theta)\to(x',u'_{(p)},\theta')$
which is projectable on the space of $(x,u_{(q)})$ for any $0\le q\le p$,
and $\Psi|_{(x,u_{(q)})}$ being the $q$-th order prolongation of $\Psi|_{(x,u)}$,
$\Psi \mathcal S=\mathcal S'$
and $\Psi|_{(x,u_{(p)})}\mathcal{L}_\theta=\mathcal{L}'_{\Psi\theta}$ for any $\theta\in\mathcal S$.
\end{definition}

Here and in what follows the action of such a point transformation $\Psi$ in the space of $(x,u_{(p)},\theta)$ on arbitrary elements from $\mathcal S$
as $p$th-order differential functions is given by the formula:
\[
\tilde\theta=\Psi\theta \quad\mbox{if}\quad
\tilde\theta(x,u_{(p)})=\Psi^\theta\smash{\Bigl(\Theta(x,u_{(p)}),\theta\bigl(\Theta(x,u_{(p)})\bigr)\Bigr)},
\]
where $\Theta=(\mathrm{pr}_p\Psi|_{(x,u)})^{-1}$ and
$\mathrm{pr}_p$ denotes the operation of standard prolongation of a point transformation
to the derivatives of orders not greater than~$p$.

Definition~\ref{DefOnSimilarClasses} is an extension of the notion of similar differential equations~\cite{Ovsiannikov1982} to 
classes of such equations. Moreover, similar classes consist of similar equations with the same similarity transformation. 

The set of transformations used in Definition~\ref{DefOnSimilarClasses}
can be extended via admitting different kinds of dependence on arbitrary elements
as in the case of equivalence groups below.
As a rule, similar classes of systems have similar properties from the group analysis point of view.
In the case of point similarity transformations, the properties really are the same up to similarity.
If~$\Psi$ is a point transformation in the space of $(x,u_{(p)},\theta)$ then these classes practically have the same transformational properties.

If the transformation~$\Psi$ is identical with respect to $x$ and $u$ then
$\mathcal{L}'_{\Psi\theta}=\mathcal{L}_\theta$ for any $\theta\in\mathcal S$, i.e., in fact
the classes $\mathcal L|_{\mathcal S}$ and $\mathcal{L}'|_{\mathcal S'}$ coincide as sets of manifolds in a jet space.
We will say that the class $\mathcal{L}'|_{\mathcal S'}$ is a \emph{re-parametrization} of the class $\mathcal L|_{\mathcal S}$,
associated with the re-parametrizing transformation~$\Psi$.
In the most general approach, $\Psi$ can be assumed an arbitrary one-to-one mapping from~$\mathcal S$ to $\mathcal S'$,
satisfying the condition $\mathcal{L}'_{\Psi\theta}=\mathcal{L}_\theta$ for any $\theta\in\mathcal S$.
Note that the number of arbitrary elements in $\mathcal S'$ might not coincide with the one in~$\mathcal S$.
Transformational properties may be broken under generalized re-parametrizations.

An example of non-point re-parametrization often applied in group analysis is given
by the classes $\{\mathcal I=\theta(\mathcal I',\mathcal J)\}$ and $\{\mathcal I'=\theta'(\mathcal I,\mathcal J)\}$,
where $\mathcal I$ and $\mathcal I'$ (resp.\ $\mathcal J$) are $k$-tuples (resp.\ an $l$-tuple) 
of fixed functionally independent expressions of~$x$ and~$u_{(p)}$, 
$\theta$ and $\theta'$ are arbitrary $(k+l)$-ary $k$-vector functions having nonzero Jacobians with respect first $k$ arguments, 
$k,l\in\mathbb N$.
The corresponding mapping between the sets of arbitrary elements is to take the inverse function to each set element.
Fortunately, such re-parametrization preserves transformational properties of classes well.

Similarity of classes implies a one-to-one correspondence between the associated sets of arbitrary elements. 
If this feature is neglected, we arrive at the more general notion of mapping between classes of differential equations. 

\begin{definition}\label{DefOnImageClasses}
The class $\mathcal{L}'|_{\mathcal S'}$ is called a \emph{point-transformation image} of 
the class $\mathcal L|_{\mathcal S}$ if
$n=n'$, $m=m'$, $p=p'$,  $k=k'$ and
there exists a family of point transformations $\varphi_\theta\colon(x,u)\to(x',u')$ parametrized by $\theta\in\mathcal S$ 
and satisfying the following conditions. 
For any $\theta\in\mathcal S$ there exists $\theta'\in\mathcal S'$ 
and, conversely, for any $\theta'\in\mathcal S'$ there exists $\theta\in\mathcal S$
such that $\mathrm{pr}_p\varphi_\theta\mathcal{L}_\theta=\mathcal{L}'_{\theta'}$. 
\end{definition}

A point-transformation image inherits certain transformational properties from its class-preimage. 
There is also a converse connection. 
For example, equations from the class-preimage are point-transformation equivalent iff 
their images are.

\emph{Subclasses} are singled out in the class $\mathcal L|_{\mathcal S}$
with additional auxiliary systems of equations and/or non-vanish conditions
which are attached to the main auxiliary system.

The intersection of a finite number of subclasses of $\mathcal L|_{\mathcal S}$
always is a subclass of $\mathcal L|_{\mathcal S}$:
\[
\mathcal L|_{\mathcal S^1\!}\cap\dots\cap\mathcal L|_{\mathcal S^j\!}=\mathcal L|_{\mathcal S^\cap}, \quad
\mathcal S^1,\dots,\mathcal S^k\subset \mathcal S, \quad \mathcal S^\cap=\mathcal S^1\cap\dots\cap\mathcal S^j.
\]
The additional auxiliary condition for the intersection~$\mathcal S^\cap$ is the system
consisting of all the additional auxiliary conditions for the intersecting sets $\mathcal S^1$, \dots, $\mathcal S^j$.

The situation concerning complements, unions and differences of subclasses of $\mathcal L|_{\mathcal S}$ is more complicated.
They will be subclasses of $\mathcal L|_{\mathcal S}$ in the above defined sense
only under special restrictions on the additional auxiliary conditions.
These difficulties arise from the simultaneous consideration of equations and 
inequalities as subclass constraints.

Thus, the complement $\overline{\mathcal L|_{\mathcal S'\!}}=\mathcal L|_{\overline{\mathcal S'\!}\,}$
of the subclass $\mathcal L|_{\mathcal S'\!}$ in the class $\mathcal L|_{\mathcal S}$ also
is a subclass of $\mathcal L|_{\mathcal S}$ if
the additional system of equations or the additional system of inequalities is empty.
If $\mathcal S'$ is determined by the system $S'_1=0$, \dots, $S'_{s'}=0$
then the additional auxiliary condition for $\overline{\mathcal S'}$ is $|S'_1|^2+\dots+|S'_{s'}|^2\ne0$.
If $\mathcal S'$ is defined by the inequalities  $\Sigma'_1\ne0$, \dots, $\Sigma'_{\sigma'\!}\ne0$
then the additional auxiliary condition for $\overline{\mathcal S'}$ is $\Sigma'_1\cdots\Sigma'_{\sigma'\!}=0$.

The union  $\mathcal L|_{\mathcal S'\!}\cup\mathcal L|_{\mathcal S''\!}=\mathcal L|_{\mathcal S'\cup\mathcal S''}$
and the difference $\mathcal L|_{\mathcal S'\!}\setminus\mathcal L|_{\mathcal S''\!}=\mathcal L|_{\mathcal S'\setminus\mathcal S''}$
of the subclasses $\mathcal L|_{\mathcal S'\!}$ and $\mathcal L|_{\mathcal S''\!}$ in the class $\mathcal L|_{\mathcal S}$ also
are subclasses of $\mathcal L|_{\mathcal S}$ if
the additional systems of equations or the additional system of inequalities of these subclasses coincide.
The corresponding additional auxiliary conditions are respectively constructed from the additional auxiliary conditions for
$\mathcal S'$ ($S'_1=0$, \dots, $S'_{s'\!}=0$, $\Sigma'_1\ne0$, \dots, $\Sigma'_{\sigma'\!}\ne0$) and
$\mathcal S''$ ($S''_1=0$, \dots, $S''_{s''\!}=0$, $\Sigma''_1\ne0$, \dots, $\Sigma''_{\sigma''\!}\ne0$)
in the following way (the coinciding part is preserved):
\begin{gather*}
\mathcal S'\cup\mathcal S''\colon\quad
|\Sigma'_1\cdots\Sigma'_{\sigma'\!}|^2+|\Sigma''_1\cdots\Sigma''_{\sigma''\!}|^2\ne0
\quad\mbox{resp.}\quad
S'_iS''_j=0,\ i=1,\dots,s'\!,\ j=1,\dots,s''\!,
\\
\mathcal S'\setminus\mathcal S''\colon\quad
\Sigma'_1\cdots\Sigma'_{\sigma'\!}\ne0,\ \Sigma''_1\cdots\Sigma''_{\sigma''\!}=0
\qquad\!\mbox{resp.}\quad
S'=0,\ |S''_1|^2+\dots+|S''_{s''\!}|^2\ne0.
\end{gather*}

For sets of subclasses to be closed with respect to all set operations, the notion of subclasses should be extended with
assuming that the union of any finite number of subclasses of the class $\mathcal L|_{\mathcal S}$
also is a subclass of $\mathcal L|_{\mathcal S}$.
It is not yet understood whether such an extension is needed for the purposes of group analysis.

\subsection{Admissible transformations}

For $\theta,\tilde\theta\in\mathcal S$ we call the set of point transformations
which map the system~$\mathcal{L}_\theta$ into the system~$\mathcal{L}_{\tilde\theta}$
the \emph{set of admissible transformations from~$\mathcal{L}_\theta$ into~$\mathcal{L}_{\tilde\theta}$}
and denote it by $\mathrm{T}(\theta,\tilde\theta)$.
The maximal point symmetry group~$G_\theta$ of the system~$\mathcal{L}_\theta$
coincides with~$\mathrm{T}(\theta,\theta)$.
If the systems~$\mathcal{L}_\theta$ and $\mathcal{L}_{\tilde\theta}$ are equivalent with respect to
point transformations then
$\mathrm{T}(\theta,\tilde\theta)=\varphi^0\circ G_\theta=G_{\tilde\theta}\circ\varphi^0$,
where $\varphi^0$ is a fixed transformation from~$\mathrm{T}(\theta,\tilde\theta)$.
Otherwise, $\mathrm{T}(\theta,\tilde\theta)=\varnothing$.

Analogously, the set
$\mathrm{T}(\theta,\mathcal L|_{\mathcal S})=\{\,(\tilde\theta,\varphi)\mid
\tilde\theta\in\mathcal S,\,
\varphi\in\mathrm{T}(\theta,\tilde\theta)\, \}$
is called the {\em set of admissible transformations of the system~$\mathcal{L}_\theta$
into the class~$\mathcal L|_{\mathcal S}$}.

\begin{definition}\label{DefOfSetOfAdmTrans}
$\mathrm{T}(\mathcal L|_{\mathcal S})=\{(\theta,\tilde\theta,\varphi)\mid
\theta,\tilde\theta\in\mathcal S,\,
\varphi\in\mathrm{T}(\theta,\tilde\theta)\}$
is called the {\em set of admissible transformations in~$\mathcal L|_{\mathcal S}$}.
\end{definition}

\begin{note}
The set of admissible transformations was first described by Kingston and Sophocleous
for a class of generalized Burgers equations~\cite{Kingston&Sophocleous1991}. 
These authors call transformations of such type {\em form-preserving}
\cite{Kingston&Sophocleous1991,Kingston&Sophocleous1998,Kingston&Sophocleous2001}.
The notion of admissible transformations can be considered as a formalization of their approach.
\end{note}

\begin{note}
Notions and results obtained in this and the following sections can be reformulated in infinitesimal terms
by using the notions of vector fields, Lie algebras instead of point transformations, Lie groups etc.
For instance, see~\cite{Borovskikh2004} for the definition of ``cones of tangent equivalences'', which is
the infinitesimal analogue of the definition of~$\mathrm{T}(\theta,\mathcal L|_{\mathcal S})$.
Ibidem a non-trivial example of semi-normalized classes of differential equations (see Definition~\ref{DefOfSemiNormalizedClasses})
is investigated in the framework of the infinitesimal approach.
\end{note}

\begin{note}
In the case of one dependent variable ($m=1$) we can extend the previous and the subsequent notions to
contact transformations.
\end{note}

An element $(\theta,\tilde\theta,\varphi)$ from $\mathrm{T}(\mathcal L|_{\mathcal S})$ is called
\emph{a gauge admissible transformation in~$\mathcal L|_{\mathcal S}$}
if $\theta\mathrel{\smash{\stackrel{\mathrm{g}}{\sim}}}\tilde\theta$ and $\varphi$ is the identical transformation.

\begin{proposition}
Similar classes have similar sets of admissible transformations.
Namely, a similarity transformation $\Psi$ from the class $\mathcal L|_{\mathcal S}$ into the class $\mathcal{L}'|_{\mathcal S'}$
generates a one-to-one mapping $\Psi^\mathrm{T}$
from $\mathrm{T}(\mathcal L|_{\mathcal S})$ into $\mathrm{T}(\mathcal{L}'|_{\mathcal S'})$
via the rule $(\theta'\!,\tilde\theta'\!,\varphi')=\Psi^\mathrm{T}(\theta,\tilde\theta,\varphi)$
if $\theta'=\Psi\theta$, $\tilde\theta'=\Psi\tilde\theta$ and
$\varphi'=\Psi|_{(x,u)}\circ\varphi\circ\Psi|_{(x,u)}{}^{-1}$.
Here $(\theta,\tilde\theta,\varphi)\in\mathrm{T}(\mathcal L|_{\mathcal S})$,
$(\theta'\!,\tilde\theta'\!,\varphi')\in\mathrm{T}(\mathcal{L}'|_{\mathcal S'})$.
\end{proposition}

\begin{proposition}
A point-transformation mapping between classes of differential equations induces a mapping between 
the corresponding sets of admissible transformations. Namely, 
if the class $\mathcal{L}'|_{\mathcal S'}$ is the point-transformation image of 
the class $\mathcal L|_{\mathcal S}$ under the family of point transformations 
$\varphi_\theta\colon(x,u)\to(x',u')$, $\theta\in\mathcal S$, then
the image of any $(\theta,\tilde\theta,\varphi)\in\mathrm{T}(\mathcal L|_{\mathcal S})$ is 
$(\theta'\!,\tilde\theta'\!,\varphi')\in\mathrm{T}(\mathcal{L}'|_{\mathcal S'})$, where 
$\mathcal{L}'_{\theta'}=\mathrm{pr}_p\varphi_\theta\mathcal{L}_\theta$,  
$\mathcal{L}'_{\tilde\theta'}=\mathrm{pr}_p\varphi_{\tilde\theta}\mathcal{L}_\theta$ and   
$\varphi'=\varphi_{\tilde\theta}\circ\varphi\circ(\varphi_\theta)^{-1}$.
\end{proposition}

Moreover, the set of admissible transformations of the initial class 
$\mathcal L|_{\mathcal S}$ is reconstructed from 
the one of its point-transformation image $\mathcal{L}'|_{\mathcal S'}$. 
Indeed, let $(\theta'\!,\tilde\theta'\!,\varphi')\in\mathrm{T}(\mathcal{L}'|_{\mathcal S'})$ and let
$\mathcal{L}_\theta$ and $\mathcal{L}_{\tilde\theta}$ be some equations mapped 
to $\mathcal{L}'_{\theta'}$ and $\smash{\mathcal{L}'_{\tilde\theta'}}$, respectively.  
Then $(\theta,\tilde\theta,\varphi)\in\mathrm{T}(\mathcal L|_{\mathcal S})$, where 
$\varphi=(\varphi_{\tilde\theta})^{-1}\circ\varphi\circ\varphi_\theta$. 
Each admissible transformation of $\mathcal L|_{\mathcal S}$ is obtainable in the above way.

\begin{proposition}
$\mathrm{T}(\mathcal L|_{\mathcal S'\!})\subset\mathrm{T}(\mathcal L|_{\mathcal S})$ for
any subclass~$\mathcal L|_{\mathcal S'\!}$ of the class~$\mathcal L|_{\mathcal S}$.
If $\mathcal L|_{\mathcal S''\!}$ is another subclass of~$\mathcal L|_{\mathcal S}$ then
$\mathrm{T}(\mathcal L|_{\mathcal S'\!})\cap\mathrm{T}(\mathcal L|_{\mathcal S''\!})=
\mathrm{T}(\mathcal L|_{\mathcal S'\cap\mathcal S''\!})$.
\end{proposition}

A number of notions connected with admissible transformations in classes 
of differential equations can be reformulated in terms of  category theory~\cite{Prokhorova2005}.
Note that in addition to admissible transformations,
the categories of parabolic partial differential equations constructed in~\cite{Prokhorova2005}
also include reduction mappings between equations with different numbers of independent variables.

\subsection{Equivalence groups}

Equivalence groups of different kinds, acting on classes of differential equations, are defined in a rigorous way
via the notion of admissible transformations.

Thus, any element~$\Phi$ from the \emph{usual equivalence group} $G^{\sim}=G^{\sim}(\mathcal L|_{\mathcal S})$
of the class~$\mathcal L|_{\mathcal S}$
is a point transformation in the space of $(x,u_{( p)},\theta)$,
which is projectable on the space of $(x,u_{( p')})$ for any $0\le p'\le p$,
so that $\Phi|_{(x,u_{( p')})}$ is the $ p'$-th order prolongation of $\Phi|_{(x,u)}$,
 $\forall\theta\in\mathcal S$: $\Phi\theta\in\mathcal S$,
and $\Phi|_{(x,u)}\in\mathrm{T}(\theta,\Phi\theta)$.
The admissible transformations of the form $(\theta,\Phi\theta,\Phi|_{(x,u)})$,
where $\theta\in\mathcal S$ and $\Phi\in G^\sim$, are called induced by the transformations from
the equivalence group $G^{\sim}$.

Let us recall that
a point transformation~$\varphi$: $\tilde z=\varphi(z)$ in the space of the variables $z=(z_1,\ldots,z_k)$
is called projectable on the space of the variables $z'=(z_{i_1},\ldots,z_{i_{k'}})$,
where $1\le i_1<\cdots<i_{k'}\le k$,
if the expressions for~$\tilde z'$ depend only on~$z'$.
We denote the restriction of~$\varphi$ to the $z'$-space as $\varphi|_{z'}$: $\tilde z'=\varphi|_{z'}(z')$.

If the arbitrary elements~$\theta$ explicitly depend on $x$ and $u$ only (one can always achieve this formally,
introducing derivatives as new dependent variables), we can admit dependence of transformations of~$(x,u)$ on~$\theta$ and consider
the \emph{generalized equivalence group} $G^{\sim}_{\rm gen}=G^{\sim}_{\rm gen}(\mathcal L|_{\mathcal S})$
\cite{Meleshko1994}.
Any element~$\Phi$ from~$G^{\sim}_{\rm gen}$
is a point transformation in $(x,u,\theta)$-space
such that $\forall\theta\in\mathcal S$: $\Phi\theta\in\mathcal S$
and $\Phi(\cdot,\cdot,\theta(\cdot,\cdot))|_{(x,u)}\in\mathrm{T}(\theta,\Phi\theta)$.

The action of $\Phi\in G^{\sim}_{\rm gen}$ on arbitrary elements 
as functions of $(x,u)$ is given by the formula:
$\tilde\theta=\Phi\theta$ if
$\tilde\theta(x,u)=\Phi^\theta(\Theta(x,u),\theta(\Theta(x,u))),$
where $\Theta=(\Phi(\cdot,\cdot,\theta(\cdot,\cdot))|_{(x,u)})^{-1}$.

Roughly speaking, $G^{\sim}$ is the set of admissible transformations
which can be applied to any~$\theta\in\mathcal S$
and~$G^{\sim}_{\rm gen}$ is formed by the admissible transformations
which can be separated to classes parametrized with $\theta$ running through $\mathcal S$.

It is possible to consider other generalizations of equivalence groups, e.g. groups with
transformations which are point transformations with respect to independent and dependent variables and
include nonlocal expressions with arbitrary elements
\cite{Ivanova&Popovych&Sophocleous2004,Vaneeva&Johnpillai&Popovych&Sophocleous2006}.
Let us give the definitions of some generalizations.

\begin{definition}\label{DefOfExtendedEquivalenceGroup}
The \emph{extended equivalence group} $\bar G^{\sim}=\bar G^{\sim}(\mathcal L|_{\mathcal S})$
of the class~$\mathcal L|_{\mathcal S}$
is formed by the transformations each of which is represented by the pair $\Phi=(\check\Phi,\hat\Phi)$.
Here $\check\Phi$ is a one-to-one mapping in the space of arbitrary elements assumed 
as functions of $(x,u_{(p)})$ and 
$\hat\Phi=\Phi|_{(x,u)}$ is a point transformation of $(x,u)$ belonging to $\mathrm T(\theta,\check\Phi\theta)$ 
for any $\theta$ from $\mathcal S$. 
\end{definition}

\begin{definition}\label{DefOfExtendedGenEquivalenceGroup}
The \emph{extended generalized equivalence group} $\bar G^{\sim}_{\rm gen}=\bar G^{\sim}_{\rm gen}(\mathcal L|_{\mathcal S})$
of the class~$\mathcal L|_{\mathcal S}$ consists of transformations each of which is represented by the tuple  
$\Phi=(\check\Phi,\{\hat\Phi^\theta,\theta\!\in\!\mathcal S\})$.
Here $\check\Phi$ is a one-to-one mapping in the space of arbitrary elements assumed 
as functions of $(x,u_{(p)})$ and 
for any $\theta$ from $\mathcal S$ the element $\smash{\hat\Phi^\theta=\Phi|_{(x,u)}^\theta}$ 
is a point transformation of $(x,u)$ belonging to $\mathrm T(\theta,\check\Phi\theta)$. 
\end{definition}

The individual classes of transformations with respect to arbitrary elements should be specified
depending on the investigated classes of systems of differential equations. 
Usually transformed arbitrary elements are smooth functions of 
independent variables, derivatives of dependent variables and arbitrary elements 
and integrals of expressions including arbitrary elements. 
Whenever possible we do not specify a fixed type of equivalence group,
indicating that any of the above notions is applicable.

Similar classes of systems of differential equations have similar equivalence groups.
More precisely, if classes are similar with respect to a transformation of a certain kind
(e.g., a point transformation in the independent variables, the dependent variables, 
their derivatives and the arbitrary elements)
then equivalence groups formed by the equivalence transformations of the same kind are similar with respect to this transformation.

The equivalence group generates an equivalence relation on the set of admissible transformations.
Namely, the admissible transformations $(\theta^1,\tilde\theta^1,\varphi^1)$ and $(\theta^2,\tilde\theta^2,\varphi^2)$
from $\mathrm{T}(\mathcal L|_{\mathcal S})$ are called \emph{$G^{\sim}$-equivalent} if
there exist $\Phi\in G^{\sim}$ such that $\theta^2=\Phi\theta^1$, $\tilde\theta^2=\Phi\tilde\theta^1$ and
$\varphi^2=\Theta\circ\varphi^1\circ\Theta^{-1}$, where $\Theta=\Phi|_{(x,u)}$
(or $\Theta=\Phi(\cdot,\cdot,\theta(\cdot,\cdot))|_{(x,u)}$ in case of $G^{\sim}_{\rm gen}$).

\subsection{Group classification problems}\label{SectionOnGroupClassificationProblems}

Let us recall that for a fixed $\theta\in\mathcal S$ the maximal local (pseudo)group of point symmetries
of the system~$\mathcal{L}_\theta$ coincides with~$\mathrm{T}(\theta,\theta)$ and is denoted by~$G_\theta$.
The common part $G^\cap=G^\cap(\mathcal L|_{\mathcal S})=\bigcap_{\theta\in{\cal S}}G_\theta$ of
all $G_\theta$, $\theta\in\mathcal S$, is called
the \emph{kernel of the maximal point symmetry groups} of systems from the class $\mathcal L|_{\mathcal S}$.
Note that $G^\cap$ can naturally be embedded into $G^{\sim}$ via trivial (identical) prolongation of
the kernel transformations to the arbitrary elements.
The associated subgroup $\hat G^\cap$ of $G^{\sim}$ is normal.

The group classification problem for the class $\mathcal L|_{\mathcal S}$ is
to describe all $G^{\sim}$-inequivalent values of $\theta\in\mathcal S$ together
with the corresponding groups $G_\theta$, for which $G_\theta\ne G^\cap$.
The solution of the group classification problem is a list of pairs
$(\mathcal S_\gamma,\{G_\theta,\theta\in\mathcal S_\gamma\})$, $\gamma\in\Gamma$.
Here $\{\mathcal S_\gamma, \gamma\in\Gamma\}$ is a family of subsets of $\mathcal S$,
$\bigcup_{\gamma\in\Gamma}\mathcal S_\gamma$ contains only $G^{\sim}$-inequivalent values of $\theta$ with $G_\theta\ne G^\cap$,
and for any $\theta\in\mathcal S$ with $G_\theta\ne G^\cap$ there exists $\gamma\in\Gamma$ such that
$\theta\in\mathcal S_\gamma\bmod G^{\sim}$.
The structures of the $G_\theta$ are similar for different values of $\theta\in\mathcal S_\gamma$ under fixed $\gamma$.
In particular, $G_\theta$, $\theta\in\mathcal S_\gamma$, display the same arbitrariness of group parameters.

Group classification problems in the above formulation are very complicated and, in the general case,
are impossible to be solved since they lead to systems of functional differential equations.
That is why one usually considers only
the connected component $G^\mathrm{p}_\theta$ of unity for each $\theta$ instead of the whole group $G_\theta$.
$G^\mathrm{p}_\theta$ is called the \emph{principal (symmetry) group} of the system~$\mathcal{L}_\theta$.
The generators of one-parameter subgroups of $G^\mathrm{p}_\theta$
form a Lie algebra $A_\theta$ of vector fields in the space of $(x,u)$, which is called
the \emph{maximal Lie invariance (or principal) algebra} of infinitesimal symmetry operators of~$\mathcal{L}_\theta$.
The kernel of principal groups of the class~$\mathcal L|_{\cal S}$ is the group
$G^{\cap\mathrm{p}}=G^{\cap\mathrm{p}}(\mathcal L|_{\cal S})=\bigcap_{\theta\in{\cal S}}G^\mathrm{p}_\theta$
for which the Lie algebra is
$A^{\cap}=A^{\cap}(\mathcal L|_{\cal S})=\bigcap_{\theta\in{\cal S}}A_\theta$.

Any operator $Q=\xi^i(x,u)\p_{x_i}+\eta^a(x,u)\p_{u^a}$ from $A_\theta$ satisfies 
the \emph{infinitesimal invariance criterion}~\cite{Olver1986,Ovsiannikov1982} for the system~$\mathcal L_\theta$ 
\[
Q_{(p)} L(x,u_{(p)},\theta(x,u_{(p)}))\big|_{\mathcal L^p_\theta}=0, 
\]
i.e., the result of acting by $Q_{(p)}$ on $L$ vanishes on the manifold~$\mathcal L^p_\theta$.
In what follows we employ the summation convention for repeated indices.
The indices~$i$ and~$a$ run from~1 to~$n$ and from~1 to~$m$, respectively.
$Q_{(p)}$ denotes the standard $p$-th prolongation of the operator~$Q$, 
\[
Q_{(p)}:=Q+\sum_{0<|\alpha|{}\leqslant  p} 
\Bigl(D_1^{\alpha_1}\ldots D_n^{\alpha_n}\bigl(\eta^a(x,u)-\xi^i(x,u)u^a_i\bigr)+\xi^iu^a_{\alpha,i}\Bigr)\p_{u^a_\alpha}.
\]
$ D_i=\p_i+u^a_{\alpha,i}\p_{u^a_\alpha}$ is the operator of total differentiation with respect to the variable~$x_i$. 
The tuple $\alpha=(\alpha_1,\ldots,\alpha_n)$ is a multi-index,  
$\alpha_i\in\mathbb{N}\cup\{0\}$, $|\alpha|\mbox{:}=\alpha_1+\cdots+\alpha_n$.
The variable $u^a_\alpha$ of the jet space $J^{(r)}$ is identified with the derivative 
$\p^{|\alpha|}u^a/\p x_1^{\alpha_1}\ldots\p x_n^{\alpha_n}$, 
and $u^a_{\alpha,i}:=\p u^a_\alpha/\p x_i$. 
The infinitesimal invariance criterion implies the \emph{system of determining equations} on the coefficients of the operator~$Q$, 
where arbitrary elements are involved as parameters.
So, Lie symmetry extensions are connected, as a rule, with extensions of solution sets of this system.

Knowing $A_\theta$, one can reconstruct $G^\mathrm{p}_\theta$.
Then in the framework of the infinitesimal approach
the problem of group classification is reformulated as finding all possible
inequivalent cases of extensions for $A_\theta$, i.e., 
as listing all $G^\sim$-inequivalent values of the arbitrary parameters~$\theta$
together with $A_\theta$ satisfying the condition
$A_\theta\ne A^{\cap}$~\cite{Akhatov&Gazizov&Ibragimov1989,Ovsiannikov1982}.
More precisely, in the infinitesimal approach
the solution of the group classification problem is a list of pairs
$(\mathcal S_\gamma,\{A_\theta,\theta\in\mathcal S_\gamma\})$, $\gamma\in\Gamma$.
Here $\{\mathcal S_\gamma, \gamma\in\Gamma\}$ is a family of subsets of $\mathcal S$,
$\bigcup_{\gamma\in\Gamma}\mathcal S_\gamma$ contains only $G^{\sim}$-inequivalent values of $\theta$ with $A_\theta\ne A^\cap$,
and for any $\theta\in\mathcal S$ with $A_\theta\ne A^\cap$ there exists $\gamma\in\Gamma$ such that
$\theta\in\mathcal S_\gamma\bmod G^{\sim}$.
The structures of the $A_\theta$ are similar for different values of $\theta\in\mathcal S_\gamma$ under fixed $\gamma$.
In particular, all $A_\theta$, $\theta\in\mathcal S_\gamma$, have the same dimension or
display the same arbitrariness of algebra parameters in the infinite-dimensional case.

The procedure of group classification can be completed by finding explicit conditions (e.g., systems of differential equations)
for the arbitrary elements, providing extensions of Lie symmetry.
In other words, for any $\gamma\in\Gamma$ one should explicitly describe the subset $\bar{\mathcal S}_\gamma\subset\mathcal S$ 
that consists of the arbitrary elements which are $G^{\sim}$-equivalent to arbitrary elements from the subset ${\mathcal S}_\gamma$.
Although this step is usually omitted, it may lead to nontrivial results (see, e.g., \cite{Borovskikh2004,Borovskikh2006}).

If the class $\mathcal L|_{\mathcal S}$ is not semi-normalized, i.e., if
$\mathrm{T}(\mathcal L|_{\mathcal S})$ is not generated by $G^\sim$ and the point symmetry groups of the systems involved 
(see Section~\ref{SectionOnDefsOfNormalizedClasses} for precise definitions),
the classification list obtained may include equations which are mutually equivalent with respect to
point transformations which do not belong to $G^\sim$.
Knowledge of such \emph{additional} equivalences allows one to substantially simplify the
further investigation of~$\mathcal L|_S$.
Their explicit construction can be considered as one further step of the algorithm of group classification~\cite{Popovych&Ivanova2003a}.
To carry out this step, one can use the fact that equivalent equations have equivalent maximal invariance algebras.
A more systematical way is to describe the complete set of admissible transformations.

The classical statements of group classification problems can be extended in two main directions.

The first of these comprises possible variations in equivalence relations up to which the group classification is carried out.
Different kinds of equivalence groups (generalized, extended, generalized extended) may be involved.
In fact, all such groups consist of point transformations with respect to the independent and dependent variables.
Modifications concern only the structure of the transformations with respect to the arbitrary elements.
If a class admits a generalized/extended equivalence group essentially wider than the usual one,
the group classification in this class with respect to the usual equivalence group is, as a rule, too cumbersome
and the classification list is far from optimal or may even be unobtainable in a closed form
\cite{Ivanova&Popovych&Sophocleous2004,Vaneeva&Johnpillai&Popovych&Sophocleous2006}.

Another possibility for the modification of a group classification problem in the same direction
is to partition the class under consideration into a family of subclasses with the following properties.
Each of the subclasses is formed by equations inequivalent to equations from the other subclasses and
possesses an equivalence group which is not contained in the equivalence group of the whole class
(the so-called nontrivial conditional equivalence group, see Section~\ref{SectionOnVCondEquivGroups}).
The final classification list for the whole class will be the union of the classification lists for the subclasses.
The subclasses of the partition and their equivalence groups are necessary elements in the presentation
of the result obtained.
It is this approach that is used in the present paper for group classification of
nonlinear Schr\"odinger equations with potentials and modular nonlinearities
(Section~\ref{SectionOnGroupClassificationOfNSchEsWithMNP}).

\looseness=-1
The above inclusion of additional equivalence transformations into the group classification framework is also
a way of changing the underlying equivalence relations.
After all additional equivalences are found,
the classification up to $G^\sim(\mathcal L|_S)$-equivalence is transmuted into
the \emph{classification up to the $\mathrm{T}(\mathcal L|_S)$-equivalence}, i.e., up to
the equivalence generated by all possible point transformations.
Under the special property called semi-normalization, the class~$\mathcal L|_S$ necessarily has no additional equivalences.
Then both the classifications coincide (Corollary~\ref{CorollaryOnCoincidingEquivalences} below).

All the above equivalences are generated by point transformations in independent and dependent variables. 
In the case of one dependent variable ($m=1$), contact transformations can be used instead of point transformations.
Potential equivalence transformations arise for some classes with two independent variables \cite{Popovych&Ivanova2005PETs}. 
In investigating approximate symmetries, it is natural to apply approximate equivalence transformations.

The second direction for modifying group classification problems
is to apply selection criteria different from admitting a Lie symmetry extension.
Transformations between equations (or classes of equations)
preserve a number of their properties, induce transformations between objects related to them and, therefore,
generate different equivalence relations on sets of pairs of the form (equation, object)
resp. (classes of equations, object).
As a result, specification of such objects or properties can be used as a selection criterion for
equations (or classes of equations) up to the above equivalence relations in a way similar to Lie symmetries.
The range of such objects is quite wide. 
It includes nonclassical (conditional) symmetries \cite{Popovych2000,Popovych&Vaneeva&Ivanova2005}, 
conservation laws, associated potential systems and potential or quasi-local symmetries 
\cite{Akhatov&Gazizov&Ibragimov1989,Ivanova&Popovych&Sophocleous2004,Popovych&Ivanova2005JMP,
Popovych&Ivanova2005PETs,Popovych&Kunzinger&Ivanova2007},
generalized (Lie--B\"acklund) symmetries \cite{BroadbridgeGodfrey1991}, 
contact and approximate symmetries, admissible transformations \cite{Popovych2006b} etc.

It is evident that similar classes have similar lists of Lie symmetries under group classification
with respect to any of the above equivalence relations. 
Namely, if $\Psi$ is a transformation realizing the similarity of the class $\mathcal{L}'|_{\mathcal S'}$ to 
the class $\mathcal L|_{\mathcal S}$ (see Definition~\ref{DefOnSimilarClasses}) and 
\[\{\{(\theta,A_\theta), \theta\in\mathcal S_\gamma\}, \mathcal S_\gamma\subset\mathcal S, \gamma\in\Gamma\}\] 
is a classification list for the class $\mathcal L|_{\mathcal S}$ then 
\[
\{\{(\Psi\theta,(\Psi|_{(x,u)})_*A_\theta), \Psi\theta\in\Psi\mathcal S_\gamma\}, \Psi\mathcal S_\gamma\subset\mathcal S', \gamma\in\Gamma\}
\] 
is a classification list for the class $\mathcal{L}'|_{\mathcal S'}$. 
Here $(\Psi|_{(x,u)})_*$ is the mapping induced by the transformation~$\Psi$ 
in the set of vector fields on the space $(x,u)$ (push-forward of vector fields).

If the class $\mathcal{L}'|_{\mathcal S'}$ is only a point-transformation image of 
the class $\mathcal L|_{\mathcal S}$ in the sense of Definition~\ref{DefOnImageClasses} without the above similarity 
then the similarity of their classifications may be broken. 
Only in the case of classifications with respect to the entire sets of admissible transformations there always
exists a one-to-one correspondence to hold between the classification lists for the class-image and the class-preimage.
For such a correspondence to hold in the case of classifications with respect to the equivalence groups, 
we need to additionally require that the preimages of any arbitrary element from $\mathcal S'$ 
are $G^{\sim}(\mathcal L|_{\mathcal S})$-equivalent.
These facts can be applied for the simplification of solving group classification problems.
If one of the classes is classified in a simpler way, possessing, e.g., a set of arbitrary elements 
(resp.\ equivalence group, resp.\ set of admissible transformations, etc.) of a simpler structure then 
its group classification can be carried out first and can subsequently be used to derive the 
classification of the other class \cite{Vaneeva&Popovych&Sophocleous2007}.

\subsection{Gauge equivalence groups}

The equivalence group $G^\sim$ of the class $\mathcal L|_{\mathcal S}$
may contain transformations which act only on arbitrary elements
and do not really change systems, i.e., which generate gauge admissible transformations.
In general, transformations of this type can be considered as trivial~\cite{LisleDissertation}
(gauge) equivalence transformations and form the \emph{gauge} subgroup
$
G^{\mathrm{g}\sim}=\{\Phi\in G^{\sim}\mid \Phi x=x,\, \Phi u=u,\,
\Phi \theta\mathrel{\smash{\stackrel{{g}}{\sim}}} \theta\}
$
of the equivalence group~$G^{\sim}$.
Moreover, $G^{\mathrm{g}\sim}$ is a normal subgroup of $G^{\sim}$.

The application of gauge equivalence transformations is equivalent to rewriting systems
in another form. Contrary to regular equivalence transformations, their role in group classification
doesn't amount to a choice of representatives in equivalence classes but to a
choice of form of these representatives.
It is quite common that the gauge equivalence relation on the set of arbitrary elements
of a class of differential equations is generated by its gauge equivalence group.

We use the name ``gauge equivalence transformation'' since there exist rather trivial equivalence
transformations which do not really transform even arbitrary elements.
Such transformations arise if the auxiliary system implies functional dependence of arbitrary elements.
They form normal subgroups in the corresponding equivalence groups and in the corresponding gauge equivalence groups.
We will neglect these transformations and assume that equivalence groups coincide if
they have the same factor group with respect to the trivial equivalence subgroups.

\subsection{Conditional equivalence groups}\label{SectionOnVCondEquivGroups}

The concept of \emph{conditional equivalence} arises as an extension of the notion of conditional symmetry
transformations of a single system of differential equations \cite{Fushchych1991}
to equivalence transformations in classes of systems.\
It is even more natural than the concept of conditional symmetry
since the description of any class includes, as a necessary element, an auxiliary system (a \emph{condition})
for the arbitrary elements.
Imposing additional constraints on arbitrary elements, we may single out a subclass in the class under consideration
whose equivalence group is not contained in the equivalence group of the whole class.
Let $\mathcal L|_{\mathcal S'\!}$ be the subclass of the class
$\mathcal L|_{\mathcal S}$, which is constrained by the additional
system of equations $S'(x,u_{( p)},\theta_{(q')})=0$ and inequalities $\Sigma'(x,u_{( p)},\theta_{(q')})\ne0$
with respect to the arbitrary elements $\theta=\theta(x,u_{( p)})$.
($\Sigma'$ can be the 0-tuple.)
Here $\mathcal S'\subset\mathcal S$ is the set of solutions of the united system $S=0$, $\Sigma\ne0$, $S'=0$,  $\Sigma'\ne0$.
We assume that the united system is compatible for the subclass $\mathcal L|_{\mathcal S'\!}$ to be nonempty.

\begin{definition}
The equivalence group $G^{\sim}(\mathcal L|_{\mathcal S'\!})$ of the subclass $\mathcal L|_{\mathcal S'\!}$
is called a \emph{conditional equivalence group} of the whole class $\mathcal L|_{\mathcal S}$ under the conditions $S'=0$, $\Sigma'\ne0$.
The conditional equivalence group is called \emph{nontrivial} iff
it is not a subgroup of $G^{\sim}(\mathcal L|_{\mathcal S})$. 
\end{definition}

Conditional equivalence groups may be trivial not with
respect to the equivalence group of the whole class but with respect to other conditional equivalence groups.
Indeed, if $\mathcal S'\subset\mathcal S''$ and
$G^{\sim}(\mathcal L|_{\mathcal S'\!})\subset G^{\sim}(\mathcal L|_{\mathcal S''\!})$
then the subclass $\mathcal L|_{\mathcal S'\!}$ is not interesting from the conditional symmetry point of view.
Therefore, the set of additional conditions on the arbitrary elements 
can be reduced substantially.

\begin{definition}
The conditional equivalence group $G^{\sim}(\mathcal L|_{\mathcal S'\!})$ of the class $\mathcal L|_{\mathcal S}$
under the additional conditions $S'=0$, $\Sigma'\ne0$ is called \emph{maximal} if for any subclass $\mathcal L|_{\mathcal S''\!}$
of the class $\mathcal L|_{\mathcal S}$ containing the subclass $\mathcal L|_{\mathcal S'\!}$
we have $G^{\sim}(\mathcal L|_{\mathcal S'\!})\nsubseteq G^{\sim}(\mathcal L|_{\mathcal S''\!})$.
\end{definition}

In other words, only maximal conditional equivalence groups are interesting.
It is evident that any maximal conditional equivalence group is nontrivial.
The equivalence group $G^{\sim}(\mathcal L|_{\mathcal S})$ of the class $\mathcal L|_{\mathcal S}$
is  its conditional equivalence group associated  with the empty additional condition.

The equivalence group $G^{\sim}(\mathcal L|_{\mathcal S})$ generates an equivalence relation on the set of
pairs of additional auxiliary conditions and the corresponding conditional equivalence groups.
Namely, if a transformation from $G^{\sim}(\mathcal L|_{\mathcal S})$ transforms the system
$S'=0$, $\Sigma'\ne0$ to the system $S''=0$, $\Sigma''\ne0$
then the conditional equivalence groups
$G^{\sim}(\mathcal L|_{\mathcal S'\!})$ and $G^{\sim}(\mathcal L|_{\mathcal S''\!})$
are similar with respect to this transformation and will be called \emph{$G^{\sim}$-equivalent}.
If a conditional equivalence group is maximal then any conditional equivalence group $G^{\sim}$-equivalent to it
is also maximal.

Building on the concept of conditional equivalence,
we can formulate the problem of describing $\mathrm{T}(\mathcal L|_{\mathcal S})$
analogously to the usual group classification problem.
Nontrivial additional auxiliary conditions for arbitrary elements naturally arise when
studying~$\mathrm{T}(\mathcal L|_{\mathcal S})$.
Typically, the following steps have to be carried out:
\begin{enumerate}\itemsep=0ex
\item
Construction of~$G^{\sim}(\mathcal L|_{\mathcal S})$ (or $G^{\sim}_{\rm gen}(\mathcal L|_{\mathcal S})$, etc.).
\item
Description of conditional equivalence transformations in $\mathcal L|_{\mathcal S}$, i.e.,
searching for a complete family of $G^{\sim}$-inequivalent additional auxiliary conditions~$S_\gamma$, $\gamma\in\Gamma$, such that
$G^{\sim}(\mathcal L|_{\mathcal S_\gamma})$ is a maximal conditional equivalence group of the class $\mathcal L|_{\mathcal S}$
for any $\gamma\in\Gamma$.
\item
Finding admissible transformations which do not belong to any conditional equivalence groups.
\end{enumerate}

Actually, the proposed procedure is far from optimal. We will return to this point after
presenting more elaborate techniques.

\section{Normalized classes of differential equations}\label{SectionOnNormalizedClasses}

Solving group classification problems is considerably simpler if the class~$\mathcal L|_{\mathcal S}$
of systems of differential equations under consideration has the additional property of normalization
with respect to point transformations.
In addition, the investigation of~$\mathrm{T}(\mathcal L|_{\mathcal S})$ can be deepened
by considering conditional equivalence groups for subclasses possessing this property.

\subsection{Definition of normalized classes of differential equations}\label{SectionOnDefsOfNormalizedClasses}

\begin{definition}
The class~$\mathcal L|_{\mathcal S}$ is called {\em normalized} 
if $\forall (\theta,\tilde\theta,\varphi)\!\in\!\mathrm{T}(\mathcal L|_{\mathcal S})$
$\exists\Phi\!\in\! G^{\sim}{:}$
$\tilde\theta=\Phi\theta$ and $\varphi=\Phi|_{(x,u)}$.
It 
is called {\em normalized} 
{\em in the generalized sense}
if $\forall (\theta,\tilde\theta,\varphi)\!\in\!\mathrm{T}(\mathcal L|_{\mathcal S})$
$\exists\Phi\!\in\! G^{\sim}_{\rm gen}{:}$
$\tilde\theta=\Phi\theta$ and $\varphi=\Phi(\cdot,\cdot,\theta(\cdot,\cdot))|_{(x,u)}$.
\end{definition}

\begin{proposition}
If the class $\mathcal L|_{\mathcal S}$ is normalized (in the usual or generalized sense)
and its subclass $\mathcal L|_{\mathcal S'}$ is closed under the action of $G^{\sim}$
(or $G^{\sim}_{\rm gen}$) then the subclass $\mathcal L|_{\mathcal S'}$ is normalized in the same sense.
\end{proposition}

\begin{definition}
The class~$\mathcal L|_{\mathcal S}$ is called {\em strongly normalized} 
if it is normalized and $G^{\sim}|_{(x,u)}=\prod_{\theta\in\mathcal S}G_\theta$.
It is called
{\em strongly normalized} 
{\em in the generalized sense} if it is normalized in the generalized sense and
$\forall \theta^0\!\in\!\mathcal S{:}$
$G^{\sim}_{\rm gen}|^{\theta=\theta^0}_{(x,u)}=
\prod_{\theta\in\mathcal S_{\theta^0}}\!G_\theta$,
where
$\mathcal S_{\theta^0}=\{\theta'\in\mathcal S\,\,|\,\,
G^{\sim}_{\rm gen}|^{\theta=\theta'}_{(x,u)}=
G^{\sim}_{\rm gen}|^{\theta=\theta^0}_{(x,u)}
\}$.
\end{definition}

\begin{definition}\label{DefOfSemiNormalizedClasses}
The class~$\mathcal L|_{\mathcal S}$ is called {\em semi-normalized} 
if $\forall (\theta,\tilde\theta,\varphi)\!\in\!\mathrm{T}(\mathcal L|_{\mathcal S})$
$\exists\tilde\varphi\!\in\! G_\theta$,
$\exists\Phi\!\in\! G^{\sim}\colon\varphi=\Phi|_{(x,u)}\circ\tilde\varphi$, i.e.,
\[
\mathrm{T}(\mathcal L|_{\mathcal S})=\{(\theta,\Phi\theta,\Phi|_{(x,u)}\circ\tilde\varphi)\mid
\theta\!\in\!\mathcal S,\,\tilde\varphi\!\in\! G_\theta,\,\Phi\!\in\! G^{\sim}\}.
\]
(\,$\mathrm{T}(\mathcal L|_{\mathcal S})=\{
(\theta^0,\Phi\theta^0,\Phi|^{\theta=\theta^0}_{(x,u)}\circ\tilde\varphi)\mid
\theta^0\!\in\!\mathcal S,\,
\tilde\varphi\!\in\! G_\theta,\,\Phi\!\in\! G^{\sim}_{\rm gen}\}$ if~$\mathcal L|_{\mathcal S}$
is {\em semi-normalized in the generalized sense}.)
\end{definition}

Roughly speaking, the class~$\mathcal L|_{\mathcal S}$ is normalized if any
admissible transformation in this class is induced by a transformation from the equivalence group~$G^{\sim}$
and is strongly normalized if additionally $G^{\sim}|_{(x,u)}$
is generated by elements from~$G_\theta$, $\theta\in\mathcal S$.
The set of admissible transformations of a semi-normalized class is generated by
the transformations from the equivalence group of the whole class and the transformations
from the Lie symmetry groups of equations of this class.

\begin{proposition}
If the class $\mathcal L|_{\mathcal S}$ is normalized/semi-normalized (in the usual or generalized sense)
and the subclass $\mathcal L|_{\mathcal S}'$ is closed under the action of $G^{\sim}$
(or $G^{\sim}_{\rm gen}$) then the subclass $\mathcal L|_{\mathcal S}'$ is normalized/semi-normalized in the same sense.
\end{proposition}

The intersection of normalized subclasses of the class~$\mathcal L|_{\mathcal S}$ with the same equivalence group $G^\sim_0$
is a normalized subclass possessing $G^\sim_0$ as a subgroup of the equivalence group,
which generates the whole corresponding set of admissible transformations.
Indeed, let $\mathcal L|_{\mathcal S'\!}$ and $\mathcal L|_{\mathcal S''\!}$ be normalized subclasses
of the class~$\mathcal L|_{\mathcal S}$ and
$G^{\sim}(\mathcal L|_{\mathcal S'\!})=G^{\sim}(\mathcal L|_{\mathcal S''\!})=G^\sim_0$.
If $\Phi\in G^\sim_0$ then
$(\theta,\Phi\theta,\Phi|_{(x,u)})\in\mathrm{T}(\mathcal L|_{\mathcal S'\cap\mathcal S''\!})$
for any $\theta\in\mathcal S'\cap\mathcal S''\!$,
i.e., $\Phi\in G^{\sim}(\mathcal L|_{\mathcal S'\cap\mathcal S''\!})$.
By the normalization of $\mathcal L|_{\mathcal S'\!}$ and $\mathcal L|_{\mathcal S''\!}$,
for any $(\theta,\tilde\theta,\varphi)\in\mathrm{T}(\mathcal L|_{\mathcal S'\cap\mathcal S''\!})$
there exists $\Phi\in G^\sim_0$ such that
$\tilde\theta=\Phi\theta$ and $\varphi=\Phi|_{(x,u)}$.
Therefore, $\mathcal L|_{\mathcal S'\cap\mathcal S''\!}$ is a normalized subclass.
The proof in the case of normalization in the generalized sense is analogous.

\subsection{Examples of normalized classes}

There exist a number of obvious examples of normalized classes.
Thus, it is intuitively understandable that the extreme cases of classes formed by
either a single system of differential equations
or all systems having a fixed number of independent variables, unknown functions and differential equations
with or without restriction of order are normalized.
Let us demonstrate this within the framework of the above formal approach.

Consider a system~$L(x,u_{( p)})=0$ of $l$~differential equations
for $m$ unknown functions $u$ of $n$~independent variables $x$,
which admits the maximal point symmetry group~$G$.
We assume that the tuple~$\theta$ consists of a single arbitrary element denoted also as~$\theta$
and $L$ depends on~$\theta$ constantly.
The auxiliary system~$\mathcal S$ for the arbitrary element~$\theta$ can be chosen in different ways.
Here we discuss two possibilities.

The first one is to constrain~$\theta$ with a single (algebraic or differential) equation, for example, $\theta=0$.
Hence, $\mathcal S$ is a singleton consisting of the function identically vanishing on $J^{( p)}$,
$\mathrm{T}(\mathcal L|_{\mathcal S})=\{\,(0,0,\varphi)\mid\varphi\in G\, \}$
and
$G^{\sim}=\{\, (\tilde x,\tilde u)=\varphi(x,u),\, \tilde\theta=F(x,u_{( p)},\theta)\theta
\mid \varphi\in G,\, F(\cdot,\cdot,0)\not=0\,\}$,
i.e., in view of definition~1 the class~$\mathcal L|_{\mathcal S}$ is normalized.
It possesses the nonempty trivial equivalence group
$G^{\sim}_\mathrm{triv}=\{\, (\tilde x,\tilde u)=(x,u),\, \tilde\theta=F(x,u_{( p)},\theta)\theta\mid
F(\cdot,\cdot,0)\not=0\}$ which should be disregarded, and
$G^{\sim}/G^{\sim}_\mathrm{triv}=\{\, (\tilde x,\tilde u)=\varphi(x,u),\, \tilde\theta=\theta\mid \varphi\in G\,\}$.

The second possibility is to impose no constraints on~$\theta$,
so $\mathcal S$ is the whole set of $ p$-th order differential functions of~$(x,u)$,
$\mathrm{T}(\mathcal L|_{\mathcal S})=\{\,(\theta,\tilde\theta,\varphi)\mid
\theta,\tilde\theta\in\mathcal S,\ \varphi\in G \}$
 and
$G^{\sim}=\{\, (\tilde x,\tilde u_{( p)})=\mathop{{\rm pr}_{ p}}\varphi(x,u_{( p)}),\,
\smash{\tilde\theta}=F(x,u_{( p)},\theta)\mid\varphi\in G,\,\partial F/\partial\theta\not=0\,\}$.
Therefore, the class
$\mathcal L|_{\mathcal S}$ is normalized.
This class provides an example of classes without one-to-one correspondence between arbitrary elements
and systems of differential equations.

The class of all systems of $l$ differential equations for $m$~unknown functions of $n$~independent variables,
which have order not greater than~$ p$, (here $l$, $m$, $n$ and $ p$ are fixed integers)
can be included within the framework of the formal approach by viewing the left hand sides
of the equations themselves
as arbitrary elements and taking the empty auxiliary system~$S$, i.e., $k=l$, $L\equiv \theta$
and $\mathcal S$ is the whole set of $l$-tuples of functionally independent
$ p$-th order differential functions of~$(x,u)$.
Then $\mathrm{T}(\mathcal L|_{\mathcal S})=\{\,(\theta,\tilde\theta,\varphi)\mid 
\theta\in\mathcal S,\,\tilde\theta=\mathop{{\rm pr}_{ p}}\varphi\, F(x,u_{( p)},\theta(x,u_{( p)})),\,
|\partial\varphi/\partial(x,u)|\not=0,\linebreak\partial F/\partial\theta|_{\theta=0}\not=0\, \}$
and $G^{\sim}=\{\,\Phi=(\varphi(x,u),F(x,u_{( p)},\theta))\mid
|\partial\varphi/\partial(x,u)|\not=0,\ \partial F/\partial\theta|_{\theta=0}\not=0\, \}$.
This obviously shows normalization of this class.

The normalization property has been established (in explicit or implicit forms)
for a number of different classes of differential equations important for applications.
As examples we mention
generalized Burgers equations~\cite{Kingston&Sophocleous1991},
eikonal equations of space dimensions one, two and three \mbox{\cite{Borovskikh2004,Borovskikh2006}},
$(1+1)$-dimensional general or quasi-linear evolution equations~\cite{Abramenko&Lagno&Samojlenko2002,
Basarab-Horwath&Lahno&Zhdanov2001,Kingston&Sophocleous1998,Lagno&Samojlenko2002,Zhdanov&Lahno1999} 
and systems of such equations~\cite{Popovych&Kunzinger&Ivanova2007},
different multi-dimensional quasi-linear parabolic equations~\cite{Prokhorova2005},
$(1+1)$-dimensional generalized nonlinear wave equations~\cite{Lahno&Zhdanov&Magda2006},
different kinds of 
nonlinear Schr\"odinger equations
\cite{Gagnon&Winternitz1993,Ivanova&Popovych&Eshraghi2005,Kunzinger&Popovych2006,Popovych&Eshraghi2004Mogran,
Popovych&Ivanova&Eshraghi2004Cubic,Popovych&Ivanova&Eshraghi2004Gamma,Zhdanov&Roman2000}.

\subsection{Normalized classes and group classification problems}

The notion of normalized classes naturally arises in group analysis. 
In an  implicit form, it was often used in solving group classification problems for
many classes of system of differential equations.
Well-known examples include Lie's classical classifications
of second-order ordinary differential equations~\cite{Lie1891} and
of second-order two-dimensional linear partial differential equations~\cite{Lie1881}.
Recently similar classification methods were applied by a number of authors
(see, e.g., \cite{Basarab-Horwath&Lahno&Zhdanov2001,Borovskikh2004,Gagnon&Winternitz1993,Lahno&Spichak2007,
Lahno&Zhdanov&Magda2006,Popovych&Ivanova&Eshraghi2004Cubic,Zhdanov&Lahno1999,Zhdanov&Roman2000}). 
In fact all these methods involve the following properties of normalized classes.

\begin{proposition}
Let the class~$\mathcal L|_{\mathcal S}$ be normalized and let
$G^i$, $i=1,2$,  be local groups of point transformations in the space of $(x,u)$, for which
$\mathcal S^i=\{\theta\!\in\!\mathcal S\,|\,G^\mathrm{p}_\theta=G^i\}\ne\varnothing$.
Then $\mathcal S^1\sim \mathcal S^2\bmod G^{\sim}$ iff $G^1\sim G^2\bmod G^{\sim}$.
\end{proposition}

\begin{proposition}
Two systems from a semi-normalized class are transformed into one another by a point transformation
iff they are equivalent with respect to~the equivalence group of this class.
\end{proposition}

\begin{corollary}\label{CorollaryOnCoincidingEquivalences}
In a semi-normalized class, the classifications up to the equivalence induced by action of the equivalence group
and up to the general point-transformation equivalence coincide.
\end{corollary}

\begin{proposition}
Each normalized class of systems of differential equations is semi-normalized.
\end{proposition}

The implementation of group classification in a normalized class always leads to the construction of a tree of subclasses 
possessing normalization properties. Let us investigate this phenomenon.

\begin{proposition}
Let the class~$\mathcal L|_{\mathcal S}$ be normalized and suppose that
a subset $\mathcal S'\!$ of~$\mathcal S$ determines a subclass $\mathcal L|_{\mathcal S'\!}$
which is invariant under the action of $G^\sim(\mathcal L|_{\mathcal S})$.
Then the subclass $\mathcal L|_{\mathcal S'}$ is normalized (in the same sense).
$G^\sim(\mathcal L|_{\mathcal S})$ is a subgroup of $G^\sim(\mathcal L|_{\mathcal S'\!})$
which generates $\mathrm{T}(\mathcal L|_{\mathcal S'\!})$, and which,
if $\mathcal L|_{\mathcal S}$ is normalized in the usual sense,
coincides with $G^\sim(\mathcal L|_{\mathcal S'\!})$ up to gauge equivalence transformations
in $\mathcal L|_{\mathcal S'\!}$.
\end{proposition}

\begin{proof}
$G^\sim(\mathcal L|_{\mathcal S'\!})\supset G^\sim(\mathcal L|_{\mathcal S})$,
since for any $\Phi\in G^\sim(\mathcal L|_{\mathcal S})$ and for any $\theta\in\mathcal S'\!$
we have $\Phi\theta\in\mathcal S'\!$, i.e.,
$(\theta,\Phi\theta,\Phi|_{(x,u)})\in\mathrm{T}(\mathcal L|_{\mathcal S'\!})$,
which implies $\Phi\in G^\sim(\mathcal L|_{\mathcal S'\!})$.
Since $\mathrm{T}(\mathcal L|_{\mathcal S'\!})\subset\mathrm{T}(\mathcal L|_{\mathcal S})$,
for any $(\theta,\tilde\theta,\varphi)\in\mathrm{T}(\mathcal L|_{\mathcal S'\!})$
there exists $\Phi\in G^\sim(\mathcal L|_{\mathcal S})$ such that $\tilde\theta=\Phi\theta$ and $\varphi=\Phi|_{(x,u)}$,
i.e., the subclass $\mathcal L|_{\mathcal S'}$ is normalized.
The above part of the proof is easily extended to the generalized case.

Any $\Psi\in G^\sim(\mathcal L|_{\mathcal S'\!})$ and any $\theta\in\mathcal S'\!$ give
the admissible transformation $(\theta,\Psi\theta,\Psi|_{(x,u)})\in\mathrm{T}(\mathcal L|_{\mathcal S'\!})$.\
Therefore, there exists $\Phi\in G^\sim(\mathcal L|_{\mathcal S})$ such that
$\Psi|_{(x,u)}=\Phi|_{(x,u)}$ and $\Psi\theta=\Phi\theta$.
\end{proof}

Note that under the above assumptions
the subclass $\mathcal L|_{\mathcal S\backslash\mathcal S'\!}$ has similar properties.

Given the class~$\mathcal L|_{\mathcal S}$ and
a local (connected) group $G$ of point transformations of $(x,u)$
such that $G=G^\mathrm{p}_\theta$ for some $\theta\!\in\!\mathcal S$,
consider the subsets of~$\mathcal S$
\begin{gather*}
\mathcal S_G=\{\,\theta\!\in\!\mathcal S\mid G^\mathrm{p}_\theta\supset G\},
\quad
\bar{\mathcal S}_G=\{\,\theta\!\in\!\mathcal S\mid G^\mathrm{p}_\theta\supset G\bmod G^\sim\},
\\
\mathcal S'_G=\{\,\theta\!\in\!\mathcal S\mid G^\mathrm{p}_\theta=G\},
\quad
\bar{\mathcal S}'_G=\{\,\theta\!\in\!\mathcal S\mid G^\mathrm{p}_\theta=G\bmod G^\sim\}.
\end{gather*}

\begin{corollary}
Let the class $\mathcal L|_{\mathcal S}$ be normalized.
Then $\mathcal L|_{\bar{\mathcal S}_G}$ and $\mathcal L|_{\bar{\mathcal S}'_G}$
are normalized subclasses of~$\mathcal L|_{\mathcal S}$.
$G^\sim(\mathcal L|_{\mathcal S})$ is a subgroup of
$G^\sim(\mathcal L|_{\bar{\mathcal S}_G})$ and $G^\sim(\mathcal L|_{\bar{\mathcal S}'_G})$ and
generates $\mathrm{T}(\mathcal L|_{\bar{\mathcal S}_G})$ and $\mathrm{T}(\mathcal L|_{\bar{\mathcal S}'_G})$.
\end{corollary}

\begin{proposition}\label{PropositionOnKernelSubclassInvariance}
The subclass~$\mathcal L|_{\mathcal S_0}$ is invariant with respect to $G^\sim(\mathcal L|_{\mathcal S})$,
where $\mathcal S_0=\mathcal S'_{G^\cap}$, $G^\cap=G^{\cap\mathrm{p}}(\mathcal L|_{\mathcal S})$.
\end{proposition}

\begin{proof}
Let us fix any $\Phi\in G^\sim(\mathcal L|_{\mathcal S})$ and any $\theta\in\mathcal S_0$.
We have to show that $\Phi\theta\in\mathcal S_0$. Now
$G^\mathrm{p}_{\Phi\theta}=\mathrm{Ad}_\Phi G^\mathrm{p}_\theta=\mathrm{Ad}_\Phi G^\cap$,
where $\mathrm{Ad}_\Phi$ is the action of $\Phi$ on transformation groups:
$G\ni\psi\to \varphi\circ\psi\circ\varphi^{-1}\in\mathrm{Ad}_\Phi G$, $\varphi:=\Phi|_{(x,u)}$.
Since $\Phi\theta\in\mathcal L|_{\mathcal S}$, $G^\mathrm{p}_{\Phi\theta}\supset G^\cap$.
If $G^\mathrm{p}_{\Phi\theta}=\mathrm{Ad}_\Phi G^\cap\varsupsetneq G^\cap$ then
$G^\cap\varsupsetneq\mathrm{Ad}_{\Phi^{-1}} G^\cap$.
But $\mathrm{Ad}_{\Phi^{-1}} G^\cap=G^\mathrm{p}_{\Phi^{-1}\theta}$,
$\Phi^{-1}\theta\in\mathcal L|_{\mathcal S}$ and, therefore, $\mathrm{Ad}_{\Phi^{-1}} G^\cap\supset G^\cap$
which implies a contradiction. That is why $G^\mathrm{p}_{\Phi\theta}=G^\cap$,
i.e., $\Phi\theta\in\mathcal S_0$.
\end{proof}

\begin{proposition}\label{PropositionOnNormalizationOfS'G}
If the class $\mathcal L|_{\mathcal S}$ is normalized in the usual sense, 
the class $\mathcal L|_{\mathcal S'_G}$ has the same property.  
The set $\mathrm{T}(\mathcal L|_{\mathcal S'_G})$ is generated by
the group $G^\sim(\mathcal L|_{\mathcal S'_G})\cap G^\sim(\mathcal L|_{\mathcal S})$
whose projection onto $(x,u)$ is the normalizer of $G$ in $G^\sim(\mathcal L|_{\mathcal S})|_{(x,u)}$.
\end{proposition}

\begin{proof}
Let us fix an arbitrary $(\theta,\tilde\theta,\varphi)\in\mathrm{T}(\mathcal L|_{\mathcal S'_G})$.
Since $\mathrm{T}(\mathcal L|_{\mathcal S'_G})\subset\mathrm{T}(\mathcal L|_{\mathcal S})$,
there exists $\Phi\in G^\sim(\mathcal L|_{\mathcal S})$ such that
$\tilde\theta=\Phi\theta$ and $\varphi=\Phi|_{(x,u)}$,
$\theta,\tilde\theta\in\mathcal S'_G$, hence
$G=G^\mathrm{p}_{\tilde\theta}=\varphi\circ G^\mathrm{p}_{\theta}\circ\varphi^{-1}=\varphi\circ G\circ\varphi^{-1}$,
i.e., $\varphi=\Phi|_{(x,u)}$ belongs to the normalizer of $G$ in $G^\sim(\mathcal L|_{\mathcal S})|_{(x,u)}$.

Consider any transformation $\Phi\in G^\sim(\mathcal L|_{\mathcal S})$ such that
its projection $\varphi=\Phi|_{(x,u)}$ belongs to the normalizer of $G$ in $G^\sim(\mathcal L|_{\mathcal S})|_{(x,u)}$.
Then $(\theta,\Phi\theta,\varphi)\in\mathrm{T}(\mathcal L|_{\mathcal S'_G})$ for arbitrary
$\theta\in\mathcal S'_G$ since $\Phi\theta\in\mathcal S'_G$.
Indeed, $\Phi\theta\in\mathcal S$ and
$G^\mathrm{p}_{\Phi\theta}=\varphi\circ G^\mathrm{p}_{\theta}\circ\varphi^{-1}=\varphi\circ G\circ\varphi^{-1}=G$.
Therefore, $\Phi\in G^\sim(\mathcal L|_{\mathcal S'_G})$.
\end{proof}

\begin{proposition}
Suppose that $\mathcal S'_G\ne\varnothing$. Then
$G^{\cap\mathrm{p}}(\mathcal L|_{\mathcal S_G})=G$, 
$G^\sim(\mathcal L|_{\mathcal S_G})\subset G^\sim(\mathcal L|_{\mathcal S'_G})$, and
if $\mathcal L|_{\mathcal S}$ is normalized in the usual sense, 
the projections of these equivalence groups onto $(x,u)$ coincide.
\end{proposition}

\begin{proof}
The first statement trivially follows from the definition of $\mathcal L|_{\mathcal S_G}$.
Then in view of Proposition~\ref{PropositionOnKernelSubclassInvariance},
$\mathcal L|_{\mathcal S'_G}$ is invariant with respect to $G^\sim(\mathcal L|_{\mathcal S_G})$,
i.e., $G^\sim(\mathcal L|_{\mathcal S_G})\subset G^\sim(\mathcal L|_{\mathcal S'_G})$ and hence 
$G^\sim(\mathcal L|_{\mathcal S_G})\cap G^\sim(\mathcal L|_{\mathcal S})\subset
G^\sim(\mathcal L|_{\mathcal S'_G})\cap G^\sim(\mathcal L|_{\mathcal S})$.
At the same time, $G^\sim(\mathcal L|_{\mathcal S_G})\cap G^\sim(\mathcal L|_{\mathcal S})$ contains 
all the transformations from $G^\sim(\mathcal L|_{\mathcal S})$ 
whose projections onto $(x,u)$ belong to the normalizer of~$G$ in $G^\sim(\mathcal L|_{\mathcal S})|_{(x,u)}$.
This implies in view of Proposition~\ref{PropositionOnNormalizationOfS'G} that 
\[
G^\sim(\mathcal L|_{\mathcal S_G})|_{(x,u)}
\supset G^\sim(\mathcal L|_{\mathcal S_G})\cap G^\sim(\mathcal L|_{\mathcal S})|_{(x,u)}
=G^\sim(\mathcal L|_{\mathcal S'_G})\cap G^\sim(\mathcal L|_{\mathcal S})|_{(x,u)}
= G^\sim(\mathcal L|_{\mathcal S'_G})|_{(x,u)}.
\]
In particular,
$G^\sim(\mathcal L|_{\mathcal S_G})\cap G^\sim(\mathcal L|_{\mathcal S})=
 G^\sim(\mathcal L|_{\mathcal S'_G})\cap G^\sim(\mathcal L|_{\mathcal S})$.
\end{proof}

\begin{note}
In general, the normalization of $\mathcal L |_{\mathcal S}$ does not imply that 
the class $\mathcal L|_{\mathcal S_G}$ is normalized.
\end{note}

In view of the above propositions,
the group classification problem in any normalized class of differential equations
is reduced to the subgroup analysis of the corresponding equivalence group.
The property of strong normalization often is an indication that many subgroups will be
Lie symmetry groups of systems from the class under consideration.
Moreover, under classification a hierarchy of normalized classes corresponding to symmetry extension cases
is naturally obtained.
The possibility of characterizing certain properties of a class in terms of normalization and 
the complexity of associated normalized subclasses strongly correlate with the complexity of 
the corresponding group classification problem.

\subsection{Normalized subclasses and admissible transformations}\label{SectionOnNormSubclassesAndAdmTrans}

An investigation of the normalization of the class~$\mathcal L|_{\mathcal S}$ or its subclasses is necessary
for the description of~$\mathrm{T}(\mathcal L|_{\mathcal S})$ and can be included as a step
in studying~$\mathrm{T}(\mathcal L|_{\mathcal S})$.
Analogously to conditional equivalence groups, only a part of the normalized subclasses is significant for this.

\begin{definition}
A normalized subclass~$\mathcal L|_{\mathcal S'\!}$ of the  class $\mathcal L|_{\mathcal S}$ is called \emph{maximal}
if there are no normalized subclasses of~$\mathcal L|_{\mathcal S}$ properly containing $\mathcal L|_{\mathcal S'\!}$.
\end{definition}

The definition of maximality can be formulated in the same way for other kinds of normalization, i.e.,
for strongly normalized and semi-normalized (in the usual or generalized sense) classes.
Generally speaking, a maximal strongly normalized subclass is not necessarily a maximal normalized subclass
and a maximal normalized subclass is not necessarily a maximal semi-normalized subclass.
Moreover, a maximal normalized or semi-normalized subclass may be non-associated with a maximal conditional equivalence group.
At the same time, no proper subclass of a normalized class leads to a maximal conditional equivalence group.

The algorithm for describing sets of admissible transformations can be modified by including as an additional step
the investigation of normalization properties of subclasses associated with maximal equivalence groups and
the construction of the complete family of maximal normalized subclasses.
(The maximal normalized subclasses can by studied up to $G^\sim$-equivalence.)
Note that under solving group classification problems, non-maximal normalized subclasses arise as possible classification cases
and, therefore, should also be studied.

The problem of classifying admissible transformations can be seen as solved, e.g., in the following cases.

In view of the definition of normalized classes, the set of admissible transformations
is known if the class turns out to be normalized and its equivalence group is calculated.
Then
\[
\mathrm{T}(\mathcal L|_{\mathcal S})=\{\,(\theta,\Phi\theta,\Phi|_{(x,u)})\mid\theta\in\mathcal S,\Phi\in G^\sim\}.
\]
This is the case for all classes of nonlinear Schr\"odinger equations investigated in Section~\ref{SectionOnNestedNormalizedClassesOfNSchEs}.
Note that any normalized class has a unique maximal conditional equivalence group which obviously coincides with the
equivalence group of this class.
A subclass cannot generate a maximal conditional equivalence group of the class if it is strongly contained in a normalized subclass.

Suppose that the class~$\mathcal L|_{\mathcal S}$ is presented as a union of disjoint normalized subclasses,
and that there are no admissible transformations between systems from different subclasses.
That is, $\mathcal S=\bigcup_{\gamma\in\Gamma}\mathcal S_\gamma$,
$\mathcal L|_{\mathcal S_\gamma}$ is normalized for any $\gamma\in\Gamma$,
$\mathcal S_\gamma\cap\mathcal S_{\gamma'\!}=\varnothing$ and
$\mathrm{T}(\theta,\theta')=\varnothing$, where $\theta\in\mathcal S_\gamma$, $\theta'\in\mathcal S_{\gamma'\!}$, $\gamma\ne\gamma'$.
Then obviously $G^{\sim}(\mathcal L|_{\mathcal S})=\cap_{\gamma\in\Gamma}G^{\sim}(\mathcal L|_{\mathcal S_\gamma})$ and
$\mathrm{T}(\mathcal L|_{\mathcal S})$ is the union of the (easily constructed) sets $\mathrm{T}(\mathcal L|_{\mathcal S_\gamma})$
of admissible transformations in the subclasses:
\[
\mathrm{T}(\mathcal L|_{\mathcal S})=
\bigcup_{\gamma\in\Gamma}\{\,(\theta,\Phi\theta,\Phi|_{(x,u)})\mid
\theta\in\mathcal S_\gamma,\Phi\in G^{\sim}(\mathcal L|_{\mathcal S_\gamma})\,\}.
\]
Let us choose a subset $\Lambda$ of $\Gamma$ such that
$G^{\sim}(\mathcal L|_{\mathcal S_\lambda})\ne G^{\sim}(\mathcal L|_{\mathcal S_{\lambda'\!}})$ for any different $\lambda,\lambda'\in\Lambda$
and for any $\gamma\in\Gamma$ there exist $\lambda\in\Lambda$ satisfying the condition
$G^{\sim}(\mathcal L|_{\mathcal S_\lambda})=G^{\sim}(\mathcal L|_{\mathcal S_\gamma})$.
Denote the subset of those $\gamma$'s corresponding to a fixed $\lambda$ by $\Gamma_\lambda$.
Then the set of maximal normalized subclasses of~$\mathcal L|_{\mathcal S}$ is exhausted by the unions
$\bar{\mathcal S}_\lambda=\bigcup_{\gamma\in\Gamma_\lambda}\mathcal S_\gamma$, $\lambda\in\Lambda$.
The groups $G^{\sim}(\mathcal L|_{\mathcal S_\lambda})=G^{\sim}(\mathcal L|_{\bar{\mathcal S}_\lambda})$
exhaust the sets of maximal conditional equivalence groups of the class~$\mathcal L|_{\mathcal S}$.

The class of nonlinear Schr\"odinger equations with potentials and general modular nonlinearities has
a set of admissible transformations of the above structure for all space dimensions
\cite{Kunzinger&Popovych2006,Popovych2006a,Popovych&Eshraghi2004Mogran}.
See also Theorem~\ref{TheoremGequivNSchEPMN} and Note~\ref{NoteNSchEPMNAdmissibleTrans}.

A more nontrivial situation concerning admissible transformations is encountered 
when maximal normalized subclasses have a nonempty intersection.
Suppose that $\mathcal S'\!,\mathcal S''\subset\mathcal S$, $\mathcal S'\cap\mathcal S''\ne\varnothing$,
the subclasses $\mathcal L|_{\mathcal S'}$ and $\mathcal L|_{\mathcal S''}$ are normalized,
$\mathcal S'=G^{\sim}(\mathcal L|_{\mathcal S'\!})\ \mathcal S'\cap\mathcal S''$ and
$\mathcal S''=G^{\sim}(\mathcal L|_{\mathcal S''\!})\ \mathcal S'\cap\mathcal S''$.
The latter two conditions mean that any equation from $\mathcal S'$ (or $\mathcal S''$) is equivalent,
with respect to $G^{\sim}(\mathcal L|_{\mathcal S'\!})$ (or $G^{\sim}(\mathcal L|_{\mathcal S''\!})$\,),
to an equation from $\mathcal S'\cap\mathcal S''$.
Then any admissible transformation $(\theta'\!,\theta''\!,\varphi)$ with $\theta'\!\in\mathcal S'$ and $\theta''\!\in\mathcal S''$,
can be represented in the form $(\theta'\!,\Phi^2(\Phi^1\theta'),(\Phi^2\circ\Phi^1)|_{(x,u)})$, where
$\Phi^1\in G^{\sim}(\mathcal L|_{\mathcal S'\!})$,
$\Phi^2\in G^{\sim}(\mathcal L|_{\mathcal S''\!})$ and $\Phi^1\theta'\in\mathcal S'\cap\mathcal S''$.

A set of admissible transformations of such structure arises, e.g.,  in the investigation of  a class of
variable coefficient diffusion--reaction equations~\cite{Vaneeva&Johnpillai&Popovych&Sophocleous2006}.

It is obvious that the above cases comprise only simplest situations 
of the description of sets of admissible transformations in terms of normalized subclasses.

\section{Lie symmetries of nonlinear Schr\"odinger equations:\\ known results}%
\label{SectionOnKnownResultsOnLieSymmetriesOfNSchEs}

Nonlinear Schr\"odinger equations (NSchEs) are important objects of investigation 
due to their interesting mathematical properties and have found many applications in 
different fields of physics and other science. 
These include optics, nonlinear quantum mechanics, the theory of Bose--Einstein condensation, plasma physics, 
computer science, and geophysics among others. 
NSchEs also occur in applications as the so-called Madelung fluid equations which are connected with the standard form 
via the Madelung transformation $\psi=\sqrt{R}\,e^{i\varphi}$, where $R$ and $\varphi$ are the new real-valued unknown functions.
See, e.g., the references on Schr\"odinger equations in this paper and references therein.
The cubic Schr\"odinger equation is one of the most intensively studied integrable models of 
mathematical physics.
At the same time, the physical interpretation of some known types of nonlinear Schr\"odinger equations
is not completely clear and is a topic of ongoing research.

Schr\"odinger equations have also been intensively studied by means of symmetry methods. 
The references in this paper, related to this subject, represent mainly investigations on classical Lie symmetries. 
In fact, the first investigation of Schr\"odinger equations from the symmetry point of view was performed by Lie. 
More precisely, his classification~\cite{Lie1881} of all the linear equations
with two independent complex variables includes, in an implicit form,
the solution of the classification problem for the linear $(1+1)$-dimensional
Schr\"odinger equations with an arbitrary potential. 

The specific study of Lie symmetries for Schr\"odinger equations was begun in the 1970s
with the linear case~\cite{Miller1977,Niederer1972,Niederer1973}.
Thereafter, symmetry investigations were extended to nonlinear Schr\"odinger equations. 
All the equations from the class~$\mathscr F$, which are invariant with respect to subalgebras of 
the Lie symmetry algebra of the $(1+1)$-dimensional free Schr\"odinger equation were constructed in~\cite{BoyerSharpWinternitz1976}. 
Later the more general problem of the description of the equations from the class~$\mathscr F$, 
possessing at most three-dimensional Lie invariance algebras, was solved in~\cite{Zhdanov&Roman2000}.
It was observed~\cite{BoyerSharpWinternitz1976,Fushchych&Moskaliuk1981} 
that $(1+n)$-dimensional NSchEs with nonlinearities of the form $F=f(|\psi|)\psi$ 
are notable for their symmetry properties because any such equation is invariant with respect to a representation of the 
$(1+n)$-dimensional Galilean group. Here $n$ is the number of spatial variables.
Extensions of the invariance group are possible for logarithmical~\cite{Fushchych&Chopyk1993} and power functions~\cite{Fushchych&Moskaliuk1981} 
and in fact only for these functions~\cite{Nikitin&Popovych2001}. 
The power $\gamma=4/n$ is special since 
the free Schr\"odinger equation and the NSchE with the nonlinearity~$|\psi|^{4/n}\psi$ are distinguished from many
similar equations by possessing the complete Galilei group extended with both the scale and conformal transformations~\cite{Fushchych&Moskaliuk1981}.
This NSchE also has other exceptional properties, which 
is why now the value $\gamma=4/n$ is called the critical power of NSchEs.
The complete group classification of constant coefficient NSchEs with nonlinearities
of the general form $F=F(\psi,\psi^*)$ was performed in~\cite{Nikitin&Popovych2001}.
Lie symmetries of NSchEs with modular nonlinearities and oscillator potential were classified in~\cite{Ivanova2002}. 

\looseness=-1
Finishing a series of works~\cite{Gagnon&Winternitz1988,Gagnon&Winternitz1989,Gagnon&Winternitz1993} on
group analysis and exact solutions of NSchEs, Gagnon and Winternitz~\cite{Gagnon&Winternitz1993} investigated 
a general class of $(1+1)$-dimensional variable coefficient cubic SchEs.
Doebner and Goldin applied a symmetry approach to obtain new equations which generalize Schr\"odinger 
equations and can be applied in nonlinear quantum mechanics~\cite{Doebner&Goldin1994}.
These equations were investigated in more detail from the symmetry point of view
by a number of authors~\cite{Fushchych&Chopyk&Nattermann&Scherer1995,Doebner&Goldin&Nattermann1999,Nattermann&Doebner1996}.
Different generalizations of NSchEs also arose under the classification 
of Galilei-invariant nonlinear systems of evolution equations in~\cite{Fushchych&Cherniha1995}.
Investigations on conditional invariance and related direct reductions led to an extension of the set of NSchEs 
whose exact solutions were constructed by symmetry or closed methods~\cite{Clarkson1992,Fushchych&Chopyk1990}.
Conditional and Lie symmetries of Schr\"odinger equations were also studied, 
with mass considered as an additional variable~\cite{Stoimenov&Henkel2005}. 
A number of exact solutions of NSchEs are collected~in~\cite{Polyanin&Zaitsev2004}.
Lie symmetries of vector nonlinear Schr\"odinger systems of the form 
$i\boldsymbol\psi_t+\triangle\boldsymbol\psi+S(t,\boldsymbol x,\boldsymbol\psi\cdot\boldsymbol\psi^*)\boldsymbol\psi=\boldsymbol0$
were computed in~\cite{Wittkopf2004,Wittkopf&Reid2000,Wittkopf&Reid2001} 
with a symbolic calculation package to demonstrate its effectiveness for solving overdetermined systems of PDEs. 
Here $\boldsymbol x$ and $\boldsymbol\psi$ are $n$- and $m$-tuples of the space variables and unknown functions, respectively. 
$S$ is a real-valued smooth function of its arguments. 
The determining equations for Lie symmetry operators of certain systems from this class were proposed to be used by other researchers 
as benchmarks for testing their algorithms and program packages dealing with overdetermined systems of PDEs. 
See the discussion in Example~\ref{ExampleOnWittkopf} for details.

\section{Nested normalized classes of $\boldsymbol{(1+1)}$-dimensional\\ nonlinear Schr\"odinger equations}%
\label{SectionOnNestedNormalizedClassesOfNSchEs}

We start with the widest (for the present paper) class~$\mathscr F$ of equations having the general form~\eqref{vgNSchE},
where $F=F(t,x,\psi,\psi^*,\psi_x,\psi^*_x)$ is an arbitrary smooth complex-valued function of its arguments.
Successively prescribing more constraints on the arbitrary element~$F$, we construct a series of
nested normalized classes of $(1+1)$-dimen\-sional nonlinear Schr\"odinger equations.
The final such class still contains Schr\"odinger equations with modular nonlinearities and potentials and
possesses certain properties which allow us to continue with the investigation of
Schr\"odinger equations with modular nonlinearities and potentials.
In what follows, subscripts of functions denote differentiation with respect to the corresponding variables.

In the case of the entire class~\eqref{vgNSchE}, the auxiliary system for arbitrary elements~$F$ consists
of the equations
\[
F_{\psi_{tt}}=F_{\psi_{tx}}=F_{\psi_{xx}}=F_{\psi_{tt}^*}=F_{\psi_{tx}^*}=F_{\psi_{xx}^*}=0, \quad
F_{\psi_t}=F_{\psi_t^*}=0.
\]

Observe that we also have to take into account the conjugate equation to equation~\eqref{vgNSchE}. 
So, we actually work with a system of two conjugate equations for two conjugate unknown functions. 
All invariance and equivalence conditions need only be tested for one of the equations
(since they then automatically hold for the conjugate equation as well), yet on the manifold of the whole system.
Alternatively we could replace equation~\eqref{vgNSchE} by a system of two equations 
in two real unknown functions but this would lead to more complicated computations. 

Any point transformation~$\mathcal T$ in the space of variables of the class~$\mathscr F$ has the form
\[
\tilde t=\mathcal T^t(t,x,\psi,\psi^*),\quad
\tilde x=\mathcal T^x(t,x,\psi,\psi^*),\quad
\tilde \psi=\mathcal T^\psi(t,x,\psi,\psi^*),\quad
\tilde \psi^*=\mathcal T^{\psi^*}(t,x,\psi,\psi^*),\quad
\]
where $\mathcal T^{\psi^*}=(\mathcal T^\psi)^*$ and the Jacobian
$|\p(\mathcal T^t,\mathcal T^x,\mathcal T^\psi,\mathcal T^{\psi^*})/\p(t,x,\psi,\psi^*)|\not=0$.

\begin{lemma}\label{LemmaOnProjectibilitiTransForNSchEs}
If a point transformation~$\mathcal T$ connects two equations from the class~$\mathscr F$ then
\begin{gather*}
\mathcal T^t_x=\mathcal T^t_\psi=\mathcal T^t_{\psi^*}=0,\quad
\mathcal T^x_\psi=\mathcal T^x_{\psi^*}=0,\quad  (\mathcal T^x_x)^2=|\mathcal T^t_t|,\\
\mathcal T^\psi_{\psi^*}=0 \quad\mbox{if}\quad \mathcal T^t_t>0  \quad\mbox{and}\quad
\mathcal T^\psi_\psi=0 \quad\mbox{if}\quad \mathcal T^t_t<0, \quad\mbox{i.e.,}\quad \mathcal T^\psi_{\hat\psi^*}=0.
\end{gather*}
\end{lemma}

\begin{note}
Hereafter in case of any complex value $\beta$ we use the notation
\[
\hat\beta=\beta \quad\mbox{if}\quad \mathcal T^t_t>0  \quad\mbox{and}\quad
\hat\beta=\beta^* \quad\mbox{if}\quad \mathcal T^t_t<0.
\]
\end{note}

\begin{proof}
Let us recall that any equation of the form~\eqref{vgNSchE} is interpreted as a system of two semi-linear evolution equations
in the functions~$\psi$ and~$\psi^*$ (or in the real and imaginary parts of the function~$\psi$). 
In~\cite{Popovych&Kunzinger&Ivanova2007} more general classes of systems of $(1+1)$-dimensional evolution equations were studied 
from the viewpoint of admissible transformations and normalization properties. 
In view of Lemma~4 from~\cite{Popovych&Kunzinger&Ivanova2007}, the evolutionary character of 
such systems implies that~$\mathcal T^t$ is a function only of~$t$.
(The above conditions on transformations with respect to~$t$ were earlier proved
for single evolution equations~\cite{Kingston&Sophocleous1998}).
The condition $\mathcal T^x_\psi=\mathcal T^x_{\psi^*}=0$ follows, according to Lemma~5 of~\cite{Popovych&Kunzinger&Ivanova2007}, 
from the quasi-linearity of the systems. 
Since all the equations from~$\mathscr F$ have the same ratio of the coefficients of $\psi_t$ and $\psi_{xx}$ 
then additionally $(\mathcal T^x_x)^2=|\mathcal T^t_t|$. 
The last condition on the transformation~$\mathcal T$ is derived from the fact that this ratio differs
from the ratio of the coefficients of $\psi^*_t$ and $\psi^*_{xx}$ in the corresponding conjugate equations.
\end{proof}

Any transformation~$\mathcal T$ satisfying the constraints of Lemma~\ref{LemmaOnProjectibilitiTransForNSchEs}
results (after application to an arbitrary equation from the class~$\mathscr F$)
in an equation from the same class.
The corresponding values of the arbitrary elements are connected in a local way,
i.e., the transformation~$\mathcal T$ belongs to the equivalence group~$G^{\sim}_{\mathscr F}$ of the class~$\mathscr F$.
This allows us to reformulate and to strengthen the results of~\cite{Zhdanov&Roman2000}
on the equivalence group~$G^{\sim}_{\mathscr F}$.

\begin{theorem}\label{TheoremOnGequivOfvgNSchE}
The class~$\mathscr F$ is strongly normalized.
The equivalence group~$G^{\sim}_{\mathscr F}$ of the class~$\mathscr F$ is formed by the transformations
\begin{gather}
\tilde t=T(t), \quad
\tilde x=\varepsilon x|T_t|^{1/2}+X(t),
\quad
\tilde \psi=\Phi(t,x,\hat\psi),\nonumber
\\[1ex]
\tilde F=\dfrac1{|T_t|}\biggl( \Phi_{\hat\psi}\hat F-i\varepsilon'\Phi_t+
i\biggl(\dfrac{T_{tt}}{2|T_t|}x+\dfrac{\varepsilon\varepsilon'X_t}{|T_t|^{1/2}}\biggr)\label{vgNSchEeqtransp}
(\Phi_x+\Phi_{\hat\psi}\hat\psi_x)
\\[1ex]\nonumber\qquad
-\Phi_{xx}-2\Phi_{x\hat\psi}\hat\psi_x-\Phi_{\hat\psi\hat\psi}(\hat\psi_x)^2 \biggr),
\end{gather}
where $T$ and $X$ are arbitrary smooth real-valued functions of $t$, $T_t\not=0$,
$\Phi$ is an arbitrary smooth complex-valued function of $t$, $x$ and $\hat\psi$, $\Phi_{\hat\psi}\not=0$.
Hereafter $\varepsilon=\pm1$, $\varepsilon'=\sign T$.
\end{theorem}

\begin{note}
Analogous arguments imply that the wider class of Schr\"odinger-like equations of the general 
form $i\psi_t+G\psi_{xx}+F=0$, where $F$ and  $G$ are arbitrary smooth complex-valued functions of $(t,x,\psi,\psi^*,\psi_x,\psi^*_x)$, 
is normalized. The subclasses with arbitrary $G=G(t,x,\psi,\psi^*)$ or $G=G(t,x)$ or 
$G$ running through real-valued functions of one of the above lists of arguments are also normalized.
A number of different constraints on the arbitrary elements can be posed under which the corresponding classes are normalized.
\end{note}

\begin{note}\label{NoteGequivOfvgNSchE}
In fact, the equivalence group~$G^{\sim}_{\mathscr F}$ is generated by the continuous family of transformations
of the form~\eqref{vgNSchEeqtransp}, where $T_t>0$ and $\varepsilon=1$, and two discrete transformations:
the space reflection $I_x$
($\tilde t=t,$ $\tilde x=-x,$ $\tilde \psi=\psi,$ $\tilde F=F$)
and the Wigner time reflection $I_t$
($\tilde t=-t,$ $\tilde x=x,$ $\tilde \psi=\psi^*,$ $\tilde F=F^*$).
Similar statements are true for the equivalence groups of the classes below.
\end{note}

\begin{note}\label{NoteStrongNormalizationOfvgNSchE}
Strong normalization of the class~$\mathscr F$ is proved easily.
In view of the definition of strongly normalized classes and Note~\ref{NoteGequivOfvgNSchE},
it is enough to prove for each of the following transformations~$\mathcal T$ there exists
an equation of the form~\eqref{vgNSchE} which is invariant under the projection of $\mathcal T$: 
the discrete transformations $I_x$ and $I_t$ and the infinitesimal equivalence transformations.
The projection of~$I_x$ is a point symmetry of equation~\eqref{vgNSchE} iff $F$ is an even function  in~$(x,\psi_x,\psi^*_x)$,
i.e., $F(t,x,\psi,\psi^*,\psi_x,\psi^*_x)=F(t,-x,\psi,\psi^*,-\psi_x,-\psi^*_x)$.
Analogously, equation~\eqref{vgNSchE} is invariant with respect to the projection of~$I_t$ iff
$F(t,x,\psi,\psi^*,\psi_x,\psi^*_x)=F^*(-t,x,\psi,\psi^*,\psi_x,\psi^*_x)$.
The projection of any operator generating a one-parameter subgroup of~$G^{\sim}_{\mathscr F}$ has the form
\[
Q=\tau(t)\p_t+\left(\frac12\tau_tx+\chi(t)\right)\p_x+\eta(t,x,\psi,\psi^*)\p_\psi+\eta^*(t,x,\psi,\psi^*)\p_{\psi^*},
\]
where the coefficients $\tau$, $\chi$ and $\eta$ are arbitrary functions of their arguments.
Infinitesimal invariance  of equation~\eqref{vgNSchE} with respect to the operator~$Q$ implies
a single first-order partial differential equation in the arbitrary element~$F$ which always has a solution.
\end{note}

Let us narrow the class under consideration to the subclass~$\mathscr F'$
with the assumption that the arbitrary element~$F$ does not depend on the derivatives~$\psi_x$ and~$\psi^*_x$,
i.e., we supplement the auxiliary system on $F$ with the constraints $F_{\psi_x}=F_{\psi_x^*}=0$.
The additional constraints on $F$ in the subclass~$\mathscr F'$
imply more conditions on the components of admissible transformations with respect to~$\psi$ and~$\psi^*$.
Namely,
\[
\mathcal T^\psi_{\hat\psi\hat\psi}=0,
\quad
\mathcal T^\psi_{\hat\psi x}=\dfrac{i\varepsilon\varepsilon'\mathcal T^x_t}{2|\mathcal T^t_t|^{1/2}}
\mathcal T^\psi_{\hat\psi}.
\]
Analogously to Theorem~\ref{TheoremOnGequivOfvgNSchE} we arrive at the following statement on the class~$\mathscr F'$.
(Strong normalization of~$\mathscr F'$ is proved in a way similar to Note~\ref{NoteStrongNormalizationOfvgNSchE}.)

\begin{theorem}\label{TheoremOnGequivOfgNSchE}
The subclass~$\mathscr F'$ of $\mathscr F$ satisfying the condition that
the arbitrary element~$F$ depends only on $t$, $x$, $\psi$ and $\psi^*$
is strongly normalized.
Its equivalence group~$G^{\sim}_{\mathscr F'}$ is a subgroup of~$G^{\sim}_{\mathscr F}$
and is formed by the transformations~\eqref{vgNSchEeqtransp}
where
\[
\Phi=\hat\psi \exp\left(\dfrac i8\dfrac{T_{tt}}{|T_t|}\,x^2+
\dfrac i2\dfrac{\varepsilon\varepsilon' X_{t}}{|T_t|^{1/2}}\,x +\Theta(t)+i\Psi(t) \right)\!+\Phi^0(t,x),
\]
and $T$, $X$, $\Theta$ and $\Psi$ are arbitrary smooth real-valued functions of $t$, $T_t\ne 0$,
and $\Phi^0$ is an arbitrary smooth complex-valued function of $t$ and~$x$.
\end{theorem}

We next narrow the class further, in a more specific way:
Consider the class~$\mathscr S$ of equations
\begin{equation}\label{NSchEGMN}
i\psi_t+\psi_{xx}+S(t,x,|\psi|)\psi=0,\quad S_{|\psi|}\not=0
\end{equation}
which encompasses the class of $(1+1)$-dimensional Schr\"odinger equations with potentials and modular nonlinearity
and is more convenient, in some sense, for a preliminary group classification.
Here $S$ is an arbitrary complex-valued function depending on $t$, $x$ and $\rho=|\psi|$,
and we additionally assume $S_\rho\not=0$. The latter condition is invariant under
any point transformation which transforms a fixed equation of the form~\eqref{NSchEGMN}
to an equation of the same form.
The converse condition $S_\rho=0$ corresponds to the linear case which should be investigated separately
because of its singularity. Hence the imposed inequality is natural.

The class~$\mathscr S$ is singled out from the class~$\mathscr F$ by the representation $F=S(t,x,|\psi|)\psi$,
i.e., the arbitrary element~$F$ satisfies, additionally to the above auxiliary equations, the conditions
\[
(\psi \partial_\psi-\psi^*\partial_{\psi^*})(F/\psi)=0,\quad
(\psi \partial_\psi+\psi^*\partial_{\psi^*})(F/\psi)\not=0.
\]
For convenience we will consider $S=F/\psi$ instead of~$F$ as an arbitrary element
depending on no derivatives and also satisfying the conditions
\begin{equation}\label{NSchEGMNsystemForEquivTrans}
\psi S_\psi-\psi^*S_{\psi^*}=0, \quad \psi S_\psi+\psi^*S_{\psi^*}\not=0.
\end{equation}

\begin{theorem}\label{TheoremGequivNSchEGMN}
The class~$\mathscr S$ is strongly normalized.
The equivalence group~$G^{\sim}_{\mathscr S}$ of the class~$\mathscr S$
is the subgroup of~$G^{\sim}_{\mathscr F'}$
defined by the condition~$\Phi^0=0$, i.e., it is formed, in terms of the arbitrary element~$S$,
by the transformations
\begin{equation}\label{NSchEGMNeqtrans}
\arraycolsep=0ex\begin{array}{l}\displaystyle
\tilde t=T(t), \quad
\tilde x=\varepsilon x|T_t|^{1/2}+X(t),
\quad
\tilde \psi=\hat\psi \exp\left(\dfrac i8\dfrac{T_{tt}}{|T_t|}\,x^2+
\dfrac i2\dfrac{\varepsilon\varepsilon' X_{t}}{|T_t|^{1/2}}\,x +\Theta(t)+i\Psi(t) \right)\!,
\\[3ex]
\tilde S=\dfrac{\hat S}{|T_t|}+
\dfrac{2T_{ttt}T_t-3T_{tt}{}^2}{16\varepsilon'T_t{}^3}x^2
+\dfrac{\varepsilon\varepsilon'}{2|T_t|^{1/2}}\left(\dfrac{X_t}{T_t}\right)_{\!t}x
+\dfrac{\Psi_t-i\Theta_t}{T_t}-\dfrac{X_t{}^2+iT_{tt}}{4T_t{}^2}.
\end{array}\end{equation}
Here $T$, $X$, $\Phi$ and $\Psi$ are arbitrary smooth real-valued functions of $t$,
$\varepsilon=\pm1$, $\varepsilon'=\sign T$.
\end{theorem}

See Note~\ref{NoteGequivNSchEGMN} for a discussion of the proof that the class~$\mathscr S$ is strongly normalized.

\begin{corollary}\label{CorollaryInvValueForS}
For any equation from the class~$\mathscr{S}$ the value $\rho S_{\rho\rho}/S_\rho$ is preserved
under any transformation which maps this equation to an equation from the same class,
excluding~$I_t$. In particular, if $\rho S_{\rho\rho}/S_\rho$ is a real-valued function then
it is an invariant of the admissible transformations in the class~$\mathscr{S}$.
\end{corollary}

\begin{note}\label{NoteLocTransNSchEGMN}{\rm
It follows from Theorem~\ref{TheoremGequivNSchEGMN}
that equivalence transformations from the equivalence group of any subclass of~~$\mathscr{S}$
have the form~\eqref{NSchEGMNeqtrans}.
This statement is also true for Lie symmetry transformations of any equation from the class~$\mathscr{S}$
if we assume in~\eqref{NSchEGMNeqtrans} that $S$ is an invariant function.
}\end{note}

In particular, Theorem~\ref{TheoremGequivNSchEGMN} jointly with the infinitesimal Lie method
results in the following statement
on Lie symmetry operators of equations from the class~$\mathscr{S}$.

\begin{theorem}\label{TheoremNSchEGMNOperators}
Any operator~$Q$ from the maximal Lie invariance algebra $A(S)$
of equation~\eqref{NSchEGMN} with an arbitrary function $S$ ($S_\rho\not=0$)
can be represented in the form $Q=D(\tau)+G(\chi)+\lambda M+\zeta I$, where
\begin{gather}
D(\tau)=\tau\p_t+\dfrac12\,\tau_tx\p_x+\dfrac18\,\tau_{tt}x^2M,\quad
G(\chi)=\chi\p_x+\dfrac12\chi_txM,\label{NSchEGMNOperators1}\\[1ex]
M=i(\psi\p_{\psi}-\psi^*\p_{\psi^*}),\quad
I=\psi\p_{\psi}+\psi^*\p_{\psi^*},\nonumber
\end{gather}
where
$\chi=\chi(t),$ $\tau=\tau(t)$, $\lambda=\lambda(t)$ and $\zeta=\zeta(t)$
are arbitrary smooth real-valued functions of $t$.
Moreover, the coefficients of $Q$ have to satisfy the classifying condition
\begin{equation}\label{NSchEGMNClassifyingCondition}
\tau S_t+\left(\frac12\tau_tx+\chi\right)S_x+\zeta\rho S_\rho+\tau_tS=
\frac18\,\tau_{ttt}x^2+\frac12\chi_{tt}x+\lambda_t-i\zeta_t-i\frac14\,\tau_{tt}.
\end{equation}
\end{theorem}

Theorem~\ref{TheoremNSchEGMNOperators} can also be proved by direct application of the infinitesimal Lie method.
Namely, consider an operator from $A(S)$ in the most general form
$Q=\xi^t\partial_t+\xi^x\partial_x+\eta\partial_\psi+\eta^*\partial_{\psi^*}$,
where $\xi^t,$ $\xi^x$ and $\eta$ are smooth functions of $t,$ $x,$ $\psi$ and $\psi^*.$
The infinitesimal invariance condition~\cite{Olver1986,Ovsiannikov1982}
of equation~\eqref{NSchEGMN} with respect to the operator~$Q$
implies the following linear overdetermined system for the coefficients of~$Q$:
\begin{gather*}
\xi^t_\psi=\xi^t_{\psi^*}=\xi^x_\psi=\xi^x_{\psi^*}=0,\quad
\xi^t_x=0,\quad \xi^t_t=2\xi^x_x,\quad
\eta_{\psi^*}=\eta_{\psi\psi}=0, \quad \psi\eta_\psi=\eta,\\
2\eta_{\psi x}=i\xi^x_t,\quad
i\eta_{\psi t}+\eta_{\psi xx}+\xi^tS_t+\xi^xS_x+\rho S_\rho\mathop{\rm Re}\nolimits\eta_\psi+\xi^t_tS=0.
\end{gather*}
Solving the latter system implies Theorem~\ref{TheoremNSchEGMNOperators}.

\begin{note}\label{NoteGequivNSchEGMN}
Under the proof of strong normalization of the class~$\mathscr S$ in a way similar to Note~\ref{NoteStrongNormalizationOfvgNSchE},
the operator projections have the form $Q=D(\tau)+G(\chi)+\lambda M+\zeta I$ as operators in Theorem~\ref{TheoremNSchEGMNOperators}.
In view of the classifying condition~\eqref{NSchEGMNClassifyingCondition},
$Q$ is a Lie invariance operator of an equation of form~\eqref{NSchEGMN} iff $(\tau,\chi,\zeta)\ne(0,0,0)$ or $\lambda_t=0$.
It is enough to deduce the statement on strong normalization of the class $S$.
\end{note}

Assuming $S$ to be arbitrary and splitting~\eqref{NSchEGMNClassifyingCondition}
with respect to $S$, $S_t$, $S_x$ and $S_\rho$, we obtain that
the Lie algebra of the kernel~$G^\cap_{\mathscr{S}}$ of maximal Lie invariance groups of equations
from the class~$\mathscr{S}$ is $A^\cap_{\mathscr{S}}=\langle M \rangle.$
The complete group~$G^\cap_{\mathscr{S}}$ coincides with the projection, to $(t,x,\psi)$,
of the normal subgroup~$\hat G^\cap_{\mathscr{S}}$
of~$G^\sim_{\mathscr{S}}$, which include the transformations~\eqref{NSchEGMNeqtrans}
acting on the arbitrary element~$S$ identically (i.e., $T=t$ and $X=\Theta=\Psi_t=0$,  
see Section~\ref{SectionOnGroupClassificationProblems}).

\begin{theorem}
$\hat G^\cap_{\mathscr{S}}$ is formed by the transformations
$\tilde t=t$, $\tilde x=x$, $\tilde\psi=\psi e^{i\Phi}$, $\tilde S=S$,
where $\Phi$ is an arbitrary constant.
\end{theorem}

\begin{note}
The operators $D(\tau)$, $G(\chi)$, $\lambda M$ and $\zeta I$,
where $\tau$, $\chi$, $\lambda$ and $\zeta$ run through the whole set of smooth functions of~$t$,
generate an infinite-dimensional Lie algebra~$A^{\cup}_{\mathscr{S}}$ under the usual Lie bracket of vector fields.
The non-zero commutation relations between the basis operators of~$A^{\cup}_{\mathscr{S}}$ are the following ones:
\[\arraycolsep=0ex\begin{array}{l}\displaystyle
[D(\tau^1),D(\tau^2)]=D(\tau^1\tau^2_t-\tau^2\tau^1_t),\qquad
[D(\tau),G(\chi)]=G\Bigl(\tau\chi_t-\frac12\tau_t\chi\Bigr),\\[1.5ex]\displaystyle
[D(\tau),\lambda M]=\tau\lambda_t M,\qquad
[D(\tau),\zeta I]=\tau\zeta_t I,\qquad
[G(\chi^1),G(\chi^1)]=\frac12(\chi^1\chi^2_t-\chi^2\chi^1_t)M.
\end{array}\]
\end{note}

\begin{note}
Sometimes (e.g. for reduction and construction of solutions)
it is convenient to use the amplitude~$\rho$ and the phase~$\varphi$
instead of the wave function~$\psi=\rho e^{i \varphi}.$ Then equation~\eqref{NSchEGMN}
is replaced by the following system for the two real-valued functions~$\rho$ and~$\varphi$:
\[
\rho_t+2\rho_x\varphi_x+\rho\varphi_{xx}+\rho\mathop{\rm Im}\nolimits S=0,\quad
-\rho\varphi_t-\rho(\varphi_x)^2+\rho_{xx}+\rho\mathop{\rm Re}\nolimits S=0
\]
where $S=S(t,x,\rho)$.
The constraining system for $S$ takes the form $S_\varphi=0$, $S_\rho\not=0$.
In the variables $(\rho,\varphi)$ the operators~$D(\tau)$ and~$G(\chi)$ have the same form~\eqref{NSchEGMNOperators1},
and $M=\partial_\varphi$, $I=\rho\partial_\rho$.
Below we use the variables $(\rho,\varphi)$ and $(\psi,\psi^*)$ simultaneously.
\end{note}

\section{Group classification of $\boldsymbol{(1+1)}$-dimensional\\ nonlinear Schr\"odinger equations
with potentials\\ and modular nonlinearities}\label{SectionOnGroupClassificationOfNSchEsWithMNP}

Let us pass to the subclass~$\mathscr V$ of the class~$\mathscr S$
which consists of the equations of the general form
\begin{equation}\label{NSchEPMN}
i\psi_t+\psi_{xx}+f(|\psi|)\psi+V\psi=0,
\end{equation}
where $f$ is an arbitrary complex-valued nonlinearity depending only on $\rho=|\psi|$, $f_\rho\not=0$, and
$V$ is an arbitrary complex-valued potential depending on $t$ and $x$.
The arbitrary element $S$ is represented as $S=f(\rho)+V(t,x)$, where $f_\rho\ne 0$.
Therefore, this subclass is derived from the class~$\mathscr S$
by the condition $S_{\rho t}=S_{\rho x}=0$ and $S_\rho\ne0$ or, in terms of $\psi$ and $\psi^*$,
\begin{equation}\label{NSchEPMNaddSystemForEquivTrans}
\psi S_{\psi t}+\psi^*S_{\psi^* t}=\psi S_{\psi x}+\psi^*S_{\psi^* x}=0, \qquad \psi S_\psi+\psi^*S_{\psi^*}\ne0.
\end{equation}
The group classification problem for the class~$\mathscr V$ is solved in this section.
Although the class~$\mathscr V$ is not normalized, the approach based on normalization is still applicable to this class
due to its representation in the form of a union of normalized classes.

\subsection{Equivalence groups and admissible transformations}\label{SectionOnAdmTransOfNSchEsWithMNP}

To find the equivalence group~$G^\sim_{\mathscr V}$ of the class~$\mathscr V$
in the framework of the direct method,
we look for all point transformations in the space of the variables
$t$, $x$, $\psi$, $\psi^*$, $S$ and $S^*$
which preserve the system formed by equations~\eqref{NSchEGMN}, \eqref{NSchEGMNsystemForEquivTrans}
and~\eqref{NSchEPMNaddSystemForEquivTrans}.
Moreover, in the same way we can classify all admissible point transformations in
the class~$\mathscr V$.

\begin{theorem}\label{TheoremGequivNSchEPMN}
$G^\sim_{\mathscr V}$ is formed by the transformations~\eqref{NSchEGMNeqtrans}
where $T_{tt}=0$ and $\Psi_t=0$.
The class~$\mathscr V$ is not normalized.
The subclass~$\mathscr V'$ of~$\mathscr V$ under the additional condition
that the value $\rho f_{\rho\rho}/f_\rho$ $({}\equiv\rho S_{\rho\rho}/S_\rho)$ is not a real constant
has the same equivalence group and is normalized.
There exist two different cases for additional (conditional) equivalence
transformations in the class~$\mathscr V$ ($\sigma$ is a complex constant):

\vspace{1ex}

{\rm 1.} $\rho f_{\rho\rho}/f_\rho=-1$, i.e., $f=\sigma\ln\rho$.

\vspace{1ex}

{\rm 2.} $\rho f_{\rho\rho}/f_\rho=\gamma-1\in\mathbb{R}$ and $\gamma\ne0$, i.e., $f=\sigma\rho^\gamma$.

\vspace{1ex}

\noindent
For any real constant~$\gamma$ the subclass~$\mathscr P_\gamma$ consisting of equations~\eqref{NSchEPMN},
where $\rho f_{\rho\rho}/f_\rho=\gamma-1$, is normalized.
There are no point transformations between equations from different subclasses
taken from the set $\{\mathscr V', \mathscr P_\gamma, \gamma\in\mathbb R\}$.
\end{theorem}

\begin{note}\label{NoteNSchEPMNTrivialEquivalence}
It is possible to find equivalence transformations in another way, considering
$f$ and $V$ as arbitrary elements instead of $S$.
Then we have to look for all point transformations in the space of the variables
$t$, $x$, $\psi$, $\psi^*$, $f$, $f^*$, $V$ and $V^*$
which preserve the system formed by the equations
\[
i\psi_t+\psi_{xx}+(f+V)\psi=0,\quad f_t=f_x=0,\quad \psi f_\psi-\psi^*f_{\psi^*}=0, \quad V_\psi=V_{\psi^*}=0,
\]
and additionally $\psi f_\psi+\psi^*f_{\psi^*}\ne0$.
Due to the representation  $S=f+V$
we additionally obtain only gauge equivalence transformations of the form
$\tilde f=f+\beta$, $\tilde V=V-\beta$, where $\beta$ is an arbitrary complex number
and $t$, $x$ and $\psi$ are not changed.
We neglect these transformations, choosing $f$ in the most suitable form.
For example, this is the reason why we can assume $f=\sigma\ln\rho$ in the case $\rho f_{\rho\rho}/f_\rho=-1$.
Analogously, we put $f=\sigma\rho^\gamma$ up to gauge equivalence transformations 
if $\rho f_{\rho\rho}/f_\rho=\gamma-1\in\mathbb{R}$ and $\gamma\ne0$.
\end{note}

\begin{note}\label{NoteNSchEPMNAdmissibleTrans}
Theorem~\ref{TheoremGequivNSchEPMN} gives an exhaustive description of the set of admissible transformations
of the class~$\mathscr V$.
Indeed, the class~$\mathscr V$ is not normalized but it is presented as the union of
disjoint normalized subclasses~$\mathscr V'$ and $\mathscr P_\gamma$, $\gamma\in\mathbb{R}$,
and there are no equations from different subclasses which are equivalent with respect to point transformations.
Therefore, the set of admissible transformations of the class~$\mathscr V$ is the union of
the sets of admissible transformations in the subclasses, which are generated by the corresponding
conditional equivalence groups. This gives an example on the second simplest structure of sets of admissible transformations,
described in Subsection~\ref{SectionOnNormSubclassesAndAdmTrans}.
\end{note}

\looseness=-1
The equivalence groups $G^\sim_{\mathscr V'}$ and $\smash{G^\sim_{\mathscr P_\gamma}}$, $\gamma\in\mathbb{R}$,
exhaust the set of the maximal conditional equivalence groups of the class~$\mathscr V$.
(See the next sections for exact formulas.)
Since they are different, the subclasses~$\mathscr V'$ and $\mathscr P_\gamma$, $\gamma\in\mathbb{R}$, are all
the maximal normalized subclasses of the class~$\mathscr V$.
Admitting generalized equivalence groups, we can unite the subclasses~$\mathscr P_\gamma$, $\gamma\not=0$,
into the total subclass~$\smash{\hat{\mathscr P}}$ of equations~\eqref{NSchEPMN} with power nonlinearities.
The subclass~$\smash{\hat{\mathscr P}}$ is normalized in the generalized sense with respect to its generalized
equivalence group $\smash{G^\sim_{\mathstrut\smash{\hat{\mathscr P}}}}$ which is an interlacement
of the groups $\smash{G^\sim_{\mathscr P_\gamma}}$, $\gamma\not=0$, by the parameter~$\gamma$.
The equivalence group $\smash{G^\sim_{\mathstrut\smash{\hat{\mathscr P}}}}$ is generalized since transformations with respect to~$\psi$
depend on the arbitrary element~$\gamma$ of the class~$\hat{\mathscr P}$.
Therefore, the class~$\mathscr V$ possesses only three maximal conditional generalized equivalence groups
$G^\sim_{\mathscr V'}$, $\smash{G^\sim_{\mathstrut\smash{\hat{\mathscr P}}}}$ and $\smash{G^\sim_{\mathscr P_0}}$
corresponding to the maximal normalized (in the generalized sense) subclasses $\mathscr V'$, $\smash{\hat{\mathscr P}}$ and $\mathscr P_0$.
There are no other maximal subclasses of~$\mathscr V$, which are normalized in the generalized sense.

Since the class~$\mathscr V$ is not normalized and possesses nontrivial conditional equivalence groups,
the entire set of admissible transformations is much wider than its subset associated with the equivalence group $G^\sim_{\mathscr V}$.
Therefore, under group classification with standard techniques a number of $G^\sim_{\mathscr V}$-inequivalent but
point-transformation equivalent cases should be included in the final classification list of equations with Lie symmetry extensions and then
a number of additional equivalence transformations should be found between such cases for the list to be simplified.

We use a more efficient way.
Namely, we separately carry out the group classification in each maximal normalized subclass
with respect to the corresponding conditional equivalence group, using the approach to group classification in
normalized classes of differential equations developed above.
The proposed algorithm is implemented in the different subclasses in a maximally unified way to demonstrate its capacities.
The classification list for the whole class~$\mathscr V$ is constructed as the union of the lists obtained for the subclasses.
Due to the special structure of the set of maximal normalized subclasses of the class~$\mathscr V$,
this list is formed by all the extension cases inequivalent with respect to point transformations.
It can be expanded to the list of $G^\sim_{\mathscr V}$-inequivalent extension cases
by the conditional equivalence transformations factorized with respect to~$G^\sim_{\mathscr V}$.

\begin{note}
Theorem~\ref{TheoremGequivNSchEGMN} and the results of Sections~\ref{SectionOnGenMNP}--\ref{SectionOnPowerMNP} imply
that $G^\sim_{\mathscr{S}}$ is generated by the transformational parts of admissible transformations of class~\eqref{NSchEPMN}.
More precisely,  for any fixed $\gamma\ne0$ each transformation from $G^\sim_{\mathscr{S}}$ 
is presented as a composition of transformations from the maximal conditional equivalence groups 
$\smash{G^\sim_{\mathscr P_0}}$ and $\smash{G^\sim_{\mathscr P_\gamma}}$.
At the same time, the class~$\mathscr{S}$ is not the minimal normalized superclass of the class~\eqref{NSchEPMN}.
It is obvious that such a class is given by the subclass of~$\mathscr{S}$ which is determined by
the condition $S=R'(t)f(R(t)|\psi|)+V(t,x)$, where $R$ and~$R'$ are arbitrary smooth real-valued functions of $t$.
\end{note}

\subsection{General case of nonlinearity}\label{SectionOnGenMNP}

In this section we adduce results only for the general case
$\rho f_{\rho\rho}/f_\rho\ne\const\in\mathbb{R}$ (the subclass~$\mathscr V'$).
The cases $f=\sigma\ln\rho$ and $f=\sigma\rho^\gamma$ (the subclass~$\mathscr P_0$ and~$\mathscr P_\gamma$, $\gamma\not=0$) which admit
extensions of conditional equivalence groups are considered in the next sections in detail. 
The classifications of the subclasses are maximally unified to demonstrate 
the universality of our approach to group classification problems.

In spite of the absence of equivalence extensions, equations with nonlinearities of the general case can possess 
sufficiently large Lie invariance algebras. 
Moreover, special nonlinearities of this kind arise in applications and mathematical investigations 
\cite{Gagnon&Winternitz1988,Polyanin&Zaitsev2004}.

Theorem~\ref{TheoremGequivNSchEPMN} implies the following statement. 

\begin{corollary}\label{CorollaryNSchEGenMNVanishingPotentail}
A potential $V$ in~\eqref{NSchEPMN} with nonlinearity of the general form can be made to vanish by means of point transformations iff
it is a function which is real-valued up to gauge equivalence transformations and is linear with respect to $x$.
\end{corollary}

\begin{note}
The action of $G^\sim_{\mathscr V}$ on $f$ is only multiplication with non-zero real constants
and/or complex conjugation.
That is why the general case can be split into an infinite number of subclasses, and
each subclass is formed by equations with nonlinearities which are proportional to an arbitrary fixed function
or its conjugation with real constant coefficients and is normalized.
Moreover, we can restrict our consideration to the class of equations with
an arbitrary fixed nonlinearity~$f(\rho)$, assuming $f$ as determined up to a real multiplier or/and complex conjugation
and only $V$ as an arbitrary element.
The equivalence group of such a restricted class~$\mathscr V^f$,
where $\rho f_{\rho\rho}/f_\rho$ is not a real constant,
will be denoted by~$G^\sim_f$ and
is formed by the transformations~\eqref{NSchEGMNeqtrans} with $T_t=1$ ($T_t=\pm1$ if $f$ is a real-valued function)
and $\Psi=0$.
\end{note}

Substitution of $S=f(\rho)+V(t,x)$ into the classifying condition~\eqref{NSchEGMNClassifyingCondition} and subsequent splitting
under the condition $\rho f_{\rho\rho}/f_\rho\ne\const\in\mathbb{R}$ imply the equations $\tau_t=0$, $\zeta=0$.
In view of Theorem~\ref{TheoremNSchEGMNOperators} the following statement is true.

\begin{lemma}\label{TheoremNSchEPMNOperators}
Any operator~$Q$ from the maximal Lie invariance algebra $A_f(V)$
of equation~\eqref{NSchEPMN} in the case where $\rho f_{\rho\rho}/f_\rho$ is not a real constant
can be presented in the form $Q=c_0\p_t+G(\chi)+\lambda M$,
where $\chi=\chi(t)$ and $\lambda=\lambda(t)$ are arbitrary smooth real-valued functions of $t$, $c_0=\const$.
Moreover, the coefficients of $Q$ should satisfy the classifying condition
\begin{equation}\label{NSchEPMNGenCaseClassifyingCondition}
c_0V_t+\chi V_x=\frac12\chi_{tt}x+\lambda_t.
\end{equation}
The kernel of the maximal Lie invariance groups of equations from the class~$\mathscr V^f$ is
$G^\cap_f= G^\cap_{\mathscr V}=G^\cap_{\mathscr{S}}$,
and its Lie algebra is $A^\cap_f=\langle M \rangle$.
\end{lemma}

Let us briefly sketch a chain of statements which yields,
in view of the results of Section~\ref{SectionOnNormalizedClasses},
a complete group classification of the class~$\mathscr V^f$.

The set $A^{\cup}_f=\{Q=c_0\p_t+G(\chi)+\lambda M\}$
is an (infinite-dimensional) Lie algebra under the usual Lie bracket of vector fields.
For any $Q\in A^{\cup}_f$ where $(c_0,\chi)\ne(0,0)$
we can find $V$ satisfying condition~\eqref{NSchEPMNGenCaseClassifyingCondition}, i.e.,
$A^{\cup}_f=\langle\,\bigcup_V A_{f}(V)\,\rangle$.
Moreover, the space reflection belongs to the point symmetry groups of certain equations from $\mathscr V^f$.
The same statement is true for the Wigner time reflection if $f$ is a real-valued function.
Therefore, $\mathscr V^f$ is a strongly normalized class of differential equations.
(This was a reason for introducing the class $\mathscr V^f$ since the class $\mathscr V'$ is not strongly normalized.)

The group~$G^\sim_f$ acts on $A^{\cup}_f$
and on the set of equations of the form~\eqref{NSchEPMNGenCaseClassifyingCondition}
and, therefore, generates equivalence relations in these sets.
The automorphism group generated on $A^{\cup}_f$ and
the equivalence group generated on the set of equations~\eqref{NSchEPMNGenCaseClassifyingCondition}
are isomorphic to $G^\sim_f/\hat G^\cap_{\mathscr{S}}$.
(The transformations from $\hat G^\cap_{\mathscr{S}}$ act on~\eqref{NSchEPMNGenCaseClassifyingCondition}
as gauge equivalence transformations and can be neglected.)
$A^\cap_f=\langle M \rangle$ coincides with the center of the algebra~$A^{\cup}_f$
and is invariant with respect to $G^\sim_f$.

Let $A^1$ and $A^2$ be the maximal Lie invariance algebras of some equations from the class~$\mathscr V^f$,
and ${\cal V}^i=\{\,V\,|\,A_f(V)=A^i\},$ $i=1,2.$
Since the class~$\mathscr V^f$ is normalized
then ${\cal V}^1\sim {\cal V}^2\bmod G^\sim_f$ iff
$\,A^1\sim A^2\bmod G^\sim_f.$

A complete list of $G^\sim_f$-inequivalent one-dimensional subalgebras
of~$A^{\cup}_f$ is exhausted by the algebras
$\langle\partial_t\rangle$, $\langle G(\chi)\rangle$, $\langle \lambda M\rangle$,
where $\chi$ and $\lambda$ are arbitrary fixed functions of~$t$.
(There exist additional equivalences in $\{\langle G(\chi)\rangle\}$
and $\{\langle \lambda M\rangle\}$, which are generated by translations with respect to~$t$
and, if $f$ is real-valued, by the Wigner time reflection $I_t$.)

\begin{note}
For convenience we use below the double numeration T.N of classification cases
where T is a table number and N is a row number.
We mean that the invariance algebras for Cases~1.0, 1.\ref{NSchEPMNcase1} and
the corresponding ones from the next tables are maximal if these cases are inequivalent under the
corresponding equivalence group to the other, more specialized, cases
from the same table.
\end{note}

\begin{theorem}\label{TheoremNSchPMNGroupClassification}
A complete set of inequivalent potentials admitting extensions of the maximal Lie invariance algebra
of equation~\eqref{NSchEPMN} in the case where $\rho f_{\rho\rho}/f_\rho$ is not a real constant
is exhausted by the potentials given in Table~1.
\end{theorem}

{\samepage\begin{center}\small
Table 1. Results of classification in the case $\rho f_{\rho\rho}/f_\rho\ne\const\in\mathbb R$.
\\[1.5ex]
\setcounter{tbn}{0}
\renewcommand{\arraystretch}{1.4}
\begin{tabular}{|r|c|l|}
\hline\vspacebefore
N &$V$ &\hfill {Basis of $A_f(V)$\hfill} \\
\hline\vspacebefore
\thetbn& $V(t,x)$ & $M\;$ \\
\hline\vspacebefore
\refstepcounter{tbn}\thetbn\label{NSchEPMNcase1}&
$V(x)$ & $M,\;$ $\p_t$\\
\refstepcounter{tbn}\thetbn\label{NSchEPMNcase2}&
$v(t)x^2+iw(t)$ & $M,\;$ $G(\chi^1),\;$ $G(\chi^2)$ \\
\refstepcounter{tbn}\thetbn\label{NSchEPMNcase3}&
$0$ or $i$ & $M,\;$ $\p_t,\;$ $G(1),\;$ $G(t)$ \\
\refstepcounter{tbn}\thetbn\label{NSchEPMNcase4}&
$x^2+i\nu$ & $M,\;$ $\p_t,\;$ $G(e^{-2t}),\;$ $G(e^{2t})$ \\
\refstepcounter{tbn}\thetbn\label{NSchEPMNcase5}&
$-x^2+i\nu$ & $M,\;$ $\p_t,\;$ $G(\cos 2t),\;$ $G(\sin 2t)$ \\
\hline
\end{tabular}
\\[2ex]
\parbox{130mm}{Here $v(t),w(t),\nu\in\mathbb{R}$, $(v_t,w_t)\ne(0,0)$.
The functions~$\chi^1=\chi^1(t)$ and~$\chi^2=\chi^2(t)$ form a fundamental system of
solutions for the ordinary differential equation $\chi_{tt}=4v\chi$.}
\end{center}

}

\begin{proof}
Suppose that equation~\eqref{NSchEPMN} has an extension of the Lie invariance algebra for a potential~$V$,
i.e., $A_f(V)\ne A^\cap_f$. Then there exists an operator
$Q=c_0\p_t+G(\chi)+\lambda M\in A_f(V)$ such that $(c_0,\chi)\ne(0,0)$.

If $c_0\ne0$ then $\langle Q\rangle\sim \langle\p_t\rangle\!\!\mod\!G^\sim_f$,
i.e., we obtain Case~1.\ref{NSchEPMNcase1}.
Investigations of additional extensions are reduced to the next case.

If $c_0=0$ then $\langle Q\rangle\sim \langle G(\chi)\rangle\!\!\mod\!G^\sim_f$.
It follows from~\eqref{NSchEPMNGenCaseClassifyingCondition} that the potential $V$ has the form
$V=v(t)x^2+iw(t)+\tilde w(t)$, and $\tilde w=0\!\!\mod\!G^\sim_f$.
For $(v_t,w_t)\ne(0,0)$ we have Case~1.\ref{NSchEPMNcase2}. The condition $v,w=\const$ results
in Cases~1.\ref{NSchEPMNcase3}, 1.\ref{NSchEPMNcase4} and~1.\ref{NSchEPMNcase5}
depending on the sign of~$v$. If $v=0$ and $w=\const$ we can reduce $w$
by means of equivalence transformations to either~0 or~1.
\end{proof}

\subsection{Logarithmic modular nonlinearity}\label{SectionOnLogMNP}

Consider the first subclass~$\mathscr P_0$ of class~$\mathscr V$,
which is introduced in Theorem~\ref{TheoremGequivNSchEPMN}
and admits extensions of the equivalence and Lie symmetry groups
in comparison with the whole class~$\mathscr V$.
It is formed by the equations
\begin{equation}\label{NSchEPLogMN}
i\psi_t+\psi_{xx}+\sigma\psi\ln|\psi| +V(t,x)\psi=0.
\end{equation}
Here $\sigma$ is an arbitrary non-zero complex number and
$V$ is an arbitrary complex-valued function of~$t$ and~$x$.
This subclass is distinguished from the larger class~$\mathscr S$ of equations~\eqref{NSchEGMN} by the condition
$S_{\rho t}=S_{\rho x}=0$ and $(\rho S_\rho)_\rho=0$, i.e., 
\begin{equation}\label{NSchEPLogMNaddSystemForEquivTrans}
\psi S_{\psi t}+\psi^*S_{\psi^* t}=\psi S_{\psi x}+\psi^*S_{\psi^* x}=0, \quad
(\psi\p_\psi+\psi^*\p_{\psi^*})^2S=0.
\end{equation}

Schr\"odinger equations with logarithmic nonlinearity were first proposed by Bia\l ynicki-Birula and Mycielski 
\cite{Bialynicki-BirulaMycielski1975} as possible models of nonlinear quantum mechanics. 
Although the possibility was later called in question, due to their nice mathematical properties
these equations are applied for the description of nonlinear phenomena in many other fields of physics
(the theory of dissipative systems, nuclear physics, optics and geophysics),
see, e.g., references in~\cite{Bialynicki-BirulaSowinski2004}. 
The maximal Lie invariance algebras of such equations with the zero potential were found in~\cite{Fushchych&Chopyk1993}, 
their exact solutions were already constructed in~\cite{Bialynicki-BirulaMycielski1975}. 
Solutions for other potentials are also known~\cite{Bialynicki-BirulaSowinski2004}. 

Similarly to the previous section, we have two ways of finding
the equivalence group~$G^\sim_{\mathscr P_0}$ of the class~$\mathscr P_0$
in the framework of the direct method.
In the first approach we look for all point transformations in the space of the variables
$t,$ $x,$ $\psi,$ $\psi^*$, $S$ and $S^*$
which preserve the system formed by equations~\eqref{NSchEGMNsystemForEquivTrans}
and~\eqref{NSchEPLogMNaddSystemForEquivTrans} under the condition that $S$ does not depend on derivatives.
Moreover, in the same way we can easily classify all possible point transformations in
the class~$\mathscr P_0$ using Note~\ref{NoteLocTransNSchEGMN}.
The second, completely equivalent, way is
to consider $\sigma$ and $V$ as arbitrary elements instead of $S$
and to find all point symmetry transformations for the system
\[
i\psi_t+\psi_{xx}+\sigma\psi\ln|\psi| +V\psi=0,\quad
\sigma_t=\sigma_x=\sigma_\psi=\sigma_{\psi^*}=0, \quad V_\psi=V_{\psi^*}=0.
\]

\begin{theorem}\label{TheoremGequivNSchEPLogMN}
$G^\sim_{\mathscr P_0}$ is formed by the transformations~\eqref{NSchEGMNeqtrans} where $T_{tt}=0$.
The corresponding transformations with respect to $\sigma$ and $V$ have the form
\[
\tilde\sigma=\dfrac{\hat\sigma}{|T_t|}, \qquad
\tilde V=\dfrac{\hat V}{|T_t|}
+\dfrac{\varepsilon X_{tt}}{2|T_t|^{3/2}}x-\hat\sigma\dfrac{\Theta}{|T_t|}
-\dfrac 14\dfrac{X_t^2}{T_t^2}+\dfrac{\Psi_t}{T_t}-i\dfrac{\Theta_t}{T_t}.
\]
Moreover, the class~$\mathscr P_0$ is normalized.
\end{theorem}

\begin{corollary}\label{CorollaryNSchELogMNVanishingPotentail}
A potential $V$ can be made to vanish in~\eqref{NSchEPLogMN} by means of point transformations iff
$V_{xx}=0$, and the coefficient of~$x$ in~$V$ is a real-valued function of $t$.
\end{corollary}

\begin{note}
The action of $G^\sim_{\mathscr P_0}$ on $\sigma$ is simply multiplication with non-zero real constants
and/or complex conjugation.
That is why we can fix an arbitrary value $\sigma$, supposing e.g.\ $|\sigma|=1$ and $\sigma_2\ge 0$,
and assuming only $V$ as an arbitrary element.
The equivalence group $G^\sim_{\mathscr P_0^\sigma}$ of the corresponding restricted class~$\mathscr P_0^\sigma$
 is formed by the transformations~\eqref{NSchEGMNeqtrans}
where $T_t=1$ if $\sigma_2>0$ and $T_t=\pm 1$ if $\sigma_2=0$.
In what follows we use the notation
$\sigma_1=\mathop{\rm Re}\nolimits \sigma$, $\sigma_2=\mathop{\rm Im}\nolimits \sigma$.
\end{note}

Substituting~$S=\sigma\ln\rho+V(t,x)$ into equation~\eqref{NSchEGMNClassifyingCondition} and subsequently splitting
with respect to~$\rho$ implies the additional equation $\tau_t=0$. As a result, we obtain the following statement
in view of Theorem~\ref{TheoremNSchEGMNOperators}.

\begin{lemma}\label{TheoremNSchEPLogMNOperators}
Any operator~$Q$ from the maximal Lie invariance algebra $A_{\ln}(V)$
of equation~\eqref{NSchEPLogMN} with an arbitrary potential $V$
can be represented in the form $Q=c_0\p_t+G(\chi)+\lambda M+\zeta I$.
Moreover, the coefficients of $Q$ have to satisfy the classifying condition
\begin{equation}\label{NSchEPMNLogCaseClassifyingCondition}
c_0V_t+\chi V_x=\frac12\chi_{tt}x+\lambda_t-i\zeta_t-\sigma\zeta.
\end{equation}
The kernel~$G^\cap_{\ln}$ of the maximal Lie invariance
groups of equations from the class~$\mathscr P_0^\sigma$ is formed
by the transformations~\eqref{NSchEGMNeqtrans}, where $T=t$, $X=0$,
$\Theta=\Theta^0\sigma_2e^{-\sigma_2t}$, $\Psi=\Psi^0-\Theta^0\sigma_1e^{-\sigma_2t}$ if $\sigma_2\not=0$ and
$\Theta=\Theta^0$, $\Psi=\Psi^0+\Theta^0\sigma_1t$ if $\sigma_2=0$.
(Here $\Theta^0$ and $\Psi^0$ are arbitrary constants.)
Its Lie algebra $A^\cap_{\ln}$ is $\langle M,\, I'\rangle$, where
$I'=e^{-\sigma_2t}(\sigma_2I-\sigma_1M)$ if $\sigma_2\ne0$ and $I'=I+\sigma_1tM$ if~$\sigma_2=0$.
\end{lemma}

Hereafter we use the subscript `$\ln$' instead of `$\mathscr P_0^\sigma$' for simplicity.
It is necessary, though, to remember that all objects marked in this way are parametrized by~$\sigma$.

The following chain of statements results in a complete group classification of \eqref{NSchEPLogMN}.

The set $A^{\cup}_{\ln}=\{Q=c_0\p_t+G(\chi)+\lambda M+\zeta I\}$
is an (infinite-dimensional) Lie algebra under the usual Lie bracket of vector fields.
For any $Q\in A^{\cup}_{\ln}$ where $(c_0,\chi)\ne(0,0)$
we can find $V$ satisfying condition~\eqref{NSchEPMNLogCaseClassifyingCondition},
i.e., $A^{\cup}_{\ln}=\langle\,\bigcup_{V} A_{\ln}(V)\,\rangle$.
Moreover, the space reflection belongs to the point symmetry groups of certain equations from $\mathscr P_0^\sigma$.
The same statement is true for the Wigner time reflection if $\sigma$ is real.
Therefore, $\mathscr P_0^\sigma$ is a strongly normalized class of differential equations.
(As in the previous section, this was a reason for introducing the class $\mathscr P_0^\sigma$
since the whole class $\mathscr P_0$ is not strongly normalized.)

The group~$G^\sim_{\ln}$ acts on $A^{\cup}_{\ln}$
and on the set of equations of the form~\eqref{NSchEPMNLogCaseClassifyingCondition}
and, therefore, generates equivalence relations in these sets.
The automorphism group generated on  $A^{\cup}_{\ln}$ is isomorphic to $G^\sim_{\ln}/\hat G^\cap_{\mathscr{S}}$.
The non-trivial equivalence group generated on the set of equations~\eqref{NSchEPMNLogCaseClassifyingCondition}
is isomorphic to $G^\sim_{\ln}/\hat G^\cap_{\ln}$ where $\hat G^\cap_{\ln}$ is the normal subgroup of~$G^\sim_{\ln}$
corresponding to~$G^\cap_{\ln}$.
The transformations from~$\hat G^\cap_{\ln}$ are gauge equivalence transformations
of \eqref{NSchEPMNLogCaseClassifyingCondition}.
$A^\cap_{\ln}=\langle M,I' \rangle$ is an ideal of the algebra~$A^{\cup}_{\ln}$
and is invariant with respect to $G^\sim_{\ln}$.

Let $A^1$ and $A^2$ be the maximal Lie invariance algebras of some equations from the class $\mathscr P_0^\sigma$,
and ${\cal V}^i=\{\,V\,|\,A_{\ln}(V)=A^i\},$ $i=1,2.$
Then ${\cal V}^1\sim {\cal V}^2\bmod G^\sim_{\ln}$ iff
$\,A^1\sim A^2\bmod G^\sim_{\ln}.$

A complete list of $G^\sim_{\ln}$-inequivalent one-dimensional subalgebras
of~$A^{\cup}_{\ln}$ is exhausted by the algebras
$\langle\partial_t\rangle$, $\langle G(\chi)+\zeta I\rangle$, $\langle \lambda M+\zeta I\rangle$.
(There exist additional equivalences in $\{\langle G(\chi)+\zeta I\rangle\}$
and $\{\langle \lambda M+\zeta I\rangle\}$, which are generated by translations with respect to~$t$
and, if $\sigma_2=0$, by the Wigner time reflection $I_t$.)

\begin{theorem}\label{TheoremNSchPLogMNGroupClassification}
A complete set of inequivalent cases of $\,V$ admitting extensions of the maximal Lie invariance algebra
of equation of the form \eqref{NSchEPLogMN} is exhausted by the potentials given in Table~2.
\end{theorem}

 \setcounter{tbn}{0}
{\begin{center}\small
Table 2. Results of classification for class~\eqref{NSchEPLogMN}.
\\[1.5ex]
\setcounter{tbn}{0}
\renewcommand{\arraystretch}{1.4}
\begin{tabular}{|r|c|l|}
\hline\vspacebefore
N &$V$ &\hfill {Basis of $A_{\ln}(V)$\hfill} \\
\hline\vspacebefore
\thetbn& $V(t,x)$ & $M,\;$ $I'$\\
\hline\vspacebefore
\refstepcounter{tbn}\thetbn\label{NSchEPLogMNcase1}&
$V(x)$ & $M,\;$ $I',\;$ $\p_t$\\
\refstepcounter{tbn}\thetbn\label{NSchEPLogMNcase2}&
$v(t)x^2+iw(t)x$ & $M,\;$ $I',\;$ $G'(\chi^1),\;$ $G'(\chi^2)$ \\
\refstepcounter{tbn}\thetbn\label{NSchEPLogMNcase3}&
$i\nu x$ & $M,\;$ $I',\;$ $\p_t,\;$ $G'(1),\;$ $G'(t)$ \\
\refstepcounter{tbn}\thetbn\label{NSchEPLogMNcase4}&
$\mu^2x^2+i\nu x$ & $M,\;$ $I',\;$ $\p_t,\;$ $G'(e^{-2\mu t}),\;$ $G'(e^{2\mu t})$ \\
\refstepcounter{tbn}\thetbn\label{NSchEPLogMNcase5}&
\hspace*{-2.6mm}$-\mu^2x^2+i\nu x$ & $M,\;$ $I',\;$ $\p_t,\;$ $G'(\cos 2\mu t),\;$ $G'(\sin 2\mu t)$ \\
\hline
\end{tabular}
\\[2ex]
\parbox{140mm}{Here $v(t),w(t),\mu,\nu\!\in\!\mathbb{R}$, $(v_t,w_t)\ne(0,0)$, $\mu>0$, $\nu\ge 0$.
The functions~$\chi^1=\chi^1(t)$ and $\chi^2=\chi^2(t)$ form a fundamental system of
solutions to the ordinary differential equation $\chi_{tt}=4v\chi$.
$G'(\chi)=G(\chi)-\tilde\chi I-\sigma_1\int e^{-\sigma_2t}\tilde\chi\, dt\, M$, 
where $\tilde\chi=\int e^{\sigma_2t}w\chi\, dt$ and 
$w=\nu=\const$ in Cases 3, 4 and 5.
}
\end{center}}

\begin{proof}
Let $A_{\ln}(V)\ne A^\cap_{\ln}$,
i.e., equation~\eqref{NSchEPLogMN} has a Lie symmetry extension for the potential~$V$.
This means that $ A_{\ln}(V)$ contains an operator
$Q=c_0\p_t+G(\chi)+\lambda M+\zeta I$  with $(c_0,\chi)\ne(0,0)$.

If $c_0\ne0$ then $\langle Q\rangle\sim \langle\p_t\rangle\bmod G^\sim_{\ln}$,
i.e., we obtain Case~2.\ref{NSchEPLogMNcase1}.
The investigation of additional extensions is reduced to the next case.

\looseness=-1
If $c_0=0$ then $\langle Q\rangle\sim \langle G(\chi)+\zeta I\rangle\bmod G^\sim_{\ln}$.
It follows from~\eqref{NSchEPMNLogCaseClassifyingCondition} that the potential $V$ has the form
$V=v(t)x^2+(\tilde w(t)+iw(t))x+\tilde u(t)+iu(t)$, and $\tilde u,u,\tilde w=0\bmod G^\sim_{\ln}$.
For $(v_t,w_t)\ne(0,0)$ we have Case~2.\ref{NSchEPLogMNcase2}. The condition $v,w=\const$ results
in Cases~2.\ref{NSchEPLogMNcase3}, 2.\ref{NSchEPLogMNcase4} and~2.\ref{NSchEPLogMNcase5}
depending on the sign of~$v$, and $w=\nu$.
\end{proof}

\subsection{Power nonlinearity}\label{SectionOnPowerMNP}

The most interesting (and, at the same time, most difficult) subclass of class~\eqref{NSchEPMN}
from the point of view of group analysis is
formed by the equations with power modular nonlinearities
\begin{equation}\label{NSchEPPowerMN}
i\psi_t+\psi_{xx}+\sigma|\psi|^\gamma\psi +V(t,x)\psi=0.
\end{equation}
\looseness=-1
Here $\sigma$ and $\gamma$ are arbitrary non-zero complex and real constants, respectively, $\gamma\not=0$
and $V$ is an arbitrary complex-valued potential depending on~$t$ and~$x$.
In view of Theorem~\ref{TheoremGequivNSchEPMN},
this subclass admits extensions of the equivalence and Lie symmetry groups of its equations
in comparison with the whole class~\eqref{NSchEPMN}.

It is possible to consider the whole class~$\mathscr{P}$ of equations having the form~\eqref{NSchEPPowerMN},
where~$\gamma$ runs through $\mathbb{R}\backslash\{0\}$.
A drawback of the above approach is the need to consider the extended equivalence group
(equivalence transformations of~$x$ will depend on~$\gamma$, see Theorem~\ref{TheoremGequivNSchEPPowerMN}),
since $\mathscr{P}$ is a normalized class in the extended sense only.

In view of Corollary~\ref{CorollaryInvValueForS}, $\gamma$ is an invariant
of all admissible transformations in the superclass~\eqref{NSchEGMN}, i.e.,
equations with different values of~$\gamma$ do not transform into one another.
Therefore, it is more natural to interpret~\eqref{NSchEPPowerMN}
as a family of classes parametrized with~$\gamma$ and then to fix an arbitrary value of~$\gamma$.
We assume below that $\gamma$ is fixed and denote the class of equations~\eqref{NSchEPPowerMN}
corresponding to the fixed value~$\gamma$ as~$\mathscr{P}_\gamma$.
(See also Section~\ref{SectionOnGenMNP} where the class~$\mathscr{P}_\gamma$ is first introduced.)
It is derived from the superclass~\eqref{NSchEGMN} by imposing the conditions
$S_{\rho t}=S_{\rho x}=0$,  $(\rho S_\rho)_\rho=\gamma S_\rho$, i.e.,
\begin{equation}\label{NSchEPPowerMNaddSystemForEquivTrans}
\psi S_{\psi t}+\psi^*S_{\psi^* t}=\psi S_{\psi x}+\psi^*S_{\psi^* x}=0, \quad
(\psi\p_\psi+\psi^*\p_{\psi^*})^2S=\gamma(\psi\p_\psi+\psi^*\p_{\psi^*})S.
\end{equation}

Similarly to the previous sections, we have two ways of finding
the equivalence group~$G^\sim_{\mathscr P_\gamma}$ of the class~$\mathscr P_\gamma$
in the framework of the direct method.
In the first approach we look for all point transformations in the space of the variables
$t$, $x$, $\psi$, $\psi^*$, $S$ and $S^*$,
under which the system formed by equations~\eqref{NSchEGMNsystemForEquivTrans}
and~\eqref{NSchEPPowerMNaddSystemForEquivTrans} is invariant.
Moreover, in the same way we can easily classify all possible point transformations
in~$\mathscr{P}_\gamma$ using Note~\ref{NoteLocTransNSchEGMN}.
The second, completely equivalent, way of proceeding is
to consider $\sigma$ and $V$ as arbitrary elements instead of $S$
and to find all point symmetry transformations for the system
\begin{gather*}
i\psi_t+\psi_{xx}+\sigma|\psi|^\gamma\psi +V\psi=0,\quad
V_\psi=V_{\psi^*}=0, \quad
\sigma_t=\sigma_x=\sigma_\psi=\sigma_{\psi^*}=0.
\end{gather*}
(Under the prolongation procedure for equivalence transformations, we suppose
$\psi$ is a function of $t$ and $x$ as well as that
$\sigma$ and $V$ are functions of $t,$ $x$, $\psi$ and $\psi^*$.)

\begin{theorem}\label{TheoremGequivNSchEPPowerMN}
The class~$\mathscr P_\gamma$, where $\gamma\not=0$, is normalized.
The equivalence group $\smash{G^\sim_{\mathscr P_\gamma}}$ is formed by the transformations~\eqref{NSchEGMNeqtrans}, where
$e^\Theta=\varkappa|T_t|^{-1/\gamma}$, $\varkappa>0$.
The corresponding transformations with respect to $\sigma$ and $V$ have the form
\begin{gather*}
\tilde V=\dfrac{\hat V}{|T_t|}+
\dfrac{2T_{ttt}T_t-3T_{tt}{}^2}{16\varepsilon'T_t{}^3}x^2
+\dfrac{\varepsilon\varepsilon'}{2|T_t|^{1/2}}\left(\dfrac{X_t}{T_t}\right)_{\!t}x
+\dfrac{\Psi_t}{T_t}-\dfrac{X_t{}^2}{4T_t{}^2}+i\gamma'\dfrac{T_{tt}}{T_t{}^2},
\quad
\tilde\sigma=\dfrac{\hat\sigma}{\varkappa^\gamma}
\end{gather*}
with $\gamma'=\dfrac 1\gamma-\dfrac 14$.
\end{theorem}

\begin{corollary}\label{CorollaryNSchEPMNVanishingPotentail}
A potential~$V$ in~\eqref{NSchEPPowerMN} can be transformed to an $x$-free one iff
\begin{equation}\label{QuadraticPotential}
V=v(t)x^2+u(t)x+\tilde w(t)+iw(t),
\end{equation}
where $v$, $u$, $\tilde w$ and $w$ are real-valued functions of $t$.
In particular, if $\gamma=4$ any real-valued potential quadratic in $x$ can be made to vanish.
In the case $\gamma\not=4$ the potential~\eqref{QuadraticPotential} is equivalent to zero
iff $\,16(\gamma')^2v=2\varepsilon\gamma' w_t+w^2$.
\end{corollary}

\begin{note}
It follows from Theorem~\ref{TheoremGequivNSchEPPowerMN}
that any point transformation in~$\mathscr{P}_\gamma$
acts on $\sigma$ only as multiplication with non-zero real constants and/or complex conjugation.
Therefore, we can assume that $\sigma$ is fixed in our consideration below
under the assumption that $|\sigma|=1$ and $\sigma_2\ge 0$
and consider only $V$ as an arbitrary element.
The equivalence group $G^\sim_{\mathscr P_\gamma^\sigma}$ of the corresponding restricted class~$\mathscr P_\gamma^\sigma$
is formed by the transformations~\eqref{NSchEGMNeqtrans}
where $T_t>0$ if $\sigma_2>0$ and $\varkappa=1$.
Hereafter $\sigma_1=\mathop{\rm Re}\nolimits \sigma$, $\sigma_2=\mathop{\rm Im}\nolimits \sigma$.
\end{note}

Below we use the subscript `$\gamma$' instead of `$\mathscr P_\gamma^\sigma$' for simplicity.
Recall, however, that all objects marked in such a way may be parametrized by~$\sigma$.

The classifying condition~\eqref{NSchEGMNClassifyingCondition} with $S=\sigma\rho^\gamma+V(t,x)$ implies under splitting
with respect to~$\rho$ that $\tau_t+\gamma\zeta=0$. Therefore, we have the following statement as a
consequence of Theorem~\ref{TheoremNSchEGMNOperators}.

\begin{lemma}\label{TheoremNSchEPPowerMNOperators}
Any operator~$Q$ from the maximal Lie invariance algebra $A_{\gamma}(V)$
of equation~\eqref{NSchEPPowerMN} with an arbitrary potential $V$
can be represented in the form $Q=D^\gamma(\tau)+G(\chi)+\lambda M$, where
$
D^\gamma(\tau)=D(\tau)-\gamma^{-1}\tau_tI.
$
Moreover, the coefficients of $Q$ have to satisfy the classifying condition
\begin{equation}\label{NSchEPMNPowerCaseClassifyingCondition}
\tau V_t+\left(\frac12\tau_tx+\chi\right)V_x+\tau_tV=
\frac18\,\tau_{ttt}x^2+\frac12\chi_{tt}x+\lambda_t+i\gamma'\tau_{tt}.
\end{equation}
The kernel of maximal Lie invariance groups of equations from class~\eqref{NSchEPMN} is
$G^\cap_\gamma=\hat G^\cap_{\mathscr{S}}$,
and its Lie algebra is $A^\cap_\gamma=\langle M \rangle$.
\end{lemma}

Let us apply the framework of normalized classes for obtaining the complete group classification of
the class~$\mathscr{P}_\gamma$.

The set $A^{\cup}_{\gamma}=\{Q=D^\gamma(\tau)+G(\chi)+\lambda M\}$
is an (infinite-dimensional) Lie algebra under the usual Lie bracket of vector fields.
For any $Q\in A^{\cup}_{\gamma}$ where $(\tau,\chi)\ne(0,0)$
we can find $V$ satisfying condition~\eqref{NSchEPMNPowerCaseClassifyingCondition},
i.e., $A^{\cup}_{\gamma}=\langle\,\bigcup_{V} A_{\gamma}(V)\,\rangle$.
Moreover, the space reflection belongs to the point symmetry groups of certain equations from $\mathscr P_\gamma^\sigma$.
The same statement is true for the Wigner time reflection if $\sigma$ is real.
Therefore, $\mathscr P_\gamma^\sigma$ is a strongly normalized class of differential equations.
(Again this was a reason for introducing the class $\mathscr P_\gamma^\sigma$
since the whole class $\mathscr P_\gamma$ is not strongly normalized.)

The group~$G^\sim_\gamma$ acts on $A^{\cup}_\gamma$
and on the set of equations of the form~\eqref{NSchEPMNPowerCaseClassifyingCondition}
and, therefore, generates equivalence relations in these sets.
The automorphism group generated on $A^{\cup}_\gamma$ and
the non-trivial equivalence group generated on the set of equations~\eqref{NSchEPMNPowerCaseClassifyingCondition}
are isomorphic to $G^\sim_\gamma/\hat G^\cap_{\mathscr{S}}$.
$A^\cap_\gamma=\langle M \rangle$ coincides with the center of the algebra~$A^{\cup}_\gamma$
and is invariant with respect to $G^\sim_\gamma$.

Let $A^1$ and $A^2$ be the maximal Lie invariance algebras of some equations from~$\mathscr{P}_\gamma^\sigma$,
and ${\cal V}^i=\{\,V\,|\,A_\gamma(V)=A^i\},$ $i=1,2.$
Then ${\cal V}^1\sim {\cal V}^2\!\!\mod\!G^\sim_\gamma$ iff
$\,A^1\sim A^2\!\!\mod\!G^\sim_\gamma.$

A complete list of $G^\sim_\gamma$-inequivalent one-dimensional subalgebras
of~$A^{\cup}_\gamma$ is exhausted by the algebras
$\langle\partial_t\rangle$, $\langle\partial_x\rangle$, $\langle tM\rangle$, $\langle M\rangle$.

\begin{corollary}\label{lemma.vtvx0}
If $\,A_\gamma(V)\ne A^\cap_\gamma\,$ then $\,V_tV_x=0\bmod G^\sim_\gamma$.
\end{corollary}

\begin{proof}
Under these assumptions there exists
an operator $Q=D^\gamma(\tau)+G(\chi)+\lambda M\in A(V)$
which does not belong to $\langle M\rangle.$
Condition~\eqref{NSchEPMNPowerCaseClassifyingCondition} implies $(\tau,\chi)\ne(0,0).$
Therefore, $\;\langle Q\rangle\sim \langle\partial_t\rangle$ or
$\langle\partial_x\rangle\bmod G^\sim_\gamma,$
i.e., $\,V_tV_x=0\bmod G^\sim_\gamma$.
\end{proof}

\begin{theorem}\label{theorem.gc.pcshe}
A complete set of inequivalent cases of $\,V$
admitting extensions of the maximal Lie invariance algebra
of equations from~$\mathscr{P}_\gamma$ is exhausted by the potentials given in Table~3.
\end{theorem}

\setcounter{tbn}{0}
{\begin{center}
Table 3. Results of classification.
Here $W(t),\nu,\alpha,\beta\in\mathbb{R},$ $(\alpha,\beta)\ne(0,0).$
\\[1.5ex] \footnotesize
\setcounter{tbn}{0}
\renewcommand{\arraystretch}{1.3}
\begin{tabular}{|r|c|l|}
\hline\vspacebefore
N &$V$ &\hfill {Basis of $A_\gamma(V)$\hfill} \\
\hline\vspacebefore
\thetbn& $V(t,x)$ & $M\;$ \\
\hline\vspacebefore
\refstepcounter{tbn}\thetbn\label{pcsheV1}& $iW(t)$ & $M,\;$ $\p_x,\;$ $G(t)$\\[0.8ex]
\refstepcounter{tbn}\thetbn\label{pcsheV2}&
$2i\dfrac{\gamma't+\nu}{t^2+1}$
& $M,\;$ $\p_x,\;$ $G(t),\;$ $D^\gamma(t^2+1)$\\[1.3ex]
\refstepcounter{tbn}\thetbn\label{pcsheV3}& $i\nu t^{-1}\!,$\quad
$\nu\ne0$, $2\gamma'$
& $M,\;$ $\p_x,\;$ $G(t),\;$ $D^\gamma(t)$\\
\refstepcounter{tbn}\thetbn\label{pcsheV4}& $i$
& $M,\;$ $\p_x,\;$ $G(t),\;$ $\p_t$ \\
\refstepcounter{tbn}\thetbn\label{pcsheV5}& $0,\:$ $\gamma\ne4$
& $M,\;$ $\p_x,\;$ $G(t),\;$ $\p_t,\;$ $D^\gamma(t)$\\
& $\phantom{0,\:}$ $\gamma=4$
& $M,\;$ $\p_x,\;$ $G(t),\;$ $\p_t,\;$ $D^\gamma(t),\;$ $D^\gamma(t^2)$\\
\refstepcounter{tbn}\thetbn\label{pcsheV6}& $V(x)$
& $M,\;$ $\p_t$\\
\refstepcounter{tbn}\label{pcsheV7}\thetbn&
$(\alpha+i\beta)x^{-2},\:$ $\gamma\ne4$ & $M,\;$ $\p_t,\;$ $D^\gamma(t)$\\
&$\phantom{(\alpha+i\beta)x^{-2},\:}$ $\gamma=4$ & $M,\;$ $\p_t,\;$ $D^\gamma(t),\;$ $D^\gamma(t^2)$\\
\hline
\end{tabular}
\end{center}}

\begin{proof}
In view of Corollary~\ref{lemma.vtvx0}, for proving Theorem~\ref{theorem.gc.pcshe} it
is sufficient to study two cases: $V_x=0$ and $V_t=0$.

The subclass of equations from~$\mathscr P_\gamma^\sigma$ with potentials satisfying
the additional assumption $V_x=0$, i.e., $V=V(t)$, is also strongly normalized.
Its equivalence group is generated by transformations of the form~\eqref{NSchEGMNeqtrans},
where $T=(a_1t+a_0)/(b_1t+b_0)$, $X=c_1T+c_0$, $e^\Theta=|T_t|^{-1/\gamma}$,
and $\Psi$ is an arbitrary smooth function of $t$.
$a_i$, $b_i$ and $c_i$ are arbitrary constants such that $\,a_1b_0-b_1a_0\not=0$.
We can make the real part of any $x$-free potential  vanish with the above equivalence transformations.
The more restricted subclass~$\mathscr{W}$ of equations from~$\mathscr P_\gamma^\sigma$ with purely
imaginary $x$-free potentials
is also strongly normalized, and moreover, its equivalence group~$G^\sim_{\mathscr W}$
is a finite-parameter group and is the subgroup of the above equivalence group which is singled out by
the constraint~$4\Psi-c_1^2T=\const$.
Let us note that $G^\sim_{\mathscr W}=\hat G^\cap_{\mathscr W}\times\hat G^\mathrm{ext}_{\mathscr W}$.
Here $\hat G^\cap_{\mathscr W}$ is a normal subgroup of~$G^\sim_{\mathscr W}$.
It is formed by transformations from~$G^\sim_{\mathscr W}$ with~$T=1$, which do not change potentials,
and it is isomorphic to the kernel~$G^\cap_{\mathscr W}$ of maximal Lie symmetry groups of equations from~$\mathscr W$.
The subgroup~$\smash{\hat G^\mathrm{ext}_{\mathscr W}}$ consists of the transformations with~$X=\Psi=0$ and is isomorphic to~$SL(2,\mathbb{R})$.
Therefore, the problem of group classification in the case of $x$-free potentials is reduced to the
subgroup analysis of~$SL(2,\mathbb{R})$.

If equation~\eqref{NSchEPPowerMN} with a stationary potential admits an extension of its Lie symmetry group then
it is equivalent either to one with an $x$-free potential or to one with the potential $V=V(x)$
satisfying the condition
\[
xV_x+2V=c_2x^2+c_1x+\tilde c_0+ic_0, \quad \textrm{where}\quad
c_2,c_1,\tilde c_0,c_0=\mathrm{const}\in\mathbb{R}.
\]
For more details we refer
the reader to~\cite{Popovych&Ivanova&Eshraghi2004Gamma}.
\end{proof}

\begin{note}\label{NoteOnClassificationOfPMNSchEs}
There exists a discrete equivalence transformation $\mathcal T$
for the set of potentials $i\nu t^{-1},$ $\nu\in\mathbb{R},$
which has the form~\eqref{NSchEGMNeqtrans}
with $T=-t^{-1}$, $X=0$, $\Psi=0$, $e^\Theta=|t|^{2/\gamma}$.
It transforms $\nu$ in the following way: $\nu\to 2\gamma'-\nu.$
For the cases from Table~3 to be completely inequivalent,
we have  to assume additionally that $\nu\ge\gamma'$
(or $\nu\le\gamma'$) in Case~3.3.
Since $I_t\in G^\sim_\gamma$ if $\sigma_2=0$, then we can assume  $\nu\ge0$ in Case~3.2 and $\beta\ge0$ in Case~3.7.
Moreover, $\mathcal T$ is a discrete symmetry transformation
for Case~3.\ref{pcsheV3} ($\nu=\gamma'$) and,
as a limit of the continuous transformations is generated by the operator $D^\gamma(t^2+1),$
for Case~3.\ref{pcsheV2}.
\end{note}

\begin{note}
Only for power nonlinearities (the class~$\mathscr{P}_\gamma$ with $\gamma\ne0$)
any extension of the maximal Lie invariance algebra is equivalent to that whose maximal Lie invariance algebra is a subalgebra of
the finite-dimensional part $\mathrm{sch}(1,1)$ of the maximal Lie invariance algebra of the $(1+1)$-dimensional free Schr\"odinger equation.
All equations from the class~$\mathscr F'$, i.e., the Schr\"odinger equations of the general form
$i\psi_t+\psi_{xx}+F(t,x,\psi,\psi^*)=0$, which additionally are invariant with respect to subalgebras of
$\mathrm{sch}(1,1)$, were constructed in~\cite{BoyerSharpWinternitz1976}.
Note that in the multi-dimensional case the analogous equations do not exhaust even the extension cases
in the class of cubic Schr\"odinger equations with potential, see the next section.
\end{note}

\section{Group classification of $\boldsymbol{(1+2)}$-dimensional cubic\\ Schr\"odinger equations
with potentials}\label{SectionOnGroupClassificationOfCSchEP12D}

Consider the class~$\mathscr C$ of $(1+2)$-dimensional cubic Schr\"odinger 
equations with potentials, having the general form~\eqref{CSchEP12D}.
In what follows the indices $a$ and $b$ run from $1$ to $2$, and 
we use summation convention over repeated indices.

\begin{theorem}\label{TheoremGequivCSchEP12D}
The class~$\mathscr C$ is strongly normalized.
The equivalence group~$G^{\sim}_{\mathscr C}$ of this class is formed by the transformations
\begin{equation}\label{CSchEP12Deqtrans}
\arraycolsep=0ex\begin{array}{l}\displaystyle
\tilde t=T, \quad
\tilde x=|T_t|^{1/2}Ox+X,
\\[1.5ex]
\tilde \psi=|T_t|^{-1/2}\exp\left(\dfrac i8\dfrac{T_{tt}}{|T_t|}\,x_ax_a+
\dfrac i2\dfrac{\varepsilon_T X^b_t}{|T_t|^{1/2}}\,O^{ba}x_a +i\Psi \right)\!\hat\psi,
\\[3.5ex]
\tilde V=\dfrac{\hat V}{|T_t|}+
\dfrac{2T_{ttt}T_t-3T_{tt}{}^2}{16\varepsilon_TT_t{}^3}x_ax_a
+\dfrac{\varepsilon_T}{2|T_t|^{1/2}}\left(\dfrac{X^b_t}{T_t}\right)_{\!t}O^{ba}x_a
+\dfrac{\Psi_t}{T_t}-\dfrac{X^a_tX^a_t}{4T_t{}^2}.
\end{array}\end{equation}
Here $T$, $X=(X^1,X^2)$ and $\Psi$ are arbitrary smooth real-valued functions of $t$,
$T_t\ne 0$, $\varepsilon_T=\mathop{\rm sign}\nolimits T$ and
$O=(O^{ab})$~is an arbitrary constant two-dimensional orthogonal matrix.
\end{theorem}

\begin{note}\label{NoteGequivCSchEP12D}{\rm
The equivalence group~$G^{\sim}_{\mathscr C}$ is generated by the continuous family of transformations 
of the form~\eqref{CSchEP12Deqtrans}, where $T_t>0$ and $\det O=1$, and two discrete transformations:
the space reflection~$I_a$ for a fixed~$a$ 
($\tilde t=t,$ $\tilde x_a=-x_a,$ $\tilde x_b=-x_b,$ $b\ne a$, $\tilde \psi=\psi,$ $\tilde \Phi=\Phi$)
and the Wigner time reflection~$I_t$
($\tilde t=-t,$ $\tilde x=x,$ $\tilde \psi=\psi^*,$ $\tilde \Phi=\Phi^*$). 
}\end{note}

Any operator~$Q$ from the maximal Lie invariance algebra $A(V)$
of the equation~\eqref{CSchEP12D} with an arbitrary element $V$ 
can be represented in the form $Q=D(\tau)+\kappa J+G(\bar\sigma)+\chi M$, where
\begin{gather*}
D(\tau)=\tau\p_t+\dfrac12\,\tau_tx_a\p_a+\dfrac18\,\tau_{tt}x_ax_aM-\dfrac12\tau_tI,\\ 
G(\bar\sigma)=\sigma^a\p_a+\dfrac12\sigma^a_tx_aM,\quad 
J=x_1\partial_2-x_2\partial_1,
\end{gather*}
$\kappa$ is a constant, 
$\tau$, $\bar\sigma=(\sigma^1,\sigma^2)$ and $\chi$ are smooth real-valued functions of $t$ and 
\begin{gather*}
M=i(\psi\p_{\psi}-\psi^*\p_{\psi^*}),\quad
I=\psi\p_{\psi}+\psi^*\p_{\psi^*}.
\end{gather*}
Moreover, the coefficients of $Q$ have to satisfy the classifying condition
\begin{equation}\label{CSchEP12DClassifyingCondition}
\tau V_t+\left(\frac12\tau_tx_a+\kappa\varepsilon_{ab}x_b+\sigma^a\right)V_a+\tau_tV=
\frac18\,\tau_{ttt}x_ax_a++\frac12\sigma^a_{tt}x_a+\chi_t, 
\end{equation}
where $\varepsilon_{11}=\varepsilon_{22}=0$, $\varepsilon_{21}=-\varepsilon_{12}=1$,

The operators $D(\tau)$, $J$, $G(\bar\sigma)$ and $\chi M$,
where $\tau$, $\bar\sigma=(\sigma^1,\sigma^2)$ and $\chi$ run through the whole set of smooth functions of~$t$,
generate an infinite-dimensional Lie algebra~$A^{\cup}_{\mathscr C}$ under the usual Lie bracket of vector fields.
The non-zero commutation relations between the basis operators of~$A^{\cup}_{\mathscr C}$ are the following ones:
\begin{gather*}
[D(\tau^1),D(\tau^2)]=D(\tau^1\tau^2_t-\tau^2\tau^1_t),\qquad
[D(\tau),G(\bar\sigma)]=G\Bigl(\tau\bar\sigma_t-\frac12\tau_t\bar\sigma\Bigr),\\
[D(\tau),\chi M]=\tau\chi_t M,\qquad
[J,G(\bar\sigma)]=G(\sigma^2,-\sigma^1),\\
[G(\bar \sigma^1),G(\bar\sigma^2)]=\frac12(\sigma^{1a}\sigma^{2a}_t-\sigma^{2a}\sigma^{1a}_t)M.
\end{gather*}
Therefore, the subspaces $\langle\chi M\rangle$, $\langle G(\bar\sigma),\chi M\rangle$, 
$\langle J,G(\bar\sigma),\chi M\rangle$ and $\langle D(\tau),G(\bar\sigma),\chi M\rangle$ are ideals of~$A^{\cup}_{\mathscr C}$. 
The subspace $\langle D(\tau)\rangle$ is a subalgebra of~$A^{\cup}_{\mathscr C}$.

Assuming $V$ to be arbitrary and splitting~\eqref{CSchEP12DClassifyingCondition}
with respect to $V$, $V_t$ and $V_a$, we obtain that
the Lie algebra of the kernel~$G^\cap_{\mathscr C}$ of maximal Lie invariance groups of equations
from the class~$\mathscr C$ is $A^\cap_{\mathscr C}=\langle M \rangle.$
The complete group~$G^\cap_{\mathscr C}$ coincides with the projection, to $(t,x,\psi)$,
of the normal subgroup~$\hat G^\cap_{\mathscr C}$
of~$G^\sim_{\mathscr C}$, which include the transformations~\eqref{CSchEP12Deqtrans}
acting as identity on the arbitrary element~$V$ (i.e., $T=t$ and $X=\Psi_t=0$).

For any fixed value of the arbitrary element~$V$, 
the classifying condition~\eqref{CSchEP12DClassifyingCondition} implies, in particular, 
a linear system of ordinary differential equations in the coefficients~$\tau$, $\sigma^a$ and~$\chi$ 
of the general form 
\begin{gather*}
\tau_{ttt}=g^{00}\tau_t+g^{01}\tau+g^{0,a+1}\sigma^a+g^{04}\chi,\\
\sigma^a_{tt}=g^{a0}\tau_t+g^{a1}\tau+g^{a,a+1}\sigma^a+g^{a4}\chi,\\
\chi_t=g^{40}\tau_t+g^{41}\tau+g^{4,a+1}\sigma^a+g^{44}\chi,
\end{gather*}
where $g^{pq}$, $p,q=1,\dots,4$, are functions of~$t$ which are determined by~$V$. 
Therefore, for any~$V$ we obviously have 
\begin{gather*}
\dim A(V)\leqslant9,\quad
A(V)\cap\langle\chi M\rangle=\langle M\rangle,\quad 
\dim A(V)\cap\langle G(\bar\sigma),\chi M\rangle\leqslant5,\\
\dim \mathop{\rm pr}\nolimits_{\langle D(\tau)\rangle} A(V)\cap\langle D(\tau),G(\bar\sigma),\chi M\rangle\leqslant3
\end{gather*}

We first list all possible inequivalent cases of Lie symmetry extensions for the class~$\mathscr C$, 
then briefly describe how to derive the presented classification list, and finally provide additional explanations on the extension cases. 
A detailed proof will be the subject of a forthcoming paper. 

In what follows, $U$ is an arbitrary complex-valued function of its arguments or an arbitrary complex constant, 
the other functions and constants take real values, 
\[
|x|=\sqrt{x_1^2+x_2^2},\quad \phi=\arctan x_2/x_1,\quad
\omega=x_1\cos t+x_2\sin t,\quad \theta=-x_1\sin t+x_2\cos t.
\]
Each item of the classification list consists of a value of the arbitrary element~$V$ 
and a basis of the maximal Lie invariance algebra $A(V)$ of the corresponding equation~\eqref{CSchEP12D}. 
More precisely, the presented algebras are maximal for the values of~$V$ 
which are $G^\sim_{\mathscr C}$-inequivalent to that possessing additional extensions of the Lie symmetry algebra. 
In those cases where the associated inequivalence conditions admit a simple formulation, they are explicitly indicated 
after the classification list. 

\begin{enumerate}\setcounter{enumi}{-1}
\renewcommand{\cc}{\item}

\cc
$V=V(t,x)\colon\quad M$.

\cc
$V=U(x_1,x_2)\colon$\; $M,\ D(1)$.

\cc
$V=U(\omega,\theta)\colon$\; $M,\ D(1)+J$.

\cc
$V=|x|^{-2}U(\zeta)$, 
$\zeta=\phi-2\beta\ln|x|$, $\beta>0\colon$\;
$M,\ D(1),\ D(t)+\beta J$.

\cc
$V=|x|^{-2}U(\phi)\colon$\;
$M,\ D(1),\ D(t),\ D(t^2)$.

\cc
$V=U(t,|x|)+\varepsilon\phi$, 
$\varepsilon\in\{0,1\}\colon$\;
$M,\ J+\varepsilon tM$.

\cc
$V=U(|x|)+\varepsilon\phi$, 
$\varepsilon\in\{0,1\}\colon$\;
$M,\ J+\varepsilon tM,\ D(1)$.

\cc
$V=|x|^{-2}U$, 
$U\ne0\colon$\;
$M,\ J,\ D(1),\ D(t),\ D(t^2)$.

\cc
$V=U(t,x_1)\colon$\;
$M,\ G(0,1),\ G(0,t)$.

\cc
$V=U(\zeta)$, $\zeta=x_1\colon$\;
$M,\ G(0,1),\ G(0,t),\ D(1)$.

\cc
$V=t^{-1}U(\zeta)$, $\zeta=|t|^{-1/2}x_1\colon$\;
$M,\ G(0,1),\ G(0,t),\ D(t)$.

\cc
$V=(t^2+1)^{-1}U(\zeta)$, $\zeta=(t^2+1)^{-1/2}x_1\colon$\;
$M,\ G(0,1),\ G(0,t),\ D(t^2+1)$.

\cc
$V=x_1^{-2}U$, 
$U\ne0\colon$\;
$M,\ G(0,1),\ G(0,t),\ D(1),\ D(t),\ D(t^2)$.

\cc
$V=U(t,\omega)+\frac14(h_tt-h)h^{-1}\theta^2+h_th^{-1}\omega\theta$, 
$h=h(t)\ne0\colon$\;
$M,\ G(h\cos t,h\sin t)$.


\cc
$V=U(\omega)+\frac14(\alpha^2-1)\theta^2+\alpha\omega\theta$, $\alpha\ne0\colon$\;
$M,\ D(1)+J,\ G(e^{\alpha t}\cos t,e^{\alpha t}\sin t)$.

\cc
$V=U(\omega)-\frac14\theta^2+\beta\theta\colon$\;
$M,\ D(1)+J,\ G(\cos t,\sin t)+\beta tM$.

\cc
$V=h^{ab}(t)x_ax_b+ih^{00}(t)$, $h^{12}=h^{21}$, ($h^{12}\ne0\vee h^{11}\ne h^{22})\colon$\;
$M,\ G(\bar\sigma^p),\ p=1,\dots,4$,\\ 
where $\{\bar\sigma^p=(\sigma^{p1}(t),\sigma^{p2}(t)),\ p=1,\dots,4\}$ is a fundamental set of solutions 
of the system $\bar\sigma_{tt}=H\bar\sigma$, $H=(h^{ab})$.

\cc
$V=\frac14\alpha x_1^2+\frac14\beta x_2^2+i\gamma$, $\alpha\ne\beta\colon$\;
$M,\ G(\sigma^{11},0),\ G(\sigma^{21},0),\ G(0,\sigma^{12}),\ G(0,\sigma^{22}),\ D(1)$,\\ 
where the functions $\sigma^{11}$ and $\sigma^{21}$ (resp. $\sigma^{12}$ and $\sigma^{22}$) of~$t$
form a fundamental set of solutions 
of the equation $\sigma_{tt}=\alpha\sigma$ (resp. $\sigma_{tt}=\beta\sigma$).

\cc
$V=\frac14\alpha\omega^2+\frac14\beta\theta^2+i\gamma$, $\alpha\ne\beta\colon$\\
$M,\ G(\sigma^{p1}\cos t+\sigma^{p2}\sin t,-\sigma^{p1}\sin t+\sigma^{p2}\cos t),\ p=1,\dots,4,\ D(1)+J$,\\ 
where $\{\bar\sigma^p=(\sigma^{p1}(t),\sigma^{p2}(t)),\ p=1,\dots,4\}$ is a fundamental set of solutions 
of the system $\sigma^1_{tt}-2\sigma^2_t=(\alpha+1)\sigma^1$, $\sigma^2_{tt}+2\sigma^1_t=(\beta+1)\sigma^2$.

\cc
$V=iW(t)\colon$
$M,\ J,\ G(1,0),\ G(t,0),\ G(0,1),\ G(0,t)$.

\cc
$V=i\colon$
$M,\ J,\ G(1,0),\ G(t,0),\ G(0,1),\ G(0,t),\ D(1)$.

\cc
$V=i\nu t^{-1}$, $\nu>0\colon$
$M,\ J,\ G(1,0),\ G(t,0),\ G(0,1),\ G(0,t),\ D(t)$.

\cc
$V=2i\nu(t^2+1)^{-1}$, $\nu>0\colon$
$M,\ J,\ G(1,0),\ G(t,0),\ G(0,1),\ G(0,t),\ D(t^2+1)$.

\cc
$V=0\colon$
$M,\ J,\ G(1,0),\ G(t,0),\ G(0,1),\ G(0,t),\ D(1),\ D(t),\ D(t^2)$.

\end{enumerate}

The general case (Case~0) is included in the list for completeness. 

To single out the main cases of the classification, we introduce the $G^\sim_{\mathscr C}$-invariant values
\begin{gather*}
r_\sigma=\rank\{\bar\sigma(t)\mid \exists \chi(t)\colon G(\bar\sigma)+\chi M\in A(V)\}\in\{0,1,2\},\\
r_J=\dim(\mathop{\rm pr}\nolimits_{\langle J\rangle} A(V)\cap\langle J,G(\bar\sigma),\chi M\rangle)\in\{0,1\}
\end{gather*} 
depending on the arbitrary element~$V$.

Let $r_\sigma=r_J=0$.
The set $S^\tau(V)=\{\tau(t)\mid \exists \kappa,\bar\sigma(t),\chi(t)\colon D(\tau)+\kappa J+G(\bar\sigma)+\chi M\in A(V)\}$ 
is a linear space which is closed with respect to the usual Poisson bracket of functions, 
i.e., $\tau^1\tau^2_t-\tau^2\tau^1_t\in S^\tau(V)$ if $\tau^1,\tau^2\in S^\tau(V)$.
$\dim S^\tau(V)\leqslant3<\infty$. 
Lie's classification of realizations of finite-dimensional Lie algebras by vector fields on a line implies that, 
up to local diffeomorphisms of~$t$, $S^\tau(V)\in\{\langle1\rangle,\langle1,t\rangle,\langle1,t,t^2\rangle\}$.
Hence, up to $G^\sim_{\mathscr C}$-equivalence, 
a basis of the algebra $A(V)$ may include, additionally to the common operator~$M$, one of the sets of operators 
$\{D(1)\}$, $\{D(1)+J\}$, $\{D(1),D(t)+J\}$, $\{D(1),D(t),D(t^2)\}$ 
which correspond to Cases~1, 2, 3 and 4. 
The algebra whose basis is presented in Case~3 (resp.\ Case~4) is the maximal Lie invariance algebra 
if and only if $\beta\ne0$ and $U_\zeta\ne0$ (resp.\ $U_\phi\ne0$).

Suppose that $r_\sigma=0$ and $r_J=1$. 
Then the algebra $A(V)$ contains the operator $J+G(\bar\sigma)+\chi M$, where $\bar\sigma=0\bmod G^\sim_{\mathscr C}$. 
The general form of~$V$ for which the associated equation from the class~$\mathscr C$ admits the operator $J+\chi M$ 
is $V=U(t,|x|)+h(t)\phi$. 
We also have $(|x|^{-1}U_{|x|})_{|x|}\ne0$ or $h\ne0$ since otherwise $r_\sigma>0$.
The subclasses of~$\mathscr C$ corresponding to the sets $\{V=U(t,|x|)+h(t)\phi\mid (|x|^{-1}U_{|x|})_{|x|}\ne0\}$ 
and $\{V=U(t,|x|)+h(t)\phi\mid h\ne0\}$ as well as their union and their intersection are normalized. 
The equivalence groups of these subclasses are isomorphic and induced by the transformations of the form~\eqref{CSchEP12Deqtrans} 
with $X=0$. 
The possible inequivalent extensions are exhausted by Cases~5--7.

Analogously, Cases 8--15 are singled out by the conditions $r_\sigma=1$ and $r_J=0$. 
In Cases~\mbox{8--12} the following $G^\sim_{\mathscr C}$-invariant condition is additionally satisfied: 
If an operator $G(\bar\sigma)+\chi M$ belongs to~$A(V)$ then $\sigma^1_t\sigma^2=\sigma^1\sigma^2_t$, 
i.e., $\bar\sigma$ is proportional to a tuple of constants. 
The further consideration in this specific case is simplified due to the fact that 
any fixed operator with this property is $G^\sim_{\mathscr C}$-equivalent to the operator~$G(0,1)$.
If $G(0,1)\in A(V)$ then the potential~$V$ has the form $V=U(t,x_1)$ and also $G(0,t)\in A(V)$. 
The subclass of~$\mathscr C$ corresponding to the set $\{V=U(t,x_1)\mid U_{111}\ne0\}$ is normalized. 
Its equivalence group is induced by the transformations of the form~\eqref{CSchEP12Deqtrans} 
with $T$ fractional linear in~$t$ and $X=c_1T+c_0$, where $c_1$ and $c_0$ are arbitrary constants. 
Therefore, the inequivalent extensions in this subclass are exhausted by the inequivalent subalgebras 
of the algebra $\langle D(1),D(t),D(t^2)\rangle$, 
namely, by the subalgebras $\langle D(1)\rangle$, $\langle D(t)\rangle$, $\langle D(t^2+1)\rangle$, $\langle D(1),D(t)\rangle$
and  $\langle D(1),D(t),D(t^2)\rangle$.
The basis elements presented in Cases 9--11 give the maximal Lie invariance algebras 
for the corresponding values of the arbitrary element~$V$ if and only if $U_{\zeta\zeta\zeta}\ne0$ and $(\zeta^2U)_\zeta\ne0$.
Cases 13--15 arise under the condition that $\sigma^1_t\sigma^2\ne\sigma^1\sigma^2_t$ 
for an operator $G(\bar\sigma)+\chi M$ from~$A(V)$.
Then $\dim A(V)\cap\langle G(\bar\sigma),\chi M\rangle=2$ since $r_\sigma=1$.
Moreover, each operator from the complement of~$\langle G(\bar\sigma),\chi M\rangle$ in~$A(V)$ 
has to nontrivially involve both an operator~$D(\tau)$ and the operator~$J$, i.e., 
only one-dimensional extensions are possible. 

The values of $(r_\sigma,r_J)$ not considered so far are $(2,0)$, $(1,1)$ and $(2,1)$. 
In view of the commutation relation $[J,G(\bar\sigma)]=G(\sigma^2,-\sigma^1)$, 
the case $r_\sigma=r_J=1$ is impossible and we necessarily have $r_\sigma=2$. 
The condition $r_\sigma=2$ is equivalent to the potential~$V$ being a quadratic polynomial in~$x_1$ and~$x_2$ 
with coefficients depending on~$t$. 
More precisely,  
\[
V=h^{ab}(t)x_ax_b+h^{0b}(t)x_b+ih^{00}(t)+\tilde h^{00}(t), 
\] 
where all the functions~$h$ are real-valued, $h^{12}=h^{21}$. 
Denote by~$\mathscr C_{\rm q}$ the subclass of equations from~$\mathscr C$ with these values of the arbitrary element~$V$. 
The subclass~$\mathscr C_{\rm q}$ is normalized, 
its equivalence group coincides with the equivalence group~$G^\sim_{\mathscr C}$ of the whole class~$\mathscr C$ 
and it is partitioned by the conditions $r_J=0$ and $r_J=1$ 
into two normalized subclasses~$\mathscr C_{\rm q}^0$ and~$\mathscr C_{\rm q}^1$ 
with the same equivalence group as that admitted by~$\mathscr C$ and~$\mathscr C_{\rm q}$.
In terms of the functions~$h$, the subclass~$\mathscr C_{\rm q}^1$ (resp.\ $\mathscr C_{\rm q}^0$) 
is singled out by the condition $h^{12}=h^{21}=0\wedge h^{11}=h^{22}$ (resp.\ its negation).

The regular classification case for the class~$\mathscr C_{\rm q}^0$ is Case~16. 
Only one-dimensional extensions necessarily involving operators of the form~$D(\tau)$ are possible in this class 
(Cases~17 and~18).

Any equation from the class~$\mathscr C_{\rm q}^1$ is $G^\sim_{\mathscr C}$-equivalent to 
an equation from the same class whose potential has the form $V=iW(t)$ with a real-valued function~$W=W(t)$. 
The subclass~$\smash{\hat{\mathscr C}_{\rm q}^1}$ corresponding to the set $\{V=iW(t)\}$ is normalized. 
The regular classification case for this subclass is Case~19. 
The equivalence group of~$\smash{\hat{\mathscr C}_{\rm q}^1}$ is induced 
by the transformations of the form~\eqref{CSchEP12Deqtrans} 
with $T$ fractional linear in~$t$, $X^a=c^a_1T+c^a_0$ and $\Psi=\frac14c^a_1c^a_1T+d_0$, 
where $c^a_1$, $c^a_0$ and $d_0$ are arbitrary constants. 
Therefore, the inequivalent extensions in~$\smash{\hat{\mathscr C}_{\rm q}^1}$ are exhausted by the inequivalent subalgebras 
of the algebra $\langle D(1),D(t),D(t^2)\rangle$, 
namely, by the subalgebras $\langle D(1)\rangle$, $\langle D(t)\rangle$, $\langle D(t^2+1)\rangle$
and  $\langle D(1),D(t),D(t^2)\rangle$ (Cases 20, 21, 22 and 23, respectively).
The algebra $\langle D(1),D(t)\rangle$ does not give a proper extension since 
the invariance of an equation from~$\smash{\hat{\mathscr C}_{\rm q}^1}$ with respect to $D(1)$ and $D(t)$ 
implies vanishing $V$ and, therefore, the invariance with respect to~$D(t^2)$.
There exists a discrete equivalence transformation $\mathcal T$
for the set of potentials $i\nu t^{-1},$ $\nu\in\mathbb{R},$
which has the form~\eqref{CSchEP12Deqtrans} with $T=-t^{-1}$, $X^a=0$, $\Psi=0$.
Its action on $\nu$ is equivalent to alternating the sign of~$\nu$. 
Analogously, the Wigner time reflection $I_t$ induces a change of sign of~$\nu$ in the potential $i\nu(t^2+1)^{-1}$.
Therefore, up to $G^\sim_{\mathscr C}$-equivalence we can set $\nu>0$ in Cases~21 and~22.

\begin{note}
Different potentials corresponding to the same case of the classification may be equivalent. 
This kind of equivalence can be used to additionally constrain parameters. 
In particular, we can set $h^{11}+h^{22}=0$ in Case 16. 
In Cases~17 and~18 the parameters~$\alpha$ and~$\beta$ can be scaled up to permutation, 
i.e., we can assume that either $\alpha=1$ and~$\beta\in[-1;1)$ or $\alpha=-1$ and~$\beta\in(-1;1)$.
\end{note}

\section{Possible applications}\label{SectionOnApplicationsOfEquivTransForNschEs}

Classical Lie symmetries and point equivalent transformations are traditionally used for the construction of
exact solutions by different methods~\cite{Anderson&Fels&Torre2000,Olver1986,Ovsiannikov1982}, 
finding conservation laws~\cite{Olver1986} etc.
The implementation of this machinery for nonlinear Schr\"odinger equations 
with potentials and modular nonlinearities will be the subject of a forthcoming paper \cite{Kunzinger&Popovych2006}.
Here we only give a few examples on applications of equivalence transformations with the aim of
establishing connections between known results and the derivation of new ones. 
(Note that we use scaling of variables and notations which may differ from those employed
in the cited papers.)

\begin{example}
Consider cubic Schr\"odinger equations with stationary potentials, 
i.e., equations~\eqref{NSchEPPowerMN} with $\gamma=2$, $V=V(x)$ and $\sigma$ real, so that it can be put equal to 1.
After the Madelung transformation $\psi=\sqrt{R}\,e^{i\varphi}$ in these equations, up to inessential signs
we obtain the so-called Madelung fluid equations
\[
i\varphi_t+(\varphi_x)^2+\frac1{\sqrt{R}}(\sqrt{R})_{xx}=R+V(x), \quad R_t+(R\varphi_x)_x=0.
\]
Here $R$ and $\varphi$ are the new real-valued unknown functions of $t$ and~$x$.
In~\cite{BaumannNonnenmacher1987} (see also \mbox{\cite[p.~214]{BlumanKumei1989}})
the Lie symmetry properties of the Madelung fluid equations were investigated
for the case $V=c_1x$. It was found that the equation with $V=c_1x$ possesses a five-dimensional maximal Lie invariance algebra.
Then the Lie symmetry operations were used for the standard Lie reduction of the equations 
under consideration and for constructing their exact solutions.
At the same time, it is evident from Theorem~\ref{TheoremGequivNSchEPPowerMN} and Corollary~\ref{CorollaryNSchEPMNVanishingPotentail}
that the equation $i\psi_t+\psi_{xx}+|\psi|^2\psi+c_1x\psi=0$ is reduced to the usual cubic Schr\"odinger equation with $V=0$
by means of the transformation~\eqref{NSchEGMNeqtrans}
with $T=t$, $X=-c_1t^2$, $\varepsilon=1$ and $\Psi=c_1^2t^3/3$, i.e., the transformation
\[
\tilde t=t, \quad
\tilde x=x-c_1t^2,\quad
\tilde \psi=\psi \exp\left(ic_1tx +\frac i3 c_1^2t^3 \right)\!.
\]
In the term of the probability density~$R$ and the phase function~$\varphi$ the transformation is
$\tilde t=t$,
$\tilde x=x-c_1t^2$,
$\tilde R=R$,
$\tilde \varphi=\varphi+c_1tx +c_1^2t^3/3$.
Therefore, all results on Lie symmetry and exact solutions of the Madelung equations with $V=c_1x$ can be
derived from the analogous results for the usual cubic Schr\"odinger equation.
\end{example}

\begin{example}
The cubic Schr\"odinger equations with heterogeneity and dielectric loss have the general form
\begin{equation}\label{EqCSchEsWithHeterogeneityAndDielectricLoss}
i\psi_t+\psi_{xx}+R(t,x)|\psi|^2\psi+V(t,x)\psi=0,
\end{equation}
where $V$ and $R$ are complex- and real-valued functions of~$t$ and~$x$.
The heterogeneity is implemented via the dependence of~$V$ and~$R$ on~$t$ and~$x$ 
($t$ is a spatial variable in some applications).
The imaginary part of the potential~$V$ represents the dielectric loss.
In~\cite{BroadbridgeGodfrey1991} all the equations~\eqref{EqCSchEsWithHeterogeneityAndDielectricLoss}
possessing third-order generalized symmetries were found.
They are exhausted by the equations with $R$ depending only on~$t$ and
\[
V=\frac{(\chi^2)_{tt}}{4\chi^2}x^2
+i\left(\frac{\chi_t}\chi+\frac12\frac{R_t}{R}\right)+\zeta x+\varsigma,
\]
where $\chi$, $\zeta$ and $\varsigma$ are arbitrary functions of~$t$.
It was proved in~\cite{BroadbridgeGodfrey1991} that any such equation is reduced by point transformations to
the standard cubic Schr\"odinger equations ($R=\pm1$, $V=0$). This evidently follows from results of our paper.
Indeed, class~\eqref{EqCSchEsWithHeterogeneityAndDielectricLoss} is a subclass of the class~$\mathscr S$
from Section~\ref{SectionOnNestedNormalizedClassesOfNSchEs}. For equations from this subclass, $S=R(t,x)|\psi|^2+V(t,x)$.
Since the class~$\mathscr S$ is normalized, any admissible transformation in class~\eqref{EqCSchEsWithHeterogeneityAndDielectricLoss}
is generated by a transformation from the equivalence group~$G^{\sim}_{\mathscr S}$ of the class~$\mathscr S$.
Moreover, the transformations from~$G^{\sim}_{\mathscr S}$, rewritten in terms of~$R$ and~$V$ instead of~$S$, form
the equivalence group of class~\eqref{EqCSchEsWithHeterogeneityAndDielectricLoss}.
Therefore, this class itself is normalized as well.
It is obvious that any equation~\eqref{EqCSchEsWithHeterogeneityAndDielectricLoss} possessing third-order generalized symmetries
is reduced to the standard cubic Schr\"odinger equations by the transformation~\eqref{NSchEGMNeqtrans} with
the functions $T$, $X$, $\Psi$ and $\Theta$ of~$t$, determined by the equations
$T_t=\chi^{-4}$, $(\chi^4X_t)_t=-2\chi^2\zeta$, $\Psi_t-\chi^4X_t{}^2/4=-\varsigma$, $e^{\Theta}=\chi^2|R|^{1/2}$.
Another strategy is to carry out the transformation step by step.
At~first, the parameter-function~$R$ is rendered equal to $\pm1$ by the transformation $\tilde\psi=\psi|R|^{1/2}$.
The resulting equation belongs to the class~$\mathscr P_2$ and can
be further simplified according to Theorem~\ref{TheoremGequivNSchEPPowerMN} and Corollary~\ref{CorollaryNSchEPMNVanishingPotentail}.
\end{example}

\begin{example}\label{ExampleOnWittkopf} 
As mentioned in the introduction, in the works~\cite{Wittkopf2004,Wittkopf&Reid2000,Wittkopf&Reid2001} 
Wittkopf and Reid compute Lie symmetries of vector nonlinear Schr\"odinger systems of the form 
\begin{equation}\label{EqVectorNSchE}
i\boldsymbol\psi_t+\triangle\boldsymbol\psi+S(t,\boldsymbol x,\boldsymbol\psi\cdot\boldsymbol\psi^*)\boldsymbol\psi=\boldsymbol0
\end{equation}
to test their programs for the symbolic solution of overdetermined systems of PDEs.  
Here $S$ is a real-valued smooth function of its arguments, 
and its derivative with respect to $\boldsymbol\psi\cdot\boldsymbol\psi^*$ does not vanish. 
The tuples $\boldsymbol x$ and $\boldsymbol\psi$ consist of $n$ and $m$ components, respectively. 

For $m=n=1$ the systems become equations and form the (normalized) subclass~$\mathscr S'$ of the class~$\mathscr S$ 
defined in Section~\ref{SectionOnNestedNormalizedClassesOfNSchEs}. 
The additional auxiliary condition on the arbitrary element~$S$ is $\mathop{\rm Im}S=0$.
It was proved in~\cite{Wittkopf2004,Wittkopf&Reid2001} that the dimensions of the Lie invariance algebras of equations from the class~$\mathscr S'$ 
do not exceed six; and the equations possessing six-dimensional Lie invariance algebras correspond to the arbitrary elements of the form 
$S=\sigma|\psi|^4+v(t)x^2+u(t)x+w(t)$ (and, therefore, belong to the class~$\mathscr P_4$).
Hereafter $v$, $u$ and $w$ are arbitrary (real-valued) functions of~$t$ and $\sigma$ is an arbitrary (real) nonzero constant. 
In view of Corollary~\ref{CorollaryNSchEPMNVanishingPotentail} of our paper, 
any such arbitrary element is point-transformation equivalent to $S=\pm|\psi|^4$. 
Since the classes~$\mathscr S'$ and~$\mathscr P_4$ are normalized, 
the point-transformation equivalence coincides with the equivalences generated by the corresponding equivalence groups.
Therefore, the first statement of Theorem~1 from~\cite{Wittkopf&Reid2001} can be reformulated in the following way. 
\emph{Any equation from class~$\mathscr S'$, possessing a Lie invariance algebra of the maximal
dimension for this class (i.e., dimension~6), is equivalent to the quintic Schr\"odinger equation.} 
The above theorem contains also the statement that the equations associated with the arbitrary elements 
of the form $S=\sigma|\psi|^\gamma+u(t)x+w(t)$, where $\gamma\ne0,4$, admit five-dimensional Lie invariance algebras. 
According to Corollary~\ref{CorollaryNSchEPMNVanishingPotentail}, they are reduced to the potential-free equation with $S=\pm|\psi|^\gamma$. 

For general values of~$m$ and~$n$, the arbitrary elements of the form 
\[
S=\sigma|\boldsymbol\psi|^\gamma+v(t)|\boldsymbol x|^2+\boldsymbol u(t)\cdot\boldsymbol x+w(t)
\]
were considered in~\cite{Wittkopf2004,Wittkopf&Reid2000,Wittkopf&Reid2001}. 
The problem of finding Lie symmetries of the corresponding systems was proposed as a benchmark due to 
the evident dependence of its computational complexity on $m$ and $n$.  
In calculations for low values of $n$ and $m$, in particular, the equations with $\gamma=4/n$ were singled out 
as the ones having Lie invariance algebras of the maximal possible dimension in this class.
It was conjectured that this dimension equals $m^2+n(n+3)/2+3$. 

Note that in the case $\gamma=4/n$ the parameter-functions $v$, $\boldsymbol u$ and~$w$ can be made to vanish 
by means of an obvious generalization of the transformations from Theorem~\ref{TheoremGequivNSchEPPowerMN} of our paper. 
The system $i\boldsymbol\psi_t+\triangle\boldsymbol\psi+|\boldsymbol\psi|^{4/n}\boldsymbol\psi=\boldsymbol0$ 
admits a Lie invariance algebra~$\mathfrak g$ isomorphic to $\mathrm{sch}_n\oplus\mathrm{su}_m$, 
where $\mathrm{su}_m$ is the special unitary algebra of order~$m$ and 
$\mathrm{sch}_n$ is the Lie invariance algebra of the corresponding single equation. 
(The particular case $n=2$ is presented in~\cite{SciarrinoWinternitz1997}.)
The algebra~$\mathfrak g$ is generated by the operators 
\begin{gather*}
\p_t, \quad 
t\p_t+\frac12x_j\p_{x_j}-\frac n4 I, \quad 
t^2\p_t+tx_j\p_{x_j}-\frac n2 tI+\frac14 x_jx_jM, \\
\p_{x_j}, \quad 
t\p_{x_j}+\frac12 x_jM, \quad 
x_k\p_{x_j}-x_j\p_{x_k}, \ k<j, \\[1ex] 
\psi^a\p_{\psi^b}-\psi^b\p_{\psi^a}+\psi^{a*}\p_{\psi^{b*}}-\psi^{b*}\p_{\psi^{a*}},\ a<b, \\[1.5ex] 
i\psi^a\p_{\psi^b}+i\psi^b\p_{\psi^a}-i\psi^{a*}\p_{\psi^{b*}}-i\psi^{b*}\p_{\psi^{a*}},\ a\leqslant b.
\end{gather*}
where $I:=\psi^a\p_{\psi^a}+\psi^{a*}\p_{\psi^{a*}}$, $M:=i\psi^a\p_{\psi^a}-i\psi^{a*}\p_{\psi^{a*}}$. 
The indices~$j$ and~$k$ run from~1 to~$n$. The indices~$a$ and~$b$ run from~1 to~$m$ and summation
over repeated indices is understood.

This implies that the above conjecture on the dimension is true. 
\end{example}

\begin{example}
In~\cite{Sakhnovich2006} the matrix cubic Schr\"odinger equations with the potentials $V=\frac i2(t+b)^{-1}$, i.e., the equations of the form 
\begin{equation}\label{EqSakhnovich2006}
i\Psi_t+\Psi_{xx}\pm \Psi\Psi^*\Psi+\frac i{2(t+b)}\Psi=0
\end{equation}
were investigated. Here $\Psi=\Psi(t,x)$ is an $m_1\times m_2$ matrix-function and 
$\Psi^*$ is the corresponding conjugate (i.e., complex conjugate and transpose) matrix. 
The constant~$b$ is complex.  
The~zero curvature representations of equations~\eqref{EqSakhnovich2006} were found and families of exact solutions 
were constructed by extending a version of the B\"acklund--Darboux transformation first introduced in~\cite{Sakhnovich1993}. 
In the scalar case ($m_1=m_2=1$), equations~\eqref{EqSakhnovich2006} belong to the class $\mathscr{P}_2$ 
(i.e., the class~\eqref{NSchEPPowerMN} with $\gamma=2$). 
Moreover, in this case for real values of~$b$ the potential $V$ satisfies the conditions of Corollary~\ref{CorollaryNSchEPMNVanishingPotentail} 
and, therefore, can be made to vanish by the transformation described in Note~\ref{NoteOnClassificationOfPMNSchEs} up to translations.
The explicit form of this transformation is 
\[
\tilde t=-\frac1{t+b}, \quad \tilde x=\frac x{t+b}, \quad 
\tilde \Psi= (t+b)\exp\left(-\frac i4\frac{x^2}{t+b}\right)\Psi.
\]
This transformation can be formally treated if $\mathop{\rm Im}b\ne0$.
Also, it extends to the general matrix case without changing its form.  
Therefore, the transformation of equations~\eqref{EqSakhnovich2006} to the standard matrix cubic equation 
can be considered as a way of deriving results of~\cite{Sakhnovich2006} from analogous results of~\cite{Sakhnovich1993}. 
See also~\cite{Sakhnovich2008} for a generalization of this example. 
\end{example}

\section{Conclusion}

The approach to group classification problems, proposed in~\cite{Popovych&Eshraghi2004Mogran,Popovych2006b} and developed in this paper, 
seems to be quite universal. 
It is based on the notion of normalized classes of differential equations, which can be considered as a core 
for the further enhancement of group classification methods. 

Depending on normalization properties of classes of differential equations, 
different strategies of group classification can be implemented. 
For a normalized class, the group classification problem is reduced to subgroup analysis of its equivalence group 
(or to subalgebra analysis of the corresponding algebra in the infinitesimal approach). 
No modifications of the classical formulation of group classification problems are necessary.
A non-normalized class can be embedded into a normalized class which, possibly, is not minimal among the normalized superclasses. 
Another way is to partition the non-normalized class into a family of normalized subclasses and then to classify each subclass separately. 
In fact, both strategies are simultaneously applied in our paper to 
the class~$\mathscr V$ of $(1+1)$-dimensional nonlinear Schr\"odinger equations with modular nonlinearities and potentials.
The normalized superclass is the class~$\mathscr S$. The partition considered is formed by the subclasses of equations with logarithmical, 
power and general nonlinearities. 
If a partition into normalized subclasses is difficult to construct due to the complicated structure of 
the set of admissible transformations, conditional equivalence groups and additional equivalence transformations may be involved in 
the group classification. 

Employing the machinery of normalized classes, we are enabled to effectively pose and solve
new kinds of classification problems in classes of differential equations, 
e.g., to classify conditional equivalence groups or to describe the corresponding sets of admissible transformations.

The investigation of admissible transformations and normalization properties of classes of multidimensional partial differential equations 
is much more complicated than in the case of two independent variables. 
Nevertheless such investigations are possible and constitute an effective tool for
solving group classification problems in the multidimensional case. 
The group classification of $(1+2)$-dimensional cubic Schr\"odinger equations with potentials 
has been presented in this paper as an example for the applicability of classification technique based on normalization properties 
in the case of more than two independent variables. 

An interesting subject for further study is to derive results on existence and uniqueness of solutions
of some boundary or initial value problems from known ones by means of 
equivalence transformations~\cite{Carles2002} or to prove existence and uniqueness for similarity solutions~\cite{KavianWeissler1994}.

\subsection*{Acknowledgements}

The authors are grateful to V.~Boyko, N.~Ivanova, A.~Nikitin, D.~Popovych, A.~Sakhnovich and A.~Sergyeyev 
for productive and helpful discussions.
MK was supported by START-project Y237 of the Austrian Science Fund.
The research of ROP was supported by the Austrian Science Fund (FWF), Lise Meitner project M923-N13 and project P20632.

\end{document}